\documentclass[longauth]{aa}

\usepackage{graphicx}
\usepackage{natbib}
\usepackage{scalerel}
\usepackage{dcolumn}
\usepackage{bm}
\usepackage[rightcaption]{sidecap}
\sidecaptionvpos{figure}{t}
\usepackage{tabularx}

\usepackage{tikz}
\usetikzlibrary{patterns.meta}

\usepackage[version=4]{mhchem}
\usepackage[nolist,printonlyused]{acronym}
\usepackage[table]{xcolor}
\usepackage{txfonts}

\usepackage[pdfencoding=auto,psdextra]{hyperref}
\hypersetup{
    colorlinks=true,
    linkcolor=blue,
    filecolor=magenta,      
    urlcolor=blue,
    citecolor=blue
}
\usepackage{xurl}
\makeatletter
\renewcommand*\aa@pageof{, page \thepage{} of \pageref*{LastPage}}
\makeatother

\usepackage[utf8]{inputenc}
\usepackage[switch, modulo]{lineno}
              
\usepackage{euclid}

\usepackage[nameinlink,capitalise]{cleveref}
\crefname{section}{Sect.}{Sects.}
\Crefname{section}{Section}{Sections}
\crefname{figure}{Fig.}{Figs.}
\Crefname{figure}{Figure}{Figures}
\crefname{equation}{Eq.}{Eqs.}
\Crefname{equation}{Equation}{Equations}
\crefname{table}{Table}{Tables}
\crefname{appendix}{Appendix}{Appendices}
\Crefname{appendix}{Appendix}{Appendices}

\newcommand{\CalF}{{\tt CalF}}
\newcommand{\CALBLOCK}{{\tt CALBLOCK}}
\newcommand{\CALPRODUCT}{{\tt CALPRODUCT}}

\definecolor{lightgreen}{HTML}{CCFFCC}
\definecolor{lightblue}{HTML}{CCCCFF}
\definecolor{lightyellow}{HTML}{FFFFCC}

\makeatletter
\renewcommand*\aa@pageof{, page \thepage{} of \pageref*{LastPage}}
\newcommand{\itemlabel}[1]{
  \phantomsection
  \protected@edef\@currentlabel{#1}
  \label{#1}
  \textbf{#1}
}
\makeatother

\begin{document}

\acrodef{AA}{azimuth angle}
\acrodef{ACE}{Advanced Composition Explorer}
\acrodef{AOI}{angle of incidence}
\acrodef{AR}{anti-reflection}
\acrodef{ASIC}{application specific integrated circuit}
\acrodef{BFE}{brighter-fatter effect}
\acrodef{CaLA}{camera-lens assembly}
\acrodef{CalCD}{calibration concept document}
\acrodef{CCD}{charge-coupled device}
\acrodef{CMU}{compensation mechanism unit}
\acrodef{CoLA}{corrector-lens assembly}
\acrodef{CDS}{Correlated Double Sampling}
\acrodef{CFC}{cryo-flex cable}
\acrodef{CFHT}{Canada-France-Hawaii Telescope}
\acrodef{CFRP}{carbon-fibre reinforced plastic}
\acrodef{CGH}{computer-generated hologram}
\acrodef{CIL}{charge-injection line}
\acrodef{CME}{coronal mass ejection}
\acrodef{CNES}{Centre National d'Etude Spacial}
\acrodef{CPPM}{Centre de Physique des Particules de Marseille}
\acrodef{CPU}{central processing unit}
\acrodef{CR}{cosmic ray}
\acrodef{CTE}{coefficient of thermal expansion}
\acrodef{CME}{coronal mass ejection}
\acrodef{CSST}{Chinese Space-station Survey Telescope}
\acrodef{CTI}{charge-transfer inefficiency}
\acrodef{CVZ}{continuous viewing zone}
\acrodef{DCU}{Detector Control Unit}
\acrodef{DES}{Dark Energy Survey}
\acrodef{DPU}{Data Processing Unit}
\acrodef{DS}{Detector System}
\acrodef{EAF}{Euclid Auxiliary Field}
\acrodef{EDS}{Euclid Deep Survey}
\acrodef{EDF}{Euclid Deep Fields}
\acrodef{EE}{encircled energy}
\acrodef{EA}{Enterprise Architect}
\acrodef{ESA}{European Space Agency}
\acrodef{ESP}{Emission of Solar Protons}
\acrodef{ECSS}{European Cooperation for Space Standardization}
\acrodef{EWS}{Euclid Wide Survey}
\acrodef{EUDF}{Euclid Ultra Deep Field}
\acrodef{EDF-N}{Euclid Deep Field North}
\acrodef{EDF-S0}{Euclid Deep Field South}
\acrodef{EDF-S}{EDF South}
\acrodef{EMDS}{Euclid Medium Deep Survey}
\acrodef{FDIR}{Fault Detection, Isolation and Recovery}
\acrodef{FGS}{fine guidance sensor}
\acrodef{FOM}{figure of merit}
\acrodef{FoM3}{folding mirror 3}
\acrodef{FoV}{field of view}
\acrodef{FPA}{focal-plane array}
\acrodef{FWA}{filter-wheel assembly}
\acrodef{FWC}{full-well capacity}
\acrodef{FWHM}{full width at half maximum}
\acrodef{GDPRD}{Ground Data Processing Requirement Document}
\acrodef{GOES}{Geostationary Operational Environmental Satellites}
\acrodef{GCR}{Galactic cosmic ray}
\acrodef{GWA}{grism-wheel assembly}
\acrodef{H2RG}{HAWAII-2RG}
\acrodef{HST}{\textit{Hubble} Space Telescope}
\acrodef{ICRS}{International Celestial Reference System}
\acrodef{IPC}{interpixel capacitance}
\acrodef{IP2I}{Institut de Physique des 2 Infinis de Lyon}
\acrodef{IPSV}{intrapixel sensitivity variation}
\acrodef{JWST}{{\em James Webb} Space Telescope}
\acrodef{IAD}{ion-assisted deposition}
\acrodef{ICU}{Instrument Control Unit}
\acrodef{IDL}{initial drain level}
\acrodef{IPC}{inter-pixel capacitance}
\acrodef{ISES}{International Space Environmental Services}
\acrodef{JWST}{\textit{James Webb} Space Telescope}
\acrodef{KiDS}{Kilo-Degree Survey}
\acrodef{LAM}{Laboratoire d'Astrophysique de Marseille}
\acrodef{LED}{light-emitting diode}
\acrodef{LMC}{Large Magellanic Cloud}
\acrodef{LSB}{low surface brightness}
\acrodef{LSF}{large-scale flat}
\acrodef{MACC}{multiple accumulated}
\acrodef{M2M}{secondary-mirror mechanism}
\acrodef{MER}{merge data set}
\acrodef{MLI}{multi-layer insulation}
\acrodef{MMU}{Mass Memory Unit}
\acrodef{MPE}{Max-Planck-Institut für extraterrestrische Physik}
\acrodef{MOCD}{Mission Operation Concept Document}
\acrodef{MOS}{metal-oxide semi-conductor}
\acrodef{MRD}{Mission Requirements Document}
\acrodef{MPIA}{Max-Planck-Institut für Astronomie}
\acrodef{NA}{numerical aperture}
\acrodef{NASA}{National Aeronautic and Space Administration}
\acrodef{NDI}{normalised detector irradiance}
\acrodef{NEP}{North Ecliptic Pole}
\acrodef{NEXUS}{North Ecliptic Pole Extragalactic Unified Survey}
\acrodef{NIEL}{non-ionising energy loss}
\acrodef{JPL}{NASA Jet Propulsion Laboratory}
\acrodef{MZ-CGH}{multi-zonal computer-generated hologram}
\acrodef{NI-CU}{NISP calibration unit}
\acrodef{NI-OA}{near-infrared optical assembly}
\acrodef{NI-GWA}{NISP Grism Wheel Assembly}
\acrodef{NIR}{near-infrared}
\acrodef{NISP}{Near-Infrared Spectrometer and Photometer}
\acrodef{NOAA}{National Oceanic and Atmospheric Administration}
\acrodef{PARMS}{plasma-assisted reactive magnetron sputtering}
\acrodef{RMS}{root mean square}
\acrodef{PLM}{payload module}
\acrodef{PN}{planetary nebula}
\acrodef{PRNU}{pixel-response non-uniformity}
\acrodef{PTFE}{polytetrafluoroethylene}
\acrodef{PV}{performance verification}
\acrodef{PWM}{pulse-width modulation}
\acrodef{PSF}{point-spread function}
\acrodef{QCM}{quartz crystal microbalance}
\acrodef{QE}{quantum efficiency}
\acrodef{QF}{quality factor}
\acrodef{PDC}{phase-diversity campaign}
\acrodef{PDF}{probability density function}
\acrodef{PERD}{Payload Elements Requirements Document}
\acrodef{PTC}{photon transfer curve}
\acrodef{ROE}{readout electronics}
\acrodef{ROI}{region of interest}
\acrodef{ROIC}{readout-integrated circuit}
\acrodef{ROS}{reference observing sequence}
\acrodef{RSD}{reference survey design} 
\acrodef{RTN}{random telegraph noise}
\acrodef{SAA}{solar aspect angle}
\acrodef{SCA}{sensor chip array}
\acrodef{SCE}{sensor chip electronic}
\acrodef{SCS}{sensor chip system}
\acrodef{SGS}{Science Ground Segment}
\acrodef{SGPS}{Solar and Galactic Proton Sensor}
\acrodef{SHS}{Shack-Hartmann sensor}
\acrodef{SNR}[S/N]{signal-to-noise ratio}
\acrodef{STIS}{Space Telescope Imaging Spectrograph}
\acrodef{SED}{spectral energy distribution}
\acrodef{SiC}{silicon carbide}
\acrodef{SEP}{South Ecliptic Pole}
\acrodef{SSN}{Sunspot number}
\acrodef{STIX}{Spectrometer/Telescope for Imaging X-rays}
\acrodef{STP}{serial trap pumping}
\acrodef{SolO}{Solar Orbiter}
\acrodef{SOPS}{spacecraft operations}
\acrodef{TP}{trap pumping}
\acrodef{UTR}{up-the-ramp}
\acrodef{SVM}{service module}
\acrodef{VIS}{visible imager}
\acrodef{WD}{white dwarf}
\acrodef{WCS}{world coordinate system}
\acrodef{WFC3}{Wide Field Camera 3}
\acrodef{WFE}{wavefront error}
\acrodef{ZP}{zero point}

\title{Euclid Data Release 1 (DR1)} 
\subtitle{Calibration strategy and observations}
   
\newcommand{\orcid}[1]{} 
\author{Euclid Collaboration: M.~Schirmer\orcid{0000-0003-2568-9994}\thanks{\email{schirmer@mpia.de}}\inst{\ref{aff1}}
\and K.~Paterson\orcid{0000-0001-8340-3486}\inst{\ref{aff1}}
\and A.~Alavi\orcid{0000-0002-8630-6435}\inst{\ref{aff2}}
\and B.~Altieri\orcid{0000-0003-3936-0284}\inst{\ref{aff3}}
\and P.~N.~Appleton\orcid{0000-0002-7607-8766}\inst{\ref{aff2}}
\and N.~Auricchio\orcid{0000-0003-4444-8651}\inst{\ref{aff4}}
\and H.~Aussel\orcid{0000-0002-1371-5705}\inst{\ref{aff5}}
\and R.~Azzollini\orcid{0000-0002-0438-0886}\inst{\ref{aff6}}
\and P.~Battaglia\orcid{0000-0002-7337-5909}\inst{\ref{aff4}}
\and R.~Barbier\inst{\ref{aff7}}
\and L.~Betancourt\orcid{0009-0008-4616-1527}\inst{\ref{aff8}}
\and D.~R.~Boxhoorn\inst{\ref{aff9}}
\and J.~Brinchmann\orcid{0000-0003-4359-8797}\inst{\ref{aff10},\ref{aff11},\ref{aff12}}
\and G.~Buenadicha\inst{\ref{aff3}}
\and G.~P.~Candini\orcid{0000-0001-9481-8206}\inst{\ref{aff6}}
\and J.~M.~Carrasco\orcid{0000-0002-3029-5853}\inst{\ref{aff13},\ref{aff14},\ref{aff15}}
\and P.~Casenove\orcid{0009-0006-6736-1670}\inst{\ref{aff16}}
\and S.~Conseil\orcid{0000-0002-3657-4191}\inst{\ref{aff17},\ref{aff18}}
\and Y.~Copin\orcid{0000-0002-5317-7518}\inst{\ref{aff18}}
\and J.-C.~Cuillandre\orcid{0000-0002-3263-8645}\inst{\ref{aff5}}
\and J.~E.~Davies\orcid{0000-0002-5079-9098}\inst{\ref{aff1}}
\and S.~Davini\orcid{0000-0003-3269-1718}\inst{\ref{aff19}}
\and S.~E.~Deustua\orcid{0000-0003-2823-360X}\inst{\ref{aff20}}
\and J.~Dinis\orcid{0000-0001-5075-1601}\inst{\ref{aff21},\ref{aff22}}
\and C.~Dolding\orcid{0009-0003-7199-6108}\inst{\ref{aff6}}
\and X.~Dupac\inst{\ref{aff3}}
\and S.~Dusini\orcid{0000-0002-1128-0664}\inst{\ref{aff23}}
\and J.~Mart\'{i}n-Fleitas\orcid{0000-0002-8594-569X}\inst{\ref{aff24}}
\and M.~Frailis\orcid{0000-0002-7400-2135}\inst{\ref{aff25}}
\and E.~Franceschi\orcid{0000-0002-0585-6591}\inst{\ref{aff4}}
\and P.-A.~Frugier\inst{\ref{aff5}}
\and M.~Fumana\orcid{0000-0001-6787-5950}\inst{\ref{aff26}}
\and L.~Gabarra\orcid{0000-0002-8486-8856}\inst{\ref{aff27}}
\and W.~Gillard\orcid{0000-0003-4744-9748}\inst{\ref{aff28}}
\and P.~G\'omez-Alvarez\orcid{0000-0002-8594-5358}\inst{\ref{aff29},\ref{aff3}}
\and C.~Grenet\inst{\ref{aff30}}
\and J.~Gracia-Carpio\orcid{0000-0003-4689-3134}\inst{\ref{aff31}}
\and A.~Gregorio\orcid{0000-0003-4028-8785}\inst{\ref{aff32},\ref{aff25},\ref{aff33}}
\and F.~Grupp\inst{\ref{aff31},\ref{aff34}}
\and L.~Guzzo\orcid{0000-0001-8264-5192}\inst{\ref{aff35},\ref{aff36},\ref{aff37}}
\and W.~G.~Hartley\orcid{0000-0002-4563-5101}\inst{\ref{aff38}}
\and S.~Hemmati\orcid{0000-0003-2226-5395}\inst{\ref{aff2}}
\and O.~Herent\inst{\ref{aff30}}
\and J.~Hoar\inst{\ref{aff3}}
\and H.~Hoekstra\orcid{0000-0002-0641-3231}\inst{\ref{aff39}}
\and W.~Holmes\orcid{0009-0007-8554-4646}\inst{\ref{aff40}}
\and K.~Jahnke\orcid{0000-0003-3804-2137}\inst{\ref{aff1}}
\and J.~J.~E.~Kajava\orcid{0000-0002-3010-8333}\inst{\ref{aff41},\ref{aff42},\ref{aff43}}
\and F.~Keller\inst{\ref{aff44}}
\and R.~Kohley\inst{\ref{aff3}}
\and B.~Kubik\orcid{0009-0006-5823-4880}\inst{\ref{aff18}}
\and M.~K\"ummel\orcid{0000-0003-2791-2117}\inst{\ref{aff34}}
\and K.~Kuijken\orcid{0000-0002-3827-0175}\inst{\ref{aff39}}
\and C.~Laigle\orcid{0009-0008-5926-818X}\inst{\ref{aff30}}
\and R.~Laureijs\inst{\ref{aff9}}
\and V.~Le~Brun\orcid{0000-0002-5027-1939}\inst{\ref{aff45}}
\and G.~Libet\inst{\ref{aff16}}
\and P.~Liebing\orcid{0009-0005-8043-7150}\inst{\ref{aff6}}
\and T.~Maciaszek\inst{\ref{aff16}}
\and R.~J.~Massey\orcid{0000-0002-6085-3780}\inst{\ref{aff46}}
\and D.~C.~Masters\orcid{0000-0001-5382-6138}\inst{\ref{aff47}}
\and H.~J.~McCracken\orcid{0000-0002-9489-7765}\inst{\ref{aff30}}
\and E.~Medinaceli\orcid{0000-0002-4040-7783}\inst{\ref{aff4}}
\and J.~Mendes\inst{\ref{aff48}}
\and L.~Miller\orcid{0000-0002-3376-6200}\inst{\ref{aff27}}
\and A.~Mora\orcid{0000-0002-1922-8529}\inst{\ref{aff49}}
\and P.~W.~Morris\orcid{0000-0002-5186-4381}\inst{\ref{aff8}}
\and S.~Mottet\inst{\ref{aff30}}
\and R.~Nakajima\orcid{0009-0009-1213-7040}\inst{\ref{aff50}}
\and D.~Neumann\orcid{0009-0006-3926-6756}\inst{\ref{aff39}}
\and H.~N.~Nguyen-Kim\inst{\ref{aff30}}
\and J.~W.~Nightingale\orcid{0000-0002-8987-7401}\inst{\ref{aff51}}
\and K.~Okumura\inst{\ref{aff52}}
\and S.~Paltani\orcid{0000-0002-8108-9179}\inst{\ref{aff38}}
\and W.~J.~Percival\orcid{0000-0002-0644-5727}\inst{\ref{aff53},\ref{aff54},\ref{aff55}}
\and G.~Polenta\orcid{0000-0003-4067-9196}\inst{\ref{aff56}}
\and M.~Portaluppi\inst{\ref{aff44}}
\and G.~D.~Racca\orcid{0000-0002-9883-8981}\inst{\ref{aff39},\ref{aff44}}
\and M.~Radovich\orcid{0000-0002-3585-866X}\inst{\ref{aff57}}
\and I.~Risso\orcid{0000-0003-2525-7761}\inst{\ref{aff58},\ref{aff19},\ref{aff36}}
\and E.~Romelli\orcid{0000-0003-3069-9222}\inst{\ref{aff25}}
\and M.~Ruffenach\orcid{0000-0001-8772-7330}\inst{\ref{aff16}}
\and S.~Awan\inst{\ref{aff6}}
\and R.~Scaramella\orcid{0000-0003-2229-193X}\inst{\ref{aff59}}
\and C.~Scarlata\orcid{0000-0002-9136-8876}\inst{\ref{aff60}}
\and M.~Schmidt\inst{\ref{aff48}}
\and M.~Scodeggio\inst{\ref{aff26}}
\and A.~Secroun\orcid{0000-0003-0505-3710}\inst{\ref{aff28}}
\and G.~Seidel\orcid{0000-0003-2907-353X}\inst{\ref{aff1}}
\and A.~D.~Short\inst{\ref{aff44}}
\and C.~Sirignano\orcid{0000-0002-0995-7146}\inst{\ref{aff61},\ref{aff23}}
\and J.~Skottfelt\orcid{0000-0003-1310-8283}\inst{\ref{aff62}}
\and L.~C.~Smith\orcid{0000-0002-3259-2771}\inst{\ref{aff63}}
\and F.~Soldano\inst{\ref{aff30}}
\and S.~Szmolka\inst{\ref{aff44}}
\and M.~Talia\orcid{0000-0003-4352-2063}\inst{\ref{aff64},\ref{aff4}}
\and A.~N.~Taylor\inst{\ref{aff65}}
\and I.~Tereno\orcid{0000-0002-4537-6218}\inst{\ref{aff21},\ref{aff22}}
\and E.~A.~Valentijn\inst{\ref{aff9}}
\and R.~Vavrek\inst{\ref{aff3}}
\and L.~M.~Gaspar~Venancio\orcid{0000-0002-9634-2982}\inst{\ref{aff44}}
\and G.~Verdoes~Kleijn\orcid{0000-0001-5803-2580}\inst{\ref{aff9}}
\and A.~Zacchei\orcid{0000-0003-0396-1192}\inst{\ref{aff25},\ref{aff66}}
\and S.~Andreon\orcid{0000-0002-2041-8784}\inst{\ref{aff36}}
\and C.~Baccigalupi\orcid{0000-0002-8211-1630}\inst{\ref{aff66},\ref{aff25},\ref{aff33},\ref{aff67}}
\and M.~Baldi\orcid{0000-0003-4145-1943}\inst{\ref{aff68},\ref{aff4},\ref{aff69}}
\and A.~Balestra\orcid{0000-0002-6967-261X}\inst{\ref{aff57}}
\and S.~Bardelli\orcid{0000-0002-8900-0298}\inst{\ref{aff4}}
\and A.~N.~Belikov\inst{\ref{aff9},\ref{aff70}}
\and R.~Bender\orcid{0000-0001-7179-0626}\inst{\ref{aff31},\ref{aff34}}
\and A.~Biviano\orcid{0000-0002-0857-0732}\inst{\ref{aff25},\ref{aff66}}
\and A.~Bonchi\orcid{0000-0002-2667-5482}\inst{\ref{aff56}}
\and E.~Branchini\orcid{0000-0002-0808-6908}\inst{\ref{aff58},\ref{aff19},\ref{aff36}}
\and M.~Brescia\orcid{0000-0001-9506-5680}\inst{\ref{aff71},\ref{aff72}}
\and S.~Camera\orcid{0000-0003-3399-3574}\inst{\ref{aff73},\ref{aff74},\ref{aff75}}
\and G.~Ca\~nas-Herrera\orcid{0000-0003-2796-2149}\inst{\ref{aff39}}
\and V.~Capobianco\orcid{0000-0002-3309-7692}\inst{\ref{aff75}}
\and C.~Carbone\orcid{0000-0003-0125-3563}\inst{\ref{aff26}}
\and J.~Carretero\orcid{0000-0002-3130-0204}\inst{\ref{aff76},\ref{aff77}}
\and F.~J.~Castander\orcid{0000-0001-7316-4573}\inst{\ref{aff78},\ref{aff15}}
\and M.~Castellano\orcid{0000-0001-9875-8263}\inst{\ref{aff59}}
\and G.~Castignani\orcid{0000-0001-6831-0687}\inst{\ref{aff4}}
\and S.~Cavuoti\orcid{0000-0002-3787-4196}\inst{\ref{aff72},\ref{aff79}}
\and A.~Cimatti\inst{\ref{aff80}}
\and C.~Colodro-Conde\inst{\ref{aff81}}
\and G.~Congedo\orcid{0000-0003-2508-0046}\inst{\ref{aff65}}
\and C.~J.~Conselice\orcid{0000-0003-1949-7638}\inst{\ref{aff82}}
\and L.~Conversi\orcid{0000-0002-6710-8476}\inst{\ref{aff83},\ref{aff3}}
\and F.~Courbin\orcid{0000-0003-0758-6510}\inst{\ref{aff14},\ref{aff84},\ref{aff85}}
\and H.~M.~Courtois\orcid{0000-0003-0509-1776}\inst{\ref{aff7}}
\and M.~Cropper\orcid{0000-0003-4571-9468}\inst{\ref{aff6}}
\and A.~Da~Silva\orcid{0000-0002-6385-1609}\inst{\ref{aff21},\ref{aff22}}
\and R.~da~Silva\orcid{0000-0003-4788-677X}\inst{\ref{aff59},\ref{aff56}}
\and H.~Degaudenzi\orcid{0000-0002-5887-6799}\inst{\ref{aff38}}
\and G.~De~Lucia\orcid{0000-0002-6220-9104}\inst{\ref{aff25}}
\and H.~Dole\orcid{0000-0002-9767-3839}\inst{\ref{aff86}}
\and F.~Dubath\orcid{0000-0002-6533-2810}\inst{\ref{aff38}}
\and C.~A.~J.~Duncan\orcid{0009-0003-3573-0791}\inst{\ref{aff65}}
\and S.~Escoffier\orcid{0000-0002-2847-7498}\inst{\ref{aff28}}
\and M.~Farina\orcid{0000-0002-3089-7846}\inst{\ref{aff87}}
\and R.~Farinelli\inst{\ref{aff4}}
\and S.~Farrens\orcid{0000-0002-9594-9387}\inst{\ref{aff5}}
\and F.~Faustini\orcid{0000-0001-6274-5145}\inst{\ref{aff59}}
\and S.~Ferriol\inst{\ref{aff18}}
\and N.~Fourmanoit\orcid{0009-0005-6816-6925}\inst{\ref{aff28}}
\and S.~Galeotta\orcid{0000-0002-3748-5115}\inst{\ref{aff25}}
\and K.~George\orcid{0000-0002-1734-8455}\inst{\ref{aff88}}
\and B.~Gillis\orcid{0000-0002-4478-1270}\inst{\ref{aff65}}
\and C.~Giocoli\orcid{0000-0002-9590-7961}\inst{\ref{aff4},\ref{aff69}}
\and B.~R.~Granett\orcid{0000-0003-2694-9284}\inst{\ref{aff36}}
\and A.~Grazian\orcid{0000-0002-5688-0663}\inst{\ref{aff57}}
\and S.~V.~H.~Haugan\orcid{0000-0001-9648-7260}\inst{\ref{aff89}}
\and I.~M.~Hook\orcid{0000-0002-2960-978X}\inst{\ref{aff90}}
\and F.~Hormuth\inst{\ref{aff91}}
\and A.~Hornstrup\orcid{0000-0002-3363-0936}\inst{\ref{aff92},\ref{aff93}}
\and M.~Huertas-Company\orcid{0000-0002-1416-8483}\inst{\ref{aff81},\ref{aff94},\ref{aff95}}
\and M.~Jhabvala\inst{\ref{aff96}}
\and S.~Kermiche\orcid{0000-0002-0302-5735}\inst{\ref{aff28}}
\and A.~Kiessling\orcid{0000-0002-2590-1273}\inst{\ref{aff40}}
\and M.~Kunz\orcid{0000-0002-3052-7394}\inst{\ref{aff97}}
\and H.~Kurki-Suonio\orcid{0000-0002-4618-3063}\inst{\ref{aff98},\ref{aff99}}
\and A.~M.~C.~Le~Brun\orcid{0000-0002-0936-4594}\inst{\ref{aff100}}
\and D.~Le~Mignant\orcid{0000-0002-5339-5515}\inst{\ref{aff45}}
\and S.~Ligori\orcid{0000-0003-4172-4606}\inst{\ref{aff75}}
\and P.~B.~Lilje\orcid{0000-0003-4324-7794}\inst{\ref{aff89}}
\and V.~Lindholm\orcid{0000-0003-2317-5471}\inst{\ref{aff98},\ref{aff99}}
\and I.~Lloro\orcid{0000-0001-5966-1434}\inst{\ref{aff101}}
\and M.~Magliocchetti\orcid{0000-0001-9158-4838}\inst{\ref{aff87}}
\and G.~Mainetti\orcid{0000-0003-2384-2377}\inst{\ref{aff102}}
\and O.~Mansutti\orcid{0000-0001-5758-4658}\inst{\ref{aff25}}
\and S.~Marcin\inst{\ref{aff103}}
\and O.~Marggraf\orcid{0000-0001-7242-3852}\inst{\ref{aff50}}
\and K.~Markovic\orcid{0000-0001-6764-073X}\inst{\ref{aff40}}
\and M.~Martinelli\orcid{0000-0002-6943-7732}\inst{\ref{aff59},\ref{aff104}}
\and N.~Martinet\orcid{0000-0003-2786-7790}\inst{\ref{aff45}}
\and F.~Marulli\orcid{0000-0002-8850-0303}\inst{\ref{aff64},\ref{aff4},\ref{aff69}}
\and N.~Mauri\orcid{0000-0001-8196-1548}\inst{\ref{aff80},\ref{aff69}}
\and S.~Mei\orcid{0000-0002-2849-559X}\inst{\ref{aff105},\ref{aff106}}
\and M.~Melchior\inst{\ref{aff107}}
\and M.~Meneghetti\orcid{0000-0003-1225-7084}\inst{\ref{aff4},\ref{aff69}}
\and E.~Merlin\orcid{0000-0001-6870-8900}\inst{\ref{aff57}}
\and G.~Meylan\orcid{0000-0001-6503-0209}\inst{\ref{aff108}}
\and P.~Monaco\orcid{0000-0003-2083-7564}\inst{\ref{aff32},\ref{aff25},\ref{aff33},\ref{aff66}}
\and M.~Moresco\orcid{0000-0002-7616-7136}\inst{\ref{aff64},\ref{aff4}}
\and C.~Moretti\orcid{0000-0003-3314-8936}\inst{\ref{aff25},\ref{aff66},\ref{aff33}}
\and L.~Moscardini\orcid{0000-0002-3473-6716}\inst{\ref{aff64},\ref{aff4},\ref{aff69}}
\and C.~Neissner\orcid{0000-0001-8524-4968}\inst{\ref{aff109},\ref{aff77}}
\and R.~C.~Nichol\orcid{0000-0003-0939-6518}\inst{\ref{aff110}}
\and S.-M.~Niemi\orcid{0009-0005-0247-0086}\inst{\ref{aff44}}
\and C.~Padilla\orcid{0000-0001-7951-0166}\inst{\ref{aff109}}
\and F.~Pasian\orcid{0000-0002-4869-3227}\inst{\ref{aff25}}
\and V.~Pettorino\orcid{0000-0002-4203-9320}\inst{\ref{aff44}}
\and A.~Pezzotta\orcid{0000-0003-0726-2268}\inst{\ref{aff36}}
\and S.~Pires\orcid{0000-0002-0249-2104}\inst{\ref{aff5}}
\and M.~Poncet\inst{\ref{aff16}}
\and L.~A.~Popa\inst{\ref{aff111}}
\and F.~Raison\orcid{0000-0002-7819-6918}\inst{\ref{aff31}}
\and A.~Refregier\inst{\ref{aff112}}
\and A.~Renzi\orcid{0000-0001-9856-1970}\inst{\ref{aff61},\ref{aff23},\ref{aff4}}
\and J.~Rhodes\orcid{0000-0002-4485-8549}\inst{\ref{aff40}}
\and G.~Riccio\inst{\ref{aff72}}
\and F.~Rizzo\orcid{0000-0002-9407-585X}\inst{\ref{aff25}}
\and M.~Roncarelli\orcid{0000-0001-9587-7822}\inst{\ref{aff4}}
\and C.~Rosset\orcid{0000-0003-0286-2192}\inst{\ref{aff105}}
\and B.~Rusholme\orcid{0000-0001-7648-4142}\inst{\ref{aff2}}
\and R.~Saglia\orcid{0000-0003-0378-7032}\inst{\ref{aff34},\ref{aff31}}
\and Z.~Sakr\orcid{0000-0002-4823-3757}\inst{\ref{aff113},\ref{aff114},\ref{aff115}}
\and A.~G.~S\'anchez\orcid{0000-0003-1198-831X}\inst{\ref{aff31}}
\and D.~Sapone\orcid{0000-0001-7089-4503}\inst{\ref{aff116}}
\and B.~Sartoris\orcid{0000-0003-1337-5269}\inst{\ref{aff117}}
\and P.~Schneider\orcid{0000-0001-8561-2679}\inst{\ref{aff50}}
\and T.~Schrabback\orcid{0000-0002-6987-7834}\inst{\ref{aff118}}
\and E.~Sefusatti\orcid{0000-0003-0473-1567}\inst{\ref{aff25},\ref{aff66},\ref{aff33}}
\and M.~Seiffert\orcid{0000-0002-7536-9393}\inst{\ref{aff40}}
\and E.~Sihvola\orcid{0000-0003-1804-7715}\inst{\ref{aff119}}
\and P.~Simon\inst{\ref{aff50}}
\and G.~Sirri\orcid{0000-0003-2626-2853}\inst{\ref{aff69}}
\and L.~Stanco\orcid{0000-0002-9706-5104}\inst{\ref{aff23}}
\and P.~Tallada-Cresp\'{i}\orcid{0000-0002-1336-8328}\inst{\ref{aff76},\ref{aff77}}
\and H.~I.~Teplitz\orcid{0000-0002-7064-5424}\inst{\ref{aff47}}
\and N.~Tessore\orcid{0000-0002-9696-7931}\inst{\ref{aff6}}
\and S.~Toft\orcid{0000-0003-3631-7176}\inst{\ref{aff120},\ref{aff121}}
\and R.~Toledo-Moreo\orcid{0000-0002-2997-4859}\inst{\ref{aff122},\ref{aff123}}
\and F.~Torradeflot\orcid{0000-0003-1160-1517}\inst{\ref{aff77},\ref{aff76}}
\and A.~Tsyganov\inst{\ref{aff124}}
\and I.~Tutusaus\orcid{0000-0002-3199-0399}\inst{\ref{aff78},\ref{aff15},\ref{aff114}}
\and L.~Valenziano\orcid{0000-0002-1170-0104}\inst{\ref{aff4},\ref{aff125}}
\and J.~Valiviita\orcid{0000-0001-6225-3693}\inst{\ref{aff98},\ref{aff99}}
\and T.~Vassallo\orcid{0000-0001-6512-6358}\inst{\ref{aff25},\ref{aff88}}
\and A.~Veropalumbo\orcid{0000-0003-2387-1194}\inst{\ref{aff36},\ref{aff19},\ref{aff58}}
\and Y.~Wang\orcid{0000-0002-4749-2984}\inst{\ref{aff2}}
\and J.~Weller\orcid{0000-0002-8282-2010}\inst{\ref{aff34},\ref{aff31}}
\and O.~R.~Williams\orcid{0000-0003-0274-1526}\inst{\ref{aff124}}
\and G.~Zamorani\orcid{0000-0002-2318-301X}\inst{\ref{aff4}}
\and F.~M.~Zerbi\orcid{0000-0002-9996-973X}\inst{\ref{aff36}}
\and E.~Zucca\orcid{0000-0002-5845-8132}\inst{\ref{aff4}}
\and I.~T.~Andika\orcid{0000-0001-6102-9526}\inst{\ref{aff34}}
\and M.~Ballardini\orcid{0000-0003-4481-3559}\inst{\ref{aff126},\ref{aff127},\ref{aff4}}
\and P.~Bergamini\orcid{0000-0003-1383-9414}\inst{\ref{aff4}}
\and S.~Borgani\orcid{0000-0001-6151-6439}\inst{\ref{aff32},\ref{aff66},\ref{aff25},\ref{aff33},\ref{aff128}}
\and E.~Bozzo\orcid{0000-0002-8201-1525}\inst{\ref{aff38}}
\and C.~Burigana\orcid{0000-0002-3005-5796}\inst{\ref{aff129},\ref{aff125}}
\and R.~Cabanac\orcid{0000-0001-6679-2600}\inst{\ref{aff114}}
\and M.~Calabrese\orcid{0000-0002-2637-2422}\inst{\ref{aff130},\ref{aff26}}
\and A.~Cappi\inst{\ref{aff117},\ref{aff4}}
\and F.~Caro\orcid{0009-0003-1053-0507}\inst{\ref{aff59}}
\and T.~Castro\orcid{0000-0002-6292-3228}\inst{\ref{aff131},\ref{aff25},\ref{aff66}}
\and D.~Di~Ferdinando\inst{\ref{aff69}}
\and J.~A.~Escartin~Vigo\inst{\ref{aff31}}
\and G.~Fabbian\orcid{0000-0002-3255-4695}\inst{\ref{aff86}}
\and J.~Garc\'ia-Bellido\orcid{0000-0002-9370-8360}\inst{\ref{aff113}}
\and T.~Gasparetto\orcid{0000-0002-7913-4866}\inst{\ref{aff59}}
\and V.~Gautard\inst{\ref{aff52}}
\and C.~M.~Gutierrez\orcid{0000-0001-7854-783X}\inst{\ref{aff81},\ref{aff132}}
\and A.~Hall\orcid{0000-0002-3139-8651}\inst{\ref{aff65}}
\and Y.~Kang\orcid{0009-0000-8588-7250}\inst{\ref{aff118}}
\and A.~Loureiro\orcid{0000-0002-4371-0876}\inst{\ref{aff133},\ref{aff134}}
\and J.~Macias-Perez\orcid{0000-0002-5385-2763}\inst{\ref{aff135}}
\and R.~Maoli\orcid{0000-0002-6065-3025}\inst{\ref{aff136},\ref{aff59}}
\and R.~B.~Metcalf\orcid{0000-0003-3167-2574}\inst{\ref{aff64},\ref{aff4}}
\and M.~P\"ontinen\orcid{0000-0001-5442-2530}\inst{\ref{aff98}}
\and E.~Sarpa\orcid{0000-0002-1256-655X}\inst{\ref{aff25}}
\and V.~Scottez\orcid{0009-0008-3864-940X}\inst{\ref{aff137},\ref{aff138}}
\and M.~Sereno\orcid{0000-0003-0302-0325}\inst{\ref{aff4},\ref{aff69}}
\and K.~Tanidis\orcid{0000-0001-9843-5130}\inst{\ref{aff139}}
\and M.~Tenti\orcid{0000-0002-4254-5901}\inst{\ref{aff69}}
\and M.~Tucci\inst{\ref{aff38}}
\and M.~Viel\orcid{0000-0002-2642-5707}\inst{\ref{aff66},\ref{aff25},\ref{aff67},\ref{aff33},\ref{aff128}}
\and M.~Wiesmann\orcid{0009-0000-8199-5860}\inst{\ref{aff89}}
\and J.~A.~Acevedo~Barroso\orcid{0000-0002-9654-1711}\inst{\ref{aff40}}
\and Y.~Akrami\orcid{0000-0002-2407-7956}\inst{\ref{aff113},\ref{aff140}}
\and G.~Angora\orcid{0000-0002-0316-6562}\inst{\ref{aff72},\ref{aff126}}
\and S.~Anselmi\orcid{0000-0002-3579-9583}\inst{\ref{aff23},\ref{aff61},\ref{aff141}}
\and M.~Archidiacono\orcid{0000-0003-4952-9012}\inst{\ref{aff35},\ref{aff37}}
\and G.~Aric\`o\orcid{0000-0002-2802-2928}\inst{\ref{aff69}}
\and F.~Atrio-Barandela\orcid{0000-0002-2130-2513}\inst{\ref{aff142}}
\and M.~Aubert\orcid{0009-0002-7667-8814}\inst{\ref{aff18}}
\and E.~Aubourg\orcid{0000-0002-5592-023X}\inst{\ref{aff105},\ref{aff143}}
\and M.~Baes\orcid{0000-0002-3930-2757}\inst{\ref{aff144}}
\and L.~Bazzanini\orcid{0000-0003-0727-0137}\inst{\ref{aff126},\ref{aff4}}
\and J.~R.~Bermejo-Climent\inst{\ref{aff145},\ref{aff146}}
\and D.~Bertacca\orcid{0000-0002-2490-7139}\inst{\ref{aff61},\ref{aff57},\ref{aff23}}
\and M.~Bethermin\orcid{0000-0002-3915-2015}\inst{\ref{aff147}}
\and F.~Beutler\orcid{0000-0003-0467-5438}\inst{\ref{aff65}}
\and A.~Blanchard\orcid{0000-0001-8555-9003}\inst{\ref{aff114}}
\and F.~Bouch\`e\orcid{0000-0002-4663-1786}\inst{\ref{aff148},\ref{aff79}}
\and M.~L.~Brown\orcid{0000-0002-0370-8077}\inst{\ref{aff82}}
\and S.~Bruton\orcid{0000-0002-6503-5218}\inst{\ref{aff8}}
\and A.~Calabro\orcid{0000-0003-2536-1614}\inst{\ref{aff59}}
\and B.~Camacho~Quevedo\orcid{0000-0002-8789-4232}\inst{\ref{aff66},\ref{aff67},\ref{aff25}}
\and P.~Carrilho\orcid{0000-0003-1339-0194}\inst{\ref{aff149}}
\and C.~S.~Carvalho\inst{\ref{aff150}}
\and Y.~Charles\orcid{0009-0000-3636-834X}\inst{\ref{aff45}}
\and F.~Cogato\orcid{0000-0003-4632-6113}\inst{\ref{aff4}}
\and T.~E.~Collett\orcid{0000-0001-5564-3140}\inst{\ref{aff151}}
\and A.~R.~Cooray\orcid{0000-0002-3892-0190}\inst{\ref{aff152}}
\and P.~Corcho-Caballero\orcid{0000-0001-6327-7080}\inst{\ref{aff9}}
\and M.~Correnti\orcid{0000-0001-6464-3257}\inst{\ref{aff59},\ref{aff56}}
\and B.~Csizi\orcid{0000-0003-3227-6581}\inst{\ref{aff118}}
\and O.~Cucciati\orcid{0000-0002-9336-7551}\inst{\ref{aff4}}
\and T.~de~Boer\orcid{0000-0001-5486-2747}\inst{\ref{aff153}}
\and F.~De~Paolis\orcid{0000-0001-6460-7563}\inst{\ref{aff154},\ref{aff155},\ref{aff156}}
\and G.~Desprez\orcid{0000-0001-8325-1742}\inst{\ref{aff9}}
\and A.~D\'iaz-S\'anchez\orcid{0000-0003-0748-4768}\inst{\ref{aff157}}
\and S.~Di~Domizio\orcid{0000-0003-2863-5895}\inst{\ref{aff58},\ref{aff19}}
\and J.~M.~Diego\orcid{0000-0001-9065-3926}\inst{\ref{aff158}}
\and V.~Duret\orcid{0009-0009-0383-4960}\inst{\ref{aff28}}
\and Y.~Fang\orcid{0000-0002-0334-6950}\inst{\ref{aff34}}
\and A.~Farina\orcid{0009-0000-3420-929X}\inst{\ref{aff36},\ref{aff19}}
\and A.~M.~N.~Ferguson\orcid{0000-0001-7934-1278}\inst{\ref{aff65}}
\and A.~Finoguenov\orcid{0000-0002-4606-5403}\inst{\ref{aff98}}
\and F.~Fontanot\orcid{0000-0003-4744-0188}\inst{\ref{aff25},\ref{aff66}}
\and A.~Franco\orcid{0000-0002-4761-366X}\inst{\ref{aff154},\ref{aff155},\ref{aff156}}
\and A.~Fumagalli\orcid{0009-0004-0300-2535}\inst{\ref{aff25}}
\and Y.~Fu\orcid{0000-0002-0759-0504}\inst{\ref{aff39},\ref{aff9}}
\and K.~Ganga\orcid{0000-0001-8159-8208}\inst{\ref{aff105}}
\and R.~Gavazzi\orcid{0000-0002-5540-6935}\inst{\ref{aff45},\ref{aff30}}
\and E.~Gaztanaga\orcid{0000-0001-9632-0815}\inst{\ref{aff78},\ref{aff15},\ref{aff151}}
\and Z.~Ghaffari\orcid{0000-0002-6467-8078}\inst{\ref{aff25},\ref{aff66}}
\and F.~Giacomini\orcid{0000-0002-3129-2814}\inst{\ref{aff69}}
\and F.~Gianotti\orcid{0000-0003-4666-119X}\inst{\ref{aff4}}
\and G.~Gozaliasl\orcid{0000-0002-0236-919X}\inst{\ref{aff159},\ref{aff98}}
\and A.~Gruppuso\orcid{0000-0001-9272-5292}\inst{\ref{aff4},\ref{aff69}}
\and M.~Guidi\orcid{0000-0001-9408-1101}\inst{\ref{aff68},\ref{aff4}}
\and N.~A.~Hatch\orcid{0000-0001-5600-0534}\inst{\ref{aff160}}
\and C.~Hern\'andez-Monteagudo\orcid{0000-0001-5471-9166}\inst{\ref{aff132},\ref{aff81}}
\and T.~Heuberger\orcid{0009-0009-3474-8467}\inst{\ref{aff103}}
\and H.~Hildebrandt\orcid{0000-0002-9814-3338}\inst{\ref{aff161}}
\and J.~Hjorth\orcid{0000-0002-4571-2306}\inst{\ref{aff162}}
\and M.~Jauzac\orcid{0000-0003-1974-8732}\inst{\ref{aff163},\ref{aff46},\ref{aff164},\ref{aff165},\ref{aff114}}
\and V.~Kansal\orcid{0000-0002-4008-6078}\inst{\ref{aff166},\ref{aff167}}
\and D.~Karagiannis\orcid{0000-0002-4927-0816}\inst{\ref{aff168},\ref{aff126},\ref{aff169}}
\and J.~Kim\orcid{0000-0003-2776-2761}\inst{\ref{aff27}}
\and C.~C.~Kirkpatrick\inst{\ref{aff119}}
\and A.~Kov\'acs\orcid{0000-0002-5825-579X}\inst{\ref{aff146},\ref{aff145}}
\and I.~Kova{\v{c}}i{\'{c}}\orcid{0000-0001-6751-3263}\inst{\ref{aff144}}
\and K.~Koyama\orcid{0000-0001-6727-6915}\inst{\ref{aff151}}
\and S.~Kruk\orcid{0000-0001-8010-8879}\inst{\ref{aff3}}
\and M.~C.~Lam\orcid{0000-0002-9347-2298}\inst{\ref{aff65}}
\and M.~Lattanzi\orcid{0000-0003-1059-2532}\inst{\ref{aff127}}
\and F.~Leclercq\orcid{0000-0002-9339-1404}\inst{\ref{aff30}}
\and S.~Lee\orcid{0000-0002-8289-740X}\inst{\ref{aff170},\ref{aff40}}
\and J.~Le~Graet\orcid{0000-0001-6523-7971}\inst{\ref{aff28}}
\and L.~Legrand\orcid{0000-0003-0610-5252}\inst{\ref{aff171},\ref{aff86}}
\and M.~Lembo\orcid{0000-0002-5271-5070}\inst{\ref{aff30}}
\and M.~D.~Lepinzan\orcid{0000-0003-1287-9801}\inst{\ref{aff25}}
\and G.~Leroy\orcid{0009-0004-2523-4425}\inst{\ref{aff163},\ref{aff46}}
\and J.~Lesgourgues\orcid{0000-0001-7627-353X}\inst{\ref{aff172}}
\and T.~I.~Liaudat\orcid{0000-0002-9104-314X}\inst{\ref{aff143}}
\and L.~Linke\orcid{0000-0002-2622-8113}\inst{\ref{aff118}}
\and S.~J.~Liu\orcid{0000-0001-7680-2139}\inst{\ref{aff87}}
\and G.~Maggio\orcid{0000-0003-4020-4836}\inst{\ref{aff25}}
\and A.~Manj\'on-Garc\'ia\orcid{0000-0002-7413-8825}\inst{\ref{aff157}}
\and F.~Mannucci\orcid{0000-0002-4803-2381}\inst{\ref{aff173}}
\and F.~R.~Marleau\orcid{0000-0002-1442-2947}\inst{\ref{aff118}}
\and C.~J.~A.~P.~Martins\orcid{0000-0002-4886-9261}\inst{\ref{aff174},\ref{aff10}}
\and M.~Miluzio\inst{\ref{aff3},\ref{aff175}}
\and G.~Morgante\inst{\ref{aff4}}
\and S.~Nadathur\orcid{0000-0001-9070-3102}\inst{\ref{aff151}}
\and K.~Naidoo\orcid{0000-0002-9182-1802}\inst{\ref{aff151},\ref{aff1}}
\and P.~Natoli\orcid{0000-0003-0126-9100}\inst{\ref{aff126},\ref{aff127}}
\and A.~Navarro-Alsina\orcid{0000-0002-3173-2592}\inst{\ref{aff50}}
\and A.~Nersesian\orcid{0000-0001-6843-409X}\inst{\ref{aff176},\ref{aff144}}
\and S.~Nesseris\orcid{0000-0002-0567-0324}\inst{\ref{aff113}}
\and L.~Nicastro\orcid{0000-0001-8534-6788}\inst{\ref{aff4}}
\and F.~Oppizzi\orcid{0000-0003-3904-8370}\inst{\ref{aff19},\ref{aff23},\ref{aff61}}
\and F.~Pace\orcid{0000-0001-8039-0480}\inst{\ref{aff73},\ref{aff74},\ref{aff75},\ref{aff22}}
\and D.~Paoletti\orcid{0000-0003-4761-6147}\inst{\ref{aff4},\ref{aff125}}
\and F.~Passalacqua\orcid{0000-0002-8606-4093}\inst{\ref{aff23}}
\and L.~Patrizii\inst{\ref{aff69}}
\and C.~Pattison\orcid{0000-0003-3272-2617}\inst{\ref{aff151}}
\and A.~Pisani\orcid{0000-0002-6146-4437}\inst{\ref{aff28}}
\and L.~Pizzuti\orcid{0000-0001-5654-7580}\inst{\ref{aff177}}
\and D.~Potter\orcid{0000-0002-0757-5195}\inst{\ref{aff178}}
\and M.~Poulain\orcid{0000-0002-7664-4510}\inst{\ref{aff179}}
\and G.~W.~Pratt\inst{\ref{aff5}}
\and S.~Quai\orcid{0000-0002-0449-8163}\inst{\ref{aff64},\ref{aff4}}
\and G.~Rodighiero\orcid{0000-0002-9415-2296}\inst{\ref{aff61},\ref{aff57}}
\and W.~Roster\orcid{0000-0002-9149-6528}\inst{\ref{aff31}}
\and M.~Sahl\'en\orcid{0000-0003-0973-4804}\inst{\ref{aff180}}
\and D.~B.~Sanders\orcid{0000-0002-1233-9998}\inst{\ref{aff153}}
\and A.~Schneider\orcid{0000-0001-7055-8104}\inst{\ref{aff178}}
\and D.~Sciotti\orcid{0009-0008-4519-2620}\inst{\ref{aff59},\ref{aff104}}
\and E.~Sellentin\orcid{0009-0002-2655-3458}\inst{\ref{aff181},\ref{aff39}}
\and S.~Serjeant\orcid{0000-0002-0517-7943}\inst{\ref{aff182}}
\and F.~Shankar\orcid{0000-0001-8973-5051}\inst{\ref{aff183}}
\and A.~Shulevski\orcid{0000-0002-1827-0469}\inst{\ref{aff184},\ref{aff9},\ref{aff185}}
\and R.~E.~Smith\orcid{0000-0001-9989-2149}\inst{\ref{aff186}}
\and E.~Soubrie\orcid{0000-0001-9295-1863}\inst{\ref{aff86}}
\and I.~Szapudi\orcid{0000-0003-2274-0301}\inst{\ref{aff153}}
\and G.~Testera\orcid{0000-0003-2970-766X}\inst{\ref{aff19}}
\and R.~Teyssier\orcid{0000-0001-7689-0933}\inst{\ref{aff187}}
\and S.~Tosi\orcid{0000-0002-7275-9193}\inst{\ref{aff58},\ref{aff36},\ref{aff19}}
\and A.~Troja\orcid{0000-0003-0239-4595}\inst{\ref{aff25}}
\and C.~Uhlemann\orcid{0000-0001-7831-1579}\inst{\ref{aff188},\ref{aff51}}
\and M.~Urbano\orcid{0000-0001-5640-0650}\inst{\ref{aff160},\ref{aff147}}
\and C.~Valieri\inst{\ref{aff69}}
\and A.~Venhola\orcid{0000-0001-6071-4564}\inst{\ref{aff179}}
\and D.~Vergani\orcid{0000-0003-0898-2216}\inst{\ref{aff4}}
\and E.~Vilenius\orcid{0000-0002-6184-7681}\inst{\ref{aff6}}
\and S.~Vinciguerra\orcid{0009-0005-4018-3184}\inst{\ref{aff45}}
\and M.~von~Wietersheim-Kramsta\orcid{0000-0003-4986-5091}\inst{\ref{aff46},\ref{aff163}}
\and N.~A.~Walton\orcid{0000-0003-3983-8778}\inst{\ref{aff63}}
\and L.~Wang\orcid{0000-0002-6736-9158}\inst{\ref{aff189},\ref{aff9}}
\and A.~H.~Wright\orcid{0000-0001-7363-7932}\inst{\ref{aff161}}
\and H.~W.~Yeung\orcid{0000-0002-4993-9014}\inst{\ref{aff65}}
\and L.~Zalesky\orcid{0000-0001-5680-2326}\inst{\ref{aff153}}
\and D.~Scott\orcid{0000-0002-6878-9840}\inst{\ref{aff190}}}
										   
\institute{Max-Planck-Institut f\"ur Astronomie, K\"onigstuhl 17, 69117 Heidelberg, Germany\label{aff1}
\and
Caltech/IPAC, 1200 E. California Blvd., Pasadena, CA 91125, USA\label{aff2}
\and
ESAC/ESA, Camino Bajo del Castillo, s/n., Urb. Villafranca del Castillo, 28692 Villanueva de la Ca\~nada, Madrid, Spain\label{aff3}
\and
INAF-Osservatorio di Astrofisica e Scienza dello Spazio di Bologna, Via Piero Gobetti 93/3, 40129 Bologna, Italy\label{aff4}
\and
Universit\'e Paris-Saclay, Universit\'e Paris Cit\'e, CEA, CNRS, AIM, 91191, Gif-sur-Yvette, France\label{aff5}
\and
Mullard Space Science Laboratory, University College London, Holmbury St Mary, Dorking, Surrey RH5 6NT, UK\label{aff6}
\and
Universit\'e Lyon 1, CNRS/IN2P3, IUF, IP2I Lyon, Villeurbanne, France \label{aff7}
\and
California Institute of Technology, 1200 E California Blvd, Pasadena, CA 91125, USA\label{aff8}
\and
Kapteyn Astronomical Institute, University of Groningen, PO Box 800, 9700 AV Groningen, The Netherlands\label{aff9}
\and
Instituto de Astrof\'isica e Ci\^encias do Espa\c{c}o, Universidade do Porto, CAUP, Rua das Estrelas, PT4150-762 Porto, Portugal\label{aff10}
\and
Faculdade de Ci\^encias da Universidade do Porto, Rua do Campo de Alegre, 4150-007 Porto, Portugal\label{aff11}
\and
European Southern Observatory, Karl-Schwarzschild-Str.~2, 85748 Garching, Germany\label{aff12}
\and
Departament de F\'isica Qu\`antica i Astrof\'isica (FQA), Universitat de Barcelona (UB), Mart\'{\i} i Franqu\`es 1, E-08028 Barcelona, Spain\label{aff13}
\and
Institut de Ci\`{e}ncies del Cosmos (ICCUB), Universitat de Barcelona (IEEC-UB), Mart\'{i} i Franqu\`{e}s 1, 08028 Barcelona, Spain\label{aff14}
\and
Institut d'Estudis Espacials de Catalunya (IEEC),  Edifici RDIT, Campus UPC, 08860 Castelldefels, Barcelona, Spain\label{aff15}
\and
Centre National d'Etudes Spatiales -- Centre spatial de Toulouse, 18 avenue Edouard Belin, 31401 Toulouse Cedex 9, France\label{aff16}
\and
Centre de Recherche Astrophysique de Lyon, UMR5574, CNRS, Universit\'e Claude Bernard Lyon 1, ENS de Lyon, 69230, Saint-Genis-Laval, France\label{aff17}
\and
Universit\'e Lyon 1, CNRS/IN2P3, IP2I Lyon, Villeurbanne, France\label{aff18}
\and
INFN-Sezione di Genova, Via Dodecaneso 33, 16146, Genova, Italy\label{aff19}
\and
National Institute of Standards and Technology, 100 Bureau Dr., Gaithersburg, MD 20899, USA\label{aff20}
\and
Departamento de F\'isica, Faculdade de Ci\^encias, Universidade de Lisboa, Edif\'icio C8, Campo Grande, PT1749-016 Lisboa, Portugal\label{aff21}
\and
Instituto de Astrof\'isica e Ci\^encias do Espa\c{c}o, Faculdade de Ci\^encias, Universidade de Lisboa, Campo Grande, 1749-016 Lisboa, Portugal\label{aff22}
\and
INFN-Padova, Via Marzolo 8, 35131 Padova, Italy\label{aff23}
\and
Aurora Technology for European Space Agency (ESA), Camino bajo del Castillo, s/n, Urbanizacion Villafranca del Castillo, Villanueva de la Ca\~nada, 28692 Madrid, Spain\label{aff24}
\and
INAF-Osservatorio Astronomico di Trieste, Via G. B. Tiepolo 11, 34143 Trieste, Italy\label{aff25}
\and
INAF-IASF Milano, Via Alfonso Corti 12, 20133 Milano, Italy\label{aff26}
\and
Department of Physics, University of Oxford, Keble Road, Oxford OX1 3RH, UK\label{aff27}
\and
Aix-Marseille Universit\'e, CNRS/IN2P3, CPPM, Marseille, France\label{aff28}
\and
FRACTAL S.L.N.E., calle Tulip\'an 2, Portal 13 1A, 28231, Las Rozas de Madrid, Spain\label{aff29}
\and
Institut d'Astrophysique de Paris, UMR 7095, CNRS, and Sorbonne Universit\'e, 98 bis boulevard Arago, 75014 Paris, France\label{aff30}
\and
Max Planck Institute for Extraterrestrial Physics, Giessenbachstr. 1, 85748 Garching, Germany\label{aff31}
\and
Dipartimento di Fisica - Sezione di Astronomia, Universit\`a di Trieste, Via Tiepolo 11, 34131 Trieste, Italy\label{aff32}
\and
INFN, Sezione di Trieste, Via Valerio 2, 34127 Trieste TS, Italy\label{aff33}
\and
Universit\"ats-Sternwarte M\"unchen, Fakult\"at f\"ur Physik, Ludwig-Maximilians-Universit\"at M\"unchen, Scheinerstr.~1, 81679 M\"unchen, Germany\label{aff34}
\and
Dipartimento di Fisica "Aldo Pontremoli", Universit\`a degli Studi di Milano, Via Celoria 16, 20133 Milano, Italy\label{aff35}
\and
INAF-Osservatorio Astronomico di Brera, Via Brera 28, 20122 Milano, Italy\label{aff36}
\and
INFN-Sezione di Milano, Via Celoria 16, 20133 Milano, Italy\label{aff37}
\and
Department of Astronomy, University of Geneva, ch. d'Ecogia 16, 1290 Versoix, Switzerland\label{aff38}
\and
Leiden Observatory, Leiden University, Einsteinweg 55, 2333 CC Leiden, The Netherlands\label{aff39}
\and
Jet Propulsion Laboratory, California Institute of Technology, 4800 Oak Grove Drive, Pasadena, CA, 91109, USA\label{aff40}
\and
Department of Physics and Astronomy, Vesilinnantie 5, University of Turku, 20014 Turku, Finland\label{aff41}
\and
Finnish Centre for Astronomy with ESO (FINCA), Quantum, Vesilinnantie 5, University of Turku, 20014 Turku, Finland\label{aff42}
\and
Serco for European Space Agency (ESA), Camino bajo del Castillo, s/n, Urbanizacion Villafranca del Castillo, Villanueva de la Ca\~nada, 28692 Madrid, Spain\label{aff43}
\and
European Space Agency/ESTEC, Keplerlaan 1, 2201 AZ Noordwijk, The Netherlands\label{aff44}
\and
Aix-Marseille Universit\'e, CNRS, CNES, LAM, Marseille, France\label{aff45}
\and
Department of Physics, Institute for Computational Cosmology, Durham University, South Road, Durham, DH1 3LE, UK\label{aff46}
\and
Infrared Processing and Analysis Center, California Institute of Technology, Pasadena, CA 91125, USA\label{aff47}
\and
European Space Agency/ESOC, Robert-Bosch-Str. 5, 64293 Darmstadt, Germany\label{aff48}
\and
Telespazio UK S.L. for European Space Agency (ESA), Camino bajo del Castillo, s/n, Urbanizacion Villafranca del Castillo, Villanueva de la Ca\~nada, 28692 Madrid, Spain\label{aff49}
\and
Universit\"at Bonn, Argelander-Institut f\"ur Astronomie, Auf dem H\"ugel 71, 53121 Bonn, Germany\label{aff50}
\and
School of Mathematics, Statistics and Physics, Newcastle University, Herschel Building, Newcastle-upon-Tyne, NE1 7RU, UK\label{aff51}
\and
CEA Saclay, DFR/IRFU, Service d'Astrophysique, Bat. 709, 91191 Gif-sur-Yvette, France\label{aff52}
\and
Waterloo Centre for Astrophysics, University of Waterloo, Waterloo, Ontario N2L 3G1, Canada\label{aff53}
\and
Department of Physics and Astronomy, University of Waterloo, Waterloo, Ontario N2L 3G1, Canada\label{aff54}
\and
Perimeter Institute for Theoretical Physics, Waterloo, Ontario N2L 2Y5, Canada\label{aff55}
\and
Space Science Data Center, Italian Space Agency, via del Politecnico snc, 00133 Roma, Italy\label{aff56}
\and
INAF-Osservatorio Astronomico di Padova, Via dell'Osservatorio 5, 35122 Padova, Italy\label{aff57}
\and
Dipartimento di Fisica, Universit\`a di Genova, Via Dodecaneso 33, 16146, Genova, Italy\label{aff58}
\and
INAF-Osservatorio Astronomico di Roma, Via Frascati 33, 00078 Monteporzio Catone, Italy\label{aff59}
\and
Minnesota Institute for Astrophysics, University of Minnesota, 116 Church St SE, Minneapolis, MN 55455, USA\label{aff60}
\and
Dipartimento di Fisica e Astronomia "G. Galilei", Universit\`a di Padova, Via Marzolo 8, 35131 Padova, Italy\label{aff61}
\and
Centre for Electronic Imaging, Open University, Walton Hall, Milton Keynes, MK7~6AA, UK\label{aff62}
\and
Institute of Astronomy, University of Cambridge, Madingley Road, Cambridge CB3 0HA, UK\label{aff63}
\and
Dipartimento di Fisica e Astronomia "Augusto Righi" - Alma Mater Studiorum Universit\`a di Bologna, via Piero Gobetti 93/2, 40129 Bologna, Italy\label{aff64}
\and
Institute for Astronomy, University of Edinburgh, Royal Observatory, Blackford Hill, Edinburgh EH9 3HJ, UK\label{aff65}
\and
IFPU, Institute for Fundamental Physics of the Universe, via Beirut 2, 34151 Trieste, Italy\label{aff66}
\and
SISSA, International School for Advanced Studies, Via Bonomea 265, 34136 Trieste TS, Italy\label{aff67}
\and
Dipartimento di Fisica e Astronomia, Universit\`a di Bologna, Via Gobetti 93/2, 40129 Bologna, Italy\label{aff68}
\and
INFN-Sezione di Bologna, Viale Berti Pichat 6/2, 40127 Bologna, Italy\label{aff69}
\and
ATG Europe BV, Huygensstraat 34, 2201 DK Noordwijk, The Netherlands\label{aff70}
\and
Department of Physics "E. Pancini", University Federico II, Via Cinthia 6, 80126, Napoli, Italy\label{aff71}
\and
INAF-Osservatorio Astronomico di Capodimonte, Via Moiariello 16, 80131 Napoli, Italy\label{aff72}
\and
Dipartimento di Fisica, Universit\`a degli Studi di Torino, Via P. Giuria 1, 10125 Torino, Italy\label{aff73}
\and
INFN-Sezione di Torino, Via P. Giuria 1, 10125 Torino, Italy\label{aff74}
\and
INAF-Osservatorio Astrofisico di Torino, Via Osservatorio 20, 10025 Pino Torinese (TO), Italy\label{aff75}
\and
Centro de Investigaciones Energ\'eticas, Medioambientales y Tecnol\'ogicas (CIEMAT), Avenida Complutense 40, 28040 Madrid, Spain\label{aff76}
\and
Port d'Informaci\'{o} Cient\'{i}fica, Campus UAB, C. Albareda s/n, 08193 Bellaterra (Barcelona), Spain\label{aff77}
\and
Institute of Space Sciences (ICE, CSIC), Campus UAB, Carrer de Can Magrans, s/n, 08193 Barcelona, Spain\label{aff78}
\and
INFN -- Sezione di Napoli, Via Cinthia 6, 80126, Napoli, Italy\label{aff79}
\and
Dipartimento di Fisica e Astronomia "Augusto Righi" - Alma Mater Studiorum Universit\`a di Bologna, Viale Berti Pichat 6/2, 40127 Bologna, Italy\label{aff80}
\and
Instituto de Astrof\'{\i}sica de Canarias, E-38205 La Laguna, Tenerife, Spain\label{aff81}
\and
Jodrell Bank Centre for Astrophysics, Department of Physics and Astronomy, University of Manchester, Oxford Road, Manchester M13 9PL, UK\label{aff82}
\and
European Space Agency/ESRIN, Largo Galileo Galilei 1, 00044 Frascati, Roma, Italy\label{aff83}
\and
Instituci\'o Catalana de Recerca i Estudis Avan\c{c}ats (ICREA), Passeig de Llu\'{\i}s Companys 23, 08010 Barcelona, Spain\label{aff84}
\and
Institut de Ciencies de l'Espai (IEEC-CSIC), Campus UAB, Carrer de Can Magrans, s/n Cerdanyola del Vall\'es, 08193 Barcelona, Spain\label{aff85}
\and
Universit\'e Paris-Saclay, CNRS, Institut d'astrophysique spatiale, 91405, Orsay, France\label{aff86}
\and
INAF-Istituto di Astrofisica e Planetologia Spaziali, via del Fosso del Cavaliere, 100, 00100 Roma, Italy\label{aff87}
\and
University Observatory, LMU Faculty of Physics, Scheinerstr.~1, 81679 Munich, Germany\label{aff88}
\and
Institute of Theoretical Astrophysics, University of Oslo, P.O. Box 1029 Blindern, 0315 Oslo, Norway\label{aff89}
\and
School of Physics and Astronomy, Lancaster University, Lancaster, LA1 4YB, UK\label{aff90}
\and
Felix Hormuth Engineering, Goethestr. 17, 69181 Leimen, Germany\label{aff91}
\and
Technical University of Denmark, Elektrovej 327, 2800 Kgs. Lyngby, Denmark\label{aff92}
\and
Cosmic Dawn Center (DAWN), Denmark\label{aff93}
\and
Universit\'e PSL, Observatoire de Paris, Sorbonne Universit\'e, CNRS, LERMA, 75014, Paris, France\label{aff94}
\and
Universit\'e Paris-Cit\'e, 5 Rue Thomas Mann, 75013, Paris, France\label{aff95}
\and
NASA Goddard Space Flight Center, Greenbelt, MD 20771, USA\label{aff96}
\and
Universit\'e de Gen\`eve, D\'epartement de Physique Th\'eorique and Centre for Astroparticle Physics, 24 quai Ernest-Ansermet, CH-1211 Gen\`eve 4, Switzerland\label{aff97}
\and
Department of Physics, P.O. Box 64, University of Helsinki, 00014 Helsinki, Finland\label{aff98}
\and
Helsinki Institute of Physics, Gustaf H{\"a}llstr{\"o}min katu 2, University of Helsinki, 00014 Helsinki, Finland\label{aff99}
\and
Laboratoire d'etude de l'Univers et des phenomenes eXtremes, Observatoire de Paris, Universit\'e PSL, Sorbonne Universit\'e, CNRS, 92190 Meudon, France\label{aff100}
\and
SKAO, Jodrell Bank, Lower Withington, Macclesfield SK11 9FT, UK\label{aff101}
\and
Centre de Calcul de l'IN2P3/CNRS, 21 avenue Pierre de Coubertin 69627 Villeurbanne Cedex, France\label{aff102}
\and
University of Applied Sciences and Arts of Northwestern Switzerland, School of Computer Science, 5210 Windisch, Switzerland\label{aff103}
\and
INFN-Sezione di Roma, Piazzale Aldo Moro, 2 - c/o Dipartimento di Fisica, Edificio G. Marconi, 00185 Roma, Italy\label{aff104}
\and
Universit\'e Paris Cit\'e, CNRS, Astroparticule et Cosmologie, 75013 Paris, France\label{aff105}
\and
CNRS-UCB International Research Laboratory, Centre Pierre Bin\'etruy, IRL2007, CPB-IN2P3, Berkeley, USA\label{aff106}
\and
University of Applied Sciences and Arts of Northwestern Switzerland, School of Engineering, 5210 Windisch, Switzerland\label{aff107}
\and
Institute of Physics, Laboratory of Astrophysics, Ecole Polytechnique F\'ed\'erale de Lausanne (EPFL), Observatoire de Sauverny, 1290 Versoix, Switzerland\label{aff108}
\and
Institut de F\'{i}sica d'Altes Energies (IFAE), The Barcelona Institute of Science and Technology, Campus UAB, 08193 Bellaterra (Barcelona), Spain\label{aff109}
\and
School of Mathematics and Physics, University of Surrey, Guildford, Surrey, GU2 7XH, UK\label{aff110}
\and
Institute of Space Science, Str. Atomistilor, nr. 409 M\u{a}gurele, Ilfov, 077125, Romania\label{aff111}
\and
Institute for Particle Physics and Astrophysics, Dept. of Physics, ETH Zurich, Wolfgang-Pauli-Strasse 27, 8093 Zurich, Switzerland\label{aff112}
\and
Instituto de F\'isica Te\'orica UAM-CSIC, Campus de Cantoblanco, 28049 Madrid, Spain\label{aff113}
\and
Institut de Recherche en Astrophysique et Plan\'etologie (IRAP), Universit\'e de Toulouse, CNRS, UPS, CNES, 14 Av. Edouard Belin, 31400 Toulouse, France\label{aff114}
\and
Universit\'e St Joseph; Faculty of Sciences, Beirut, Lebanon\label{aff115}
\and
Departamento de F\'isica, FCFM, Universidad de Chile, Blanco Encalada 2008, Santiago, Chile\label{aff116}
\and
Universit\'e C\^{o}te d'Azur, Observatoire de la C\^{o}te d'Azur, CNRS, Laboratoire Lagrange, Bd de l'Observatoire, CS 34229, 06304 Nice cedex 4, France\label{aff117}
\and
Universit\"at Innsbruck, Institut f\"ur Astro- und Teilchenphysik, Technikerstr. 25/8, 6020 Innsbruck, Austria\label{aff118}
\and
Department of Physics and Helsinki Institute of Physics, Gustaf H\"allstr\"omin katu 2, University of Helsinki, 00014 Helsinki, Finland\label{aff119}
\and
Cosmic Dawn Center (DAWN)\label{aff120}
\and
Niels Bohr Institute, University of Copenhagen, Jagtvej 128, 2200 Copenhagen, Denmark\label{aff121}
\and
Universidad Polit\'ecnica de Cartagena, Departamento de Electr\'onica y Tecnolog\'ia de Computadoras,  Plaza del Hospital 1, 30202 Cartagena, Spain\label{aff122}
\and
European University of Technology EUt+, European Union\label{aff123}
\and
Centre for Information Technology, University of Groningen, P.O. Box 11044, 9700 CA Groningen, The Netherlands\label{aff124}
\and
INFN-Bologna, Via Irnerio 46, 40126 Bologna, Italy\label{aff125}
\and
Dipartimento di Fisica e Scienze della Terra, Universit\`a degli Studi di Ferrara, Via Giuseppe Saragat 1, 44122 Ferrara, Italy\label{aff126}
\and
Istituto Nazionale di Fisica Nucleare, Sezione di Ferrara, Via Giuseppe Saragat 1, 44122 Ferrara, Italy\label{aff127}
\and
ICSC - Centro Nazionale di Ricerca in High Performance Computing, Big Data e Quantum Computing, Via Magnanelli 2, Bologna, Italy\label{aff128}
\and
INAF, Istituto di Radioastronomia, Via Piero Gobetti 101, 40129 Bologna, Italy\label{aff129}
\and
Astronomical Observatory of the Autonomous Region of the Aosta Valley (OAVdA), Loc. Lignan 39, I-11020, Nus (Aosta Valley), Italy\label{aff130}
\and
Department of Mathematical Physics, Institute of Physics, University of S\~ao Paulo, R. do Mat\~ao 1371, 05508-090, S\~ao Paulo, SP, Brazil\label{aff131}
\and
Universidad de La Laguna, Dpto. Astrof\'\i sica, E-38206 La Laguna, Tenerife, Spain\label{aff132}
\and
Oskar Klein Centre for Cosmoparticle Physics, Department of Physics, Stockholm University, Stockholm, SE-106 91, Sweden\label{aff133}
\and
Astrophysics Group, Blackett Laboratory, Imperial College London, London SW7 2AZ, UK\label{aff134}
\and
Univ. Grenoble Alpes, CNRS, Grenoble INP, LPSC-IN2P3, 53, Avenue des Martyrs, 38000 Grenoble, France\label{aff135}
\and
Dipartimento di Fisica, Sapienza Universit\`a di Roma, Piazzale Aldo Moro 2, 00185 Roma, Italy\label{aff136}
\and
Institut d'Astrophysique de Paris, 98bis Boulevard Arago, 75014, Paris, France\label{aff137}
\and
ICL, Junia, Universit\'e Catholique de Lille, LITL, 59000 Lille, France\label{aff138}
\and
Center for Astrophysics and Cosmology, University of Nova Gorica, Nova Gorica, Slovenia\label{aff139}
\and
CERCA/ISO, Department of Physics, Case Western Reserve University, 10900 Euclid Avenue, Cleveland, OH 44106, USA\label{aff140}
\and
Laboratoire Univers et Th\'eorie, Observatoire de Paris, Universit\'e PSL, Universit\'e Paris Cit\'e, CNRS, 92190 Meudon, France\label{aff141}
\and
Departamento de F{\'\i}sica Fundamental. Universidad de Salamanca. Plaza de la Merced s/n. 37008 Salamanca, Spain\label{aff142}
\and
IRFU, CEA, Universit\'e Paris-Saclay 91191 Gif-sur-Yvette Cedex, France\label{aff143}
\and
Universiteit Gent, Department of Physics and Astronomy, Proeftuinstraat 86 N3, 9000 Ghent, Belgium
\label{aff144}
\and
Konkoly Observatory, HUN-REN CSFK, MTA Centre of Excellence, Budapest, Konkoly Thege Mikl\'os {\'u}t 15-17. H-1121, Hungary\label{aff145}
\and
MTA-CSFK Lend\"ulet Large-Scale Structure Research Group, Konkoly-Thege Mikl\'os \'ut 15-17, H-1121 Budapest, Hungary\label{aff146}
\and
Universit\'e de Strasbourg, CNRS, Observatoire astronomique de Strasbourg, UMR 7550, 67000 Strasbourg, France\label{aff147}
\and
Scuola Superiore Meridionale, Via Mezzocannone 4, 80138, Napoli, Italy\label{aff148}
\and
Department of Physics, Astronomy and Mathematics, University of Hertfordshire, College Lane, Hatfield AL10 9AB, UK\label{aff149}
\and
Instituto de Astrof\'isica e Ci\^encias do Espa\c{c}o, Faculdade de Ci\^encias, Universidade de Lisboa, Tapada da Ajuda, 1349-018 Lisboa, Portugal\label{aff150}
\and
Institute of Cosmology and Gravitation, University of Portsmouth, Portsmouth PO1 3FX, UK\label{aff151}
\and
Department of Physics \& Astronomy, University of California Irvine, Irvine CA 92697, USA\label{aff152}
\and
Institute for Astronomy, University of Hawaii, 2680 Woodlawn Drive, Honolulu, HI 96822, USA\label{aff153}
\and
Department of Mathematics and Physics E. De Giorgi, University of Salento, Via per Arnesano, CP-I93, 73100, Lecce, Italy\label{aff154}
\and
INFN, Sezione di Lecce, Via per Arnesano, CP-193, 73100, Lecce, Italy\label{aff155}
\and
INAF-Sezione di Lecce, c/o Dipartimento Matematica e Fisica, Via per Arnesano, 73100, Lecce, Italy\label{aff156}
\and
Departamento F\'isica Aplicada, Universidad Polit\'ecnica de Cartagena, Campus Muralla del Mar, 30202 Cartagena, Murcia, Spain\label{aff157}
\and
Instituto de F\'isica de Cantabria, Edificio Juan Jord\'a, Avenida de los Castros, 39005 Santander, Spain\label{aff158}
\and
Department of Computer Science, Aalto University, PO Box 15400, Espoo, FI-00 076, Finland\label{aff159}
\and
School of Physics and Astronomy, University of Nottingham, University Park, Nottingham NG7 2RD, UK\label{aff160}
\and
Ruhr University Bochum, Faculty of Physics and Astronomy, Astronomical Institute (AIRUB), German Centre for Cosmological Lensing (GCCL), 44780 Bochum, Germany\label{aff161}
\and
DARK, Niels Bohr Institute, University of Copenhagen, Jagtvej 155, 2200 Copenhagen, Denmark\label{aff162}
\and
Department of Physics, Centre for Extragalactic Astronomy, Durham University, South Road, Durham, DH1 3LE, UK\label{aff163}
\and
Astrophysics Research Centre, University of KwaZulu-Natal, Westville Campus, Durban 4041, South Africa\label{aff164}
\and
School of Mathematics, Statistics \& Computer Science, University of KwaZulu-Natal, Westville Campus, Durban 4041, South Africa\label{aff165}
\and
ARC Centre of Excellence for Dark Matter Particle Physics, Melbourne, Australia\label{aff166}
\and
Centre for Astrophysics \& Supercomputing, Swinburne University of Technology,  Hawthorn, Victoria 3122, Australia\label{aff167}
\and
Van Swinderen Institute, University of Groningen, Nijenborgh 3, 9747 AG Groningen, The Netherlands\label{aff168}
\and
Department of Physics and Astronomy, University of the Western Cape, Bellville, Cape Town, 7535, South Africa\label{aff169}
\and
Ohio University, Physics \& Astronomy Department,1 Ohio University, Athens, OH 45701, USA\label{aff170}
\and
Brazilian Center for Research in Physics (CBPF), Dr. Xavier Sigaud st. 150, zip 22290-180, Rio de Janeiro, RJ, Brazil\label{aff171}
\and
Institute for Theoretical Particle Physics and Cosmology (TTK), RWTH Aachen University, 52056 Aachen, Germany\label{aff172}
\and
INAF-Osservatorio Astrofisico di Arcetri, Largo E. Fermi 5, 50125, Firenze, Italy\label{aff173}
\and
Centro de Astrof\'{\i}sica da Universidade do Porto, Rua das Estrelas, 4150-762 Porto, Portugal\label{aff174}
\and
HE Space for European Space Agency (ESA), Camino bajo del Castillo, s/n, Urbanizacion Villafranca del Castillo, Villanueva de la Ca\~nada, 28692 Madrid, Spain\label{aff175}
\and
STAR Institute, University of Li{\`e}ge, Quartier Agora, All\'ee du six Ao\^ut 19c, 4000 Li\`ege, Belgium\label{aff176}
\and
Dipartimento di Fisica ``G. Occhialini", Universit\`a degli Studi di Milano Bicocca, Piazza della Scienza 3, 20126 Milano, Italy\label{aff177}
\and
Department of Astrophysics, University of Zurich, Winterthurerstrasse 190, 8057 Zurich, Switzerland\label{aff178}
\and
Space physics and astronomy research unit, University of Oulu, Pentti Kaiteran katu 1, FI-90014 Oulu, Finland\label{aff179}
\and
Theoretical astrophysics, Department of Physics and Astronomy, Uppsala University, Box 524, 751 20 Uppsala, Sweden\label{aff180}
\and
Mathematical Institute, University of Leiden, Einsteinweg 55, 2333 CA Leiden, The Netherlands\label{aff181}
\and
School of Physical Sciences, The Open University, Milton Keynes, MK7 6AA, UK\label{aff182}
\and
School of Physics \& Astronomy, University of Southampton, Highfield Campus, Southampton SO17 1BJ, UK\label{aff183}
\and
ASTRON, the Netherlands Institute for Radio Astronomy, Postbus 2, 7990 AA, Dwingeloo, The Netherlands\label{aff184}
\and
Center for Advanced Interdisciplinary Research, Ss. Cyril and Methodius University in Skopje, Macedonia\label{aff185}
\and
Department of Physics \& Astronomy, University of Sussex, Brighton BN1 9QH, UK\label{aff186}
\and
Department of Astrophysical Sciences, Peyton Hall, Princeton University, Princeton, NJ 08544, USA\label{aff187}
\and
Fakult\"at f\"ur Physik, Universit\"at Bielefeld, Postfach 100131, 33501 Bielefeld, Germany\label{aff188}
\and
SRON Netherlands Institute for Space Research, Landleven 12, 9747 AD, Groningen, The Netherlands\label{aff189}
\and
Department of Physics and Astronomy, University of British Columbia, Vancouver, BC V6T 1Z1, Canada\label{aff190}}    

\abstract{
This paper reviews how \Euclid's top-level calibration requirements were translated into a calibration plan, how this plan is implemented and maintained during the survey, and how it performs in the current DR1 release. We show that DR1 already meets or exceeds most end-of-mission spectrophotometric calibration requirements. The construction of spectroscopically complete and pure samples relies on ongoing data acquisition for the Euclid Deep Survey, while point-spread-function modelling and shear calibration are still converging to their stringent accuracy requirements. With all DR1 data reprocessed with the DR1 pipeline configuration, more subtle effects have been detected and characterised, and correction strategies have been developed for DR2. These include improved absolute flux calibration, astrometry, charge-persistence correction, shutter effects, wavelength calibration, and spectrophotometric corrections for water-ice contamination. Finally, synthetic photometric transformations between \Euclid and each of \textit{Roman}, the \textit{James Webb} Space Telescope, and the \ac{CSST} are presented.}


\titlerunning{Calibration strategy and observations}
\authorrunning{Euclid Collaboration: M.~Schirmer et al.: DR1}
   
   \maketitle
   \nolinenumbers

\section{Introduction}
\label{sec:intro}

\Euclid is a fundamental physics experiment designed measure cosmological parameters with unprecedented accuracy \citep{EuclidSkyOverview}. Every design choice was optimised within the mission's cost, mass, and engineering constraints. As a result, \Euclid was built with minimal movable mechanisms, and equipped with exceptionally well-characterised instruments and telescope, together forming the scientific \ac{PLM}. Achieving the required dark energy figure of merit, ${\rm FoM}\ge 400$, drives all specifications for the hardware, the survey, the calibration plan, and the \ac{SGS} that processes the data. 

DR1 \citep{DR1cite} is released in two stages: DR1-Foundation, in November 2026, and the complete DR1 in mid-2027, the latter including higher-level science products for galaxy clustering and weak lensing, whose complexity requires the additional development time. This paper equally applies to 
the full DR1, because the latter builds on derived data products.

This paper introduces the calibration strategy and the observations that support it. Both have been shaped by the anticipated and actual behaviour of \Euclid in space and its evolution there, which are presented in a companion paper \citep{DR1-TP026b}. The requirement flow and the calibration concept are described in \cref{sec:DR1_performance}, together with a performance assessment against the top-level calibration requirements. Details of the DR1 processing pipelines and calibration tasks are given in the separate papers accompanying this release \citep[see][]{DR1-TP001}. The factors that govern the development and maintenance of the calibration plan, including the sensitivity of \Euclid to solar stray light and to X-rays from solar flares, are explained in \cref{sec:practicalimplementation,sec:calplanlimitations}. The self-calibration concept, which forms the central pillar of the calibration plan, is introduced in \cref{sec:selfcal}, and the plan itself is laid out in \cref{sec:InflightCalPlan}. The limitations of the DR1 data and the improvements planned for DR2 are discussed in \cref{sec:DR2}. Synthetic transformations between the \Euclid, \textit{Roman}, \ac{JWST}, and \ac{CSST} photometric systems are presented in \cref{sec:roman_crosscalib}; they are given in the AB magnitude system \citep{oke1983}. A summary is given in \cref{sec:summary}.

\section{Requirement flow and DR1 performance\label{sec:DR1_performance}}

\subsection{Mission Requirements Document}
The \Euclid \ac{MRD} scopes and permeates the design of the space and ground segments, and is immutable; details about the \ac{MRD} verification approach are given in \cite{vavrek2016}. On its scientific side, the \ac{MRD} has 22 requirements on weak lensing, nine on galaxy clustering, seven on the \ac{EWS}, and five on the \ac{EDS}, listed in \cref{table:MRD}. They set the imaging and spectroscopic capabilities, as well as the survey area and depth  \citep{Scaramella-EP1}. These, in turn, determined the mirror size, \ac{FoV}, instrumentation \citep{EuclidSkyVIS,EuclidSkyNISP}, survey strategy, calibration plan, and processing pipelines. 

As the implementation progressed, an increasing number of properties became fixed within the boundaries of the \ac{MRD}: first the high-level architecture (for example optics, instrument mass and volume, bandpasses and dispersers, and detector layouts), followed by the lower-level hardware and software details. At the same time, performance trade-offs were managed and the calibration plan evolved in step. The \ac{MRD} scientific requirements flow down to a two-tier calibration concept (\cref{fig:calflow}) as follows.

\subsection{Calibration concept document -- part A\label{sec:calcda}}
The CalCD-A contains 15 top-level requirements on \ac{PSF} and shear calibration, spectrophotometry, photometric redshifts, and spectroscopy (\cref{table:calcda}). Like the \ac{MRD}, the CalCD-A is under ownership of the \ac{ESA} and immutable.

Formally, CalCD-A requirements must be met by DR3 at the end of the nominal mission. \Cref{table:MRD,table:calcda} show that this is already the case for many of them, with more details in the companion papers and with considerable possibilities for further improvements in DR2 (\cref{sec:DR2}). The CalCD-A requirements that are still unmet -- purity and completeness of the spectroscopic redshift sample, and the VIS \ac{PSF} model -- are discussed in \cref{sec:requirements_unmet}.

\subsection{Calibration concept document part B\label{sec:calcdb}}
The CalCD-B unfolds the CalCD-A requirements into individual error budgets and dependence trees, comprising approximately 350 entries. It is under ownership of the \ac{SGS} and was frozen in 2018 after it reached sufficient maturity to guide the -- parallel -- development of the calibration plan (\cref{sec:practicalimplementation}).

The large envelope of the CalCD-B demands simplicity in order to remain tractable and purposeful as a static reference. It treats all effects contributing to an error budget as statistically independent, and uses quadratic error propagation whenever possible. This is often unrealistic, since several effects can be entangled at the physical level, such as nonlinearity, gain, \acl{CTI} \citep[\acs{CTI},][]{massey2014}, and the `\acl{BFE}' \citep[\acs{BFE},][]{antilogus2014,plazas2018,hirata2020}.\acused{CTI}\acused{BFE}
The permissible knowledge error of such complexes can be evaluated with simulations. This, however, requires realistic models for these effects and of their target -- in this example, the \ac{PSF} -- in the simulator well before launch. Thus, the CalCD-B does not capture \ac{BFE} or \ac{CTI}, but both effects are accounted for in the calibration plan as well as the VIS and shear data-processing chains.

The CalCD-B also uses ground-characterisation data whenever possible \citep{Waczynski16,grupp2019,Schirmer-EP18,Gaspar-Venancio23,EP-Serrano,EuclidSkyVIS,EuclidSkyNISP,EP-Gillard,EU-Kubik}. However, some of these tests need to be conducted in a thermal vacuum and the facilities that offer these conditions are a limited resource. Furthermore, tests cannot necessarily replicate space conditions accurately, such as a zodiacal background, in-flight temperatures, zero gravity, non-standard instrument operations, and the irradiation by the Sun; some characteristics can be determined accurately only in flight \citep{DR1-TP026b}.

When a given error budget remained otherwise unconstrained, its noise contributors were thus assumed to contribute equally. The CalCD-B merely serves as a guideline for the \ac{SGS}, which may rebalance the error budgets accounting for pipeline performance and characteristics of real data, as long as the CalCD-A requirements are met. When constructing the in-flight calibration plan (\cref{sec:calf}), a conservative approach was followed to ensure sufficient \ac{SNR} and margins, in particular where the CalCD-B was uncertain.

\subsection{Validation and verification process\label{sec:validationandverification}}
Calibration products undergo a multi-stage quality assessment before use. An initial check is performed after the pipeline has produced the calibration product, confirming successful execution and matching expectations. In the next step, the detector and instrument scientists review the data for correct instrument-specific behaviour, and check for evolutionary trends. The final inspection by the calibration scientist closes the loop, and may include a check against the CalCD-B. Formally, the first two stages constitute validation -- assessing whether the product looks correct and behaves as expected -- while the final comparison against the CalCD-B is verification, confirming that the product meets its requirements. These stages largely overlap in practice and run on different timescales as the data become available; the flat hierarchy and redundancy are intentional, so that issues can be caught that a single reviewer might miss. Outcomes are reviewed on a weekly basis.

For every major data release (DR), once the data have been reprocessed with a frozen pipeline configuration, the \ac{SGS} does a full verification against  CalCD-A requirements. Verification may also occur between releases, for example when major development milestones have been achieved and are regression-tested on several hundred square degrees of data \citep[section 7 in][]{DR1-TP033}. \ac{ESA} may additionally call a results review to verify MRD conformance.

\subsection{Requirements not yet met by DR1\label{sec:requirements_unmet}} 
Most mission-level requirements have been met already by DR1-Foundation ahead of DR3 (\cref{sec:calcda}). The two areas where requirements are not yet met have developed more slowly, because not all necessary data have been taken for spectroscopy, and because the \ac{PSF} model has a large dependency tree that requires careful calibration, characterisation, and control of systematic errors.

\begin{figure}[t]
    \centering
    \includegraphics[width=\linewidth]{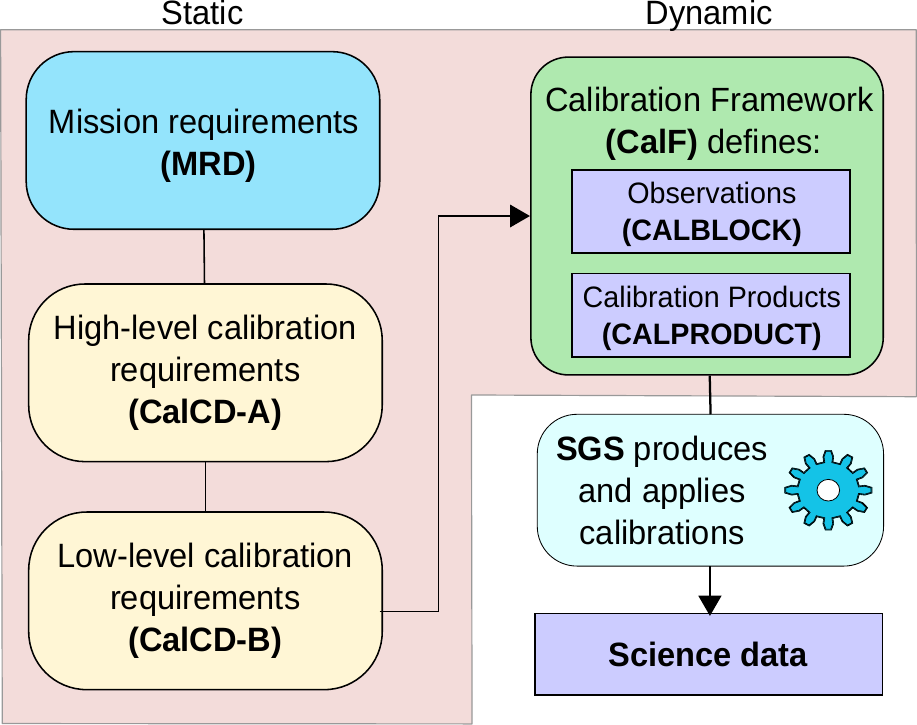}
    \caption{Flow-down of mission requirements to calibration products. The scope of this paper is indicated by the shaded area. The \ac{SGS} processing is described by \cite{DR1-TP033}.}
    \label{fig:calflow}
\end{figure}

\subsubsection{Redshift requirements\label{sec:redshift_requirements_unmet}}

\hyperlink{target:MRD-GC-006}{\tt MRD-GC-006} (\cref{table:MRD}) requires a purity of $\mathcal{P} > 99\%$ for the \ac{EDS} reference redshift sample. This cannot yet be formally assessed, because both blue- and red-grism data at full depth are needed, as well as an external redshift catalogue that is complete to the selection limits of $\HE<24$ and $f_{{\rm H}\alpha} >2\times10^{-16}$\,erg\,cm$^{-2}$\,s$^{-1}$. For a first assessment, the centre of the COSMOS field \citep{Scoville2007}, observed to full \ac{EDS} depth during \ac{PV}, was used together with around 8000 reference redshifts from \cite{khostovan2026}, which are incomplete at this depth. A purity of $\mathcal{P} = 85\%$ ($93$--$100\%$) is found for a redshift-reliability threshold of \texttt{SPE\_Z\_PROB}~$>0.99$ ($>0.999$). The purity in the \ac{EDF-N} is expected to be lower than in COSMOS, because DR1 has not yet reached full depth in this area, which lowers the fraction of correctly assigned redshifts for fainter targets. This is discussed in sections 4 and 5 of \cite{DR1-TP025}. Meeting \texttt{MRD-GC-006} is a prerequisite for assessing the following two requirements.

\hyperlink{target:MRD-GC-007}{\tt MRD-GC-007} requires $\mathcal{P} > 80\%$ for the \ac{EWS}. The current best estimate, obtained from the COSMOS and \ac{EDF-N} data, is $\mathcal{P} \approx 75\%$ at a number density of galaxies with correct redshifts of $\mathcal{D}>1700$\,deg$^{-2}$. To enable a better purity assessment, the survey plan was modified so that the \ac{EDF-N} is completed for DR2 instead of DR3 \citep[see also section 2.3 in][]{DR1-TP027}.

\hyperlink{target:MRD-GC-002}{\tt MRD-GC-002} requires a redshift completeness of $\mathcal{C} > 45\%$ for the \ac{EWS}, with a goal of $\mathcal{C} > 65\%$. This cannot yet be assessed directly, because the available deep catalogues are themselves incomplete and impure. Instead, the completeness is estimated from the number density $\mathcal{D}$ of galaxies with correct redshifts in the COSMOS data. For \texttt{SPE\_Z\_PROB}~$> 0.99$, a value of $[\mathcal{D},\, \mathcal{P}] \approx [750\,\mathrm{deg}^{-2},\, 62\%]$ is found, below the target of $[1700\,\mathrm{deg}^{-2},\, 80\%]$. The higher zodiacal and Galactic backgrounds in the COSMOS field were compensated for by approximately twice the exposure time. Preliminary estimates of $[1600\,\mathrm{deg}^{-2},\, 63\%]$ and $[1700\,\mathrm{deg}^{-2},\, 68\%]$ are obtained from the \ac{EDF-N} and \ac{EDF-S}, respectively, which suggests that the number density in COSMOS is low because of cosmic variance. A more robust assessment remains contingent on the full-depth \ac{EDF-N} data.

\subsubsection{PSF requirements\label{sec:psf_requirements_unmet}}
Shear calibration requires a highly complicated, chromatic model of the VIS \ac{PSF} that is still under development at the time of writing. Its aspects and performance are discussed in current and upcoming DR1 publications: shear overview, requirement, method, and validation (Taylor et al., in prep.); shear calibration (Martinet et al., in prep.); the \ac{PSF} model (Miller et al., in prep.); \ac{PSF} calibration (Duncan et al., in prep.); \ac{PSF} calibration fields (Kim et al., in prep.); and \Euclid's thermo-optical properties and and guiding performance \citep[][and Whittam et al., in prep.]{EU-Anselmi,DR1-TP026b}. An overview of the observational programme that supports the \ac{PSF} model is presented in \cref{sec:pdc,sec:psfmonitoring}. 

\section{Implementation of the calibration plan\label{sec:practicalimplementation}}

\subsection{At the system-engineering level\label{sec:EA}}
\cite{lorenzoalvarez2016} describe the model-based system-engineering approach chosen for the development of the \Euclid mission. The implementation was done in {\tt \ac{EA}}\footnote{\url{https://sparxsystems.com/products/ea/}}, a comprehensive tool that provides, among many other capabilities, automated flow checks and compliance matrices for the development of the CalCD-B. However, the large volume of the CalCD-B (\cref{sec:calcdb}) and the complexity of {\tt \ac{EA}} made this approach intractable for the \ac{SGS}. Furthermore, as \Euclid matured, \ac{ESA}'s system-engineering support shifted to other missions in earlier development stages.

We considered exporting the calibration- and processing-related components from this model for maintenance in simpler tools. This, however, would have required considerable time and training across a broad group of core personnel to ensure consistent use and correct operation -- changes of this depth were untenable in the run up to launch. We then descoped the calibration framework (\cref{sec:calf}) from the model, while CalCD-A and -B persist in \ac{EA} as static references for our operations (\cref{fig:calflow}).

\subsection{Calibration Framework (CalF)\label{sec:calf}}
The calibration plan has been developed and is maintained on a lighter and more pragmatic system, based on {\tt python-sphinx} and originally developed by the Kilo-Degree Survey \citep[KiDS;][]{dejong2013} team for the same purpose. This calibration framework (hereafter \CalF) provides an {\tt html} information base as the central reference for all practical calibration matters. It is flexible enough to accommodate differences between expected and actual performance, as well as unforeseen events, especially during early operations. Compliance matrices linking calibration observations (\CALBLOCK s) and calibration products (\CALPRODUCT s) are generated automatically from the {\tt sphinx} internal hyperlink targets whenever the \CalF\ is updated. The proper flow-down of all CalCD-B requirements to \CALPRODUCT s was verified in \ac{ESA}'s critical design reviews before launch. 

In 2022, around 1.5 years before launch, the \CalF\ was placed under configuration control as the survey design and \ac{PV} planning became increasingly stable and rigid. After major updates, frozen releases of the \CalF\ were issued, establishing it as the new reference for all downstream stakeholders, such as the survey computations, \ac{ESA}'s mission operations, the instrument operation teams, and the \ac{SGS} processing functions \citep[][]{DR1-TP001,DR1-TP033}. The \CalF\ does not automatically issue update notifications, nor does it have an interface to receive change requests. These are all interactive processes between team members, ultimately signed off by configuration-control boards to ensure full compliance with hardware, software, operations, the survey, and science.

\subsection{Updates since launch\label{sec:calf_updates}}
The \CalF\ absorbs all calibration-related dynamics from operations, respecting the frozen nature of the CalCD-B. Three updates were made after commissioning and \ac{PV}, to suppress parasitic stray-light paths from the Sun, and to become more robust against X-rays from solar flares. During routine operations, additional updates occur whenever instrument monitoring, product validation, space weather, or survey constraints require changes; four such updates were made over 2.7 years of science operations, for example adding more calibration-lamp configurations for flat-fields and nonlinearity measurements. No adjustments were required because of molecular contamination, which was anticipated and folded into the plan from the start, yet has posed considerable challenges at times \citep[sections 5 and 6 in][]{DR1-TP026b}. Overall, the launch and DR1 configurations of the \CalF\ are around 95\% identical, and only minor changes have occurred since toward DR2. Most of the current development concentrates on how the calibration data are processed and applied to the science data, rather than on how they are obtained in space.

\section{Limitations on the calibration plan\label{sec:calplanlimitations}}

The spacecraft, telescope, instruments, orbit, and mission lifetime pose numerous constraints on the \ac{PV} and calibration plans. As an additional layer of complication, every on-orbit operation causes a reaction in the spacecraft that must be evaluated for subtle impacts on calibration and performance. There is no ground to absorb torque or force, and any excess heat -- or lack thereof -- caused by non-standard operations eventually changes the thermo-optical equilibrium \citep[][and Whittam et al., in prep.]{DR1-TP026b,EU-Anselmi}.

\subsection{Reference observing sequence\label{sec:ROS}}
Onboard activities are orchestrated by the MIL-STD-1553B data bus protocol standard\footnote{\url{https://ecss.nl/standard/ecss-e-st-50-13c-interface-and-communication-protocol-for-mil-std-1553b-data-bus-onboard-spacecraft/}}, implemented in a deterministic serial data bus used in spacecraft to connect sub-systems. It operates with a dual-redundant, command-response architecture, prioritising reliability and fault tolerance over data rate (1\,Mbit\,s$^{-1}$); science data are transferred using a SpaceWire\footnote{\url{https://www.esa.int/Enabling_Support/Space_Engineering_Technology/Onboard_Computers_and_Data_Handling/SpaceWire}} interface with higher data rates (several MByte\,s$^{-1}$). 
At bus level, simultaneous on-sky observations with NISP and VIS are substantially more complicated than a single-instrument observation, because both instruments' timelines become interwoven.\footnote{The complexity is lower when both instruments take simultaneous off-sky calibration data, such as internal flats and biases.}

\Euclid utilises a restricted set of observing modes cast into the \ac{ROS} that exploits simultaneous observations. The \ac{ROS} was extensively tested in this regard, and is therefore used whenever possible for both science observations and on-sky calibration purposes. It comprises four dither positions, at each of which the same series of VIS and NISP science observations is repeated \citep[section 2.1 in][]{DR1-TP027}. Aside from small sky offsets, the dithers differ only in the VIS calibrations (\cref{table:ROS}) that are taken while NISP acquires its imaging exposures. At the end of the \ac{ROS}, during the slew to the next field, NISP takes a slew dark with the same exposure duration as a NISP photometric image. The slew dark is not pure, because it also contains the charge-persistence signal \citep{DR1-TP038} from the detectors' previous illumination history; its purpose is to deliver high-cadence snapshots to assess detector health. The total duration of the ROS is on average 73.9\,min, including the variable length of approximately $230$\,s for a typical field slew. 

Although the \ac{ROS} often generates data beyond the immediate calibration requirements, its use ensures high reliability, minimises development and maintenance costs, takes the calibration data in the same way as the nominal survey observations, maintains the thermal stability of the spacecraft, and increases the astrophysical value of the acquired data. Special, simultaneous on-sky NISP and VIS calibrations were implemented only for a few recurring, long-duration tasks to maximise efficiency. 

Some on-sky calibrations need only the observations performed at the first dither position, which we refer to as ROS\_D1, repeated at each node of a non-standard dither pattern. Concatenating many ROS\_D1 observations does not disturb the thermal equilibrium.

\subsection{Spacecraft restrictions\label{sec:calplan_restrictions_tel}}
Numerous restrictions apply to the design of the calibration plan.
\begin{itemize}
\item{For self-calibration purposes, a field with perennial visibility must be observed on a monthly basis to detect and absorb evolution of the \ac{PLM}. The only such field usable is near the North Ecliptic Pole (\cref{sec:selfcal}).}
\item{Any calibration observation must minimise the thermal disturbance of the \ac{PLM}, maximising \ac{PSF} stability for the survey. To this end, the spacecraft must stay within the allowed ranges of the \ac{AA} and \ac{SAA}, and the instruments must operate with a survey-like cadence. If one instrument stops taking data, for example because it is not needed for some calibration, or it observes at a much higher cadence, then the temperatures in the \ac{PLM} change and, subsequently, also the \ac{PSF}. Even changes as small as a few decimal kelvin of the baseplate carrying the primary mirror on one side and the instruments on the other will be seen as a change in the \ac{PSF} after a few hours \citep[see][]{DR1-TP026b,EU-Anselmi}.}
\item{The typical thermo-optical e-folding timescale was estimated to be 21\,h before launch \citep[see also section 4 in][]{EU-Anselmi}, meaning the telescope may need 2--10 days to reach thermo-optical stability after a major excursion \citep{DR1-TP026b,EU-Anselmi}.}
\item{The data volume created by calibration activities may exceed the downlink telemetry limit of 820\,Gbit for a few days only.}
\item{The high-gain antenna must repoint at least every 50\,min to track the designated ground station. To avoid spacecraft jitter during the exposures, antenna movements are synchronised with a dither or field slew. Hence, \Euclid cannot stare at the same point on the sky for longer than 50\,min without making an offset.}
\item{The absolute pointing accuracy is around \ang{;;3} or 30 VIS pixels, limiting system characterisations where higher pointing accuracy is needed; for example, placing a bright star exactly on a detector edge to check for unwanted reflections and glints.} 
\item{The minimum eigenslew length (dither offset) is \ang{;;73}. Offsets smaller than this threshold must be synthesised via larger, partially compensating slews. Precise sub-pixel offsets cannot be made.}
\item{The NISP filter and grism-wheel movements are not perfectly compensated by the spacecraft. Wheel movements are forbidden while VIS is taking flats or on-sky exposures.}
\item{The minimum replanning time for \Euclid is five weeks, meaning failed calibrations cannot simply be repeated back-to-back. When an outage occurs, the spacecraft rejoins the survey at the next possible point. Lost calibrations of great importance must be rescheduled, and might require an expensive recomputation of the survey. Therefore, the calibration plan must be robust enough to tolerate the loss of single instances of most calibration types without replacement. The only permissible exceptions are those that do not alter the precomputed survey timeline, such as a change of calibration-lamp flux or type.}
\item{Orbital maintenance and other \ac{SVM}-related operations occur during 6\,h-long fixed time slots every four weeks on a Monday during working hours, the \ac{SOPS} window; the cadence was decreased in January 2026 to once every eight weeks due to excellent spacecraft performance \citep[section 2.3.3 in][]{DR1-TP026b}. Instruments do not take data during this time; long calibration observations that cannot be interrupted must not fall into these windows.}
\item{The secondary-mirror focusing and tilt mechanism has a limited lifetime budget of total distance travelled. It must not be exceeded by the defocused observations taken during the phase-diversity campaign (\hyperlink{target:F-013}{\CALBLOCK\tt -F-013}, see \cref{sec:apdx_calblocks}).}
\end{itemize}

\subsection{Instrument restrictions: VIS\label{sec:calplan_restrictions_vis}}
\begin{itemize}
\item{The stray-light sensitivity of VIS \citep{EuclidSkyOverview} must be respected. The spacecraft attitude is limited to ${\rm AA}\in[\ang{-8},\ang{-3}]$ and ${\rm SAA}\in[\ang{87},\ang{112}]$, narrowing the windows during the year when specific calibration fields other than the self-calibration field become visible.}
\item{While the stray-light patterns are present in both open- and closed-shutter configurations, they are not identical.}
\item{Because the VIS shutter must remain open to receive light from the calibration lamp, astrophysical sources are inevitably superimposed on every flat field. Each flat requires a corresponding 2\,s science exposure within a few pixels of the same dither position, to identify affected pixels and exclude them during data processing.}
\item{The VIS onboard processing time depends on the number of sources in the image, and on the brightness of a flat-field exposure.}
\end{itemize}

\subsection{Instrument restrictions: NISP\label{sec:calplan_restrictions_nisp}}
\begin{itemize}
\item{The NISP wheel positioning is not perfectly deterministic, meaning the optical distortions are slightly different when restoring a previous configuration.}
\item{The onboard image processing of a single dither must be completed before the end of the first exposure of the next dither \citep{Medinaceli23}. Up to 46 co-added groups, telemetry, and ancillary data can be acquired per dither \citep{bonoli2016}.}
\item{NISP uses the \ac{MACC} mode for exposures, specified as MACC($g,r,d$),  as described in section 3.7.2 in \cite{EuclidSkyNISP}. The parameters describe, respectively, the number of groups ($g\leq15$), the number of frames per group ($f\leq16$), and the number of dropped frames between groups, with a frame exposure time of approximately 1.45\,s.}
\item{NISP can hold at most seven exposures at a single dither.}
\end{itemize}
 
\subsection{Options and design choices\label{sec:calplan_choices}}

\begin{itemize}
    \item{VIS and NISP can take darks and flats regardless of time and current spacecraft pointing, outside \ac{SOPS} windows reserved for orbital maintenance.}
    \item{On-sky calibrations and filler observations during waiting times \citep[for example, thermal stabilisation after decontamination; sections 5 and 6 in][]{DR1-TP026b} should always be taken at the self-calibration field to have a well-known reference; as a spin-off, this increases the depth of the \ac{EUDF} and supports time-domain astrophysics.}
    \item{The fluence in the NISP flats is controlled by the lamp flux, not the exposure time, so that the flats can be taken with the same MACC(4,16,4) mode as the survey imaging observations.}
    \item{Slew darks ( \cref{CALPRODUCT-NI-014}) are nominal darks that must use the MACC(4,16,4) and MACC(15,16,11) modes from imaging and spectroscopy, respectively. Slew darks are taken at the end of every \ac{ROS} for detector-monitoring and high-energy space-weather effects \citep{Gabarra2026}.}
    \item{The \ac{ROS} must be used as much as possible, in order to have calibrations taken in a manner as close as possible to the science observations, and to maximise thermal stability of the \ac{PLM}.}
\end{itemize}

Within these constraints, the calibration plan was designed for maximum synergy, extracting multiple calibration products from a single \CALBLOCK\ while simultaneously protecting science time \citep[see also section 4 in][]{EuclidSkyOverview}. Concise descriptions of the \CALBLOCK s are given in \cref{sec:apdx_calblocks}.

\section{Self-calibration\label{sec:selfcal}}

\subsection{Illumination non-uniformities
\label{sec:selfcal_principle}}
\subsubsection{Causes}
Despite \Euclid's vignetting-free design, the telescope pupil is seen at different angles from across the \ac{FPA}, producing a spatial variation in illumination \citep[geometric apodisation, a passive byproduct of the chosen optical design; see][]{reiss1945}. Mirror-coating variations introduce additional non-uniformities, through spatial variations and its dependence on angle of incidence. Although the coating reflectances were measured on the ground as a function of wavelength and angle of incidence across optical surfaces, no ray-tracing model including apodisation was made. The combined effects are expected to produce static illumination variations of 1--2\% across the \ac{FPA}, on angular scales $\gtrsim \ang{;1;}$.

VIS has a shutter-induced exposure-time gradient (\cref{sec:dr2_visshutter}), whilst for NISP there are contributions from uncorrected persistence. Considerable vignetting of 10\% occurs in the NISP dispersed images over the outermost column of four detectors, when the $\pm4^\circ$ rotated grism positions are used \citep[section 2.4.3 in][]{Q1-TP006}.

A time-dependence is caused by space weathering that slowly erodes optical surfaces \citep{engelhart2017,rodmann2019,mcelwain2023} on timescales of years, and by molecular contami\-nation \citep{Schirmer-EP29} that can act on much shorter timescales of weeks and months, see \cref{sec:DR2_nisp_ice_photometry} here, and section 6 in \cite{DR1-TP026b}. Both factors introduce chromaticity, because of the wavelength dependence of scattering and interference effects (\cref{sec:DR2_lsf_chromaticity}).

\begin{figure}[t]
\centering
\includegraphics[width=0.99\linewidth]{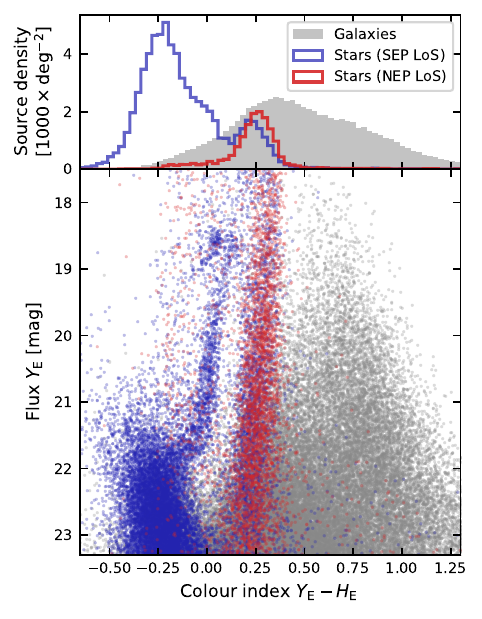}
\caption{Colour--magnitude diagram of galaxies (grey) in a typical \ac{EWS} tile, and the stellar populations toward the \ac{NEP} (red) and \ang{6.5;;} west of the \ac{SEP} (blue). The line-of-sight near the \ac{SEP} is crowded with hot main-sequence stars in the \ac{LMC}, which are considerably bluer than typical stars and galaxies in the \ac{EWS} fields, making the \ac{SEP} unsuitable for self-calibration purposes.}
\label{fig:cmd_nep_sep_histo}
\end{figure}

\subsubsection{Mitigation\label{sec:selfcal_mitigation}}
We account for these effects jointly, using repeated and widely dithered observations of the self-calibration field (\cref{sec:selfcal-location}). Each visit of the field lasts 23.5\,h and is scheduled every 25 to 35 days to sufficiently sample the expected evolutionary processes \citep[with exception of the heavy contamination period between March and June 2023, see section 6.2 in][]{DR1-TP026b}. This frequency is high enough to tolerate the loss of a single visit, for example due to solar activity.\footnote{Three self-calibration visits have been lost to radiation storms since launch.} The correcting photometric flat, or \ac{LSF}, is computed from the differential photometry of sources as dither offsets move them across the \ac{FPA} \citep[respective sections 4.4 and 3.2.2 in][]{Q1-TP002,Q1-TP003}.

Differential photometry can also be obtained from overlapping regions of adjacent survey fields. In the \ac{EWS}, however, these overlaps amount to only around 2.2\% of the field area, reflecting the tiling strategy that maximises unique sky coverage within the mission lifetime. This fraction is too small to support an uber-calibration strategy across all detectors as in \cite{padmanabhan2008}, \cite{wolf2018}, \cite{magnier2020}, and \cite{xiao2022}; we consider it as an independent sanity check for DR3.

\begin{SCfigure*}[1][h]
    \centering
    \includegraphics[width=1.5\linewidth]{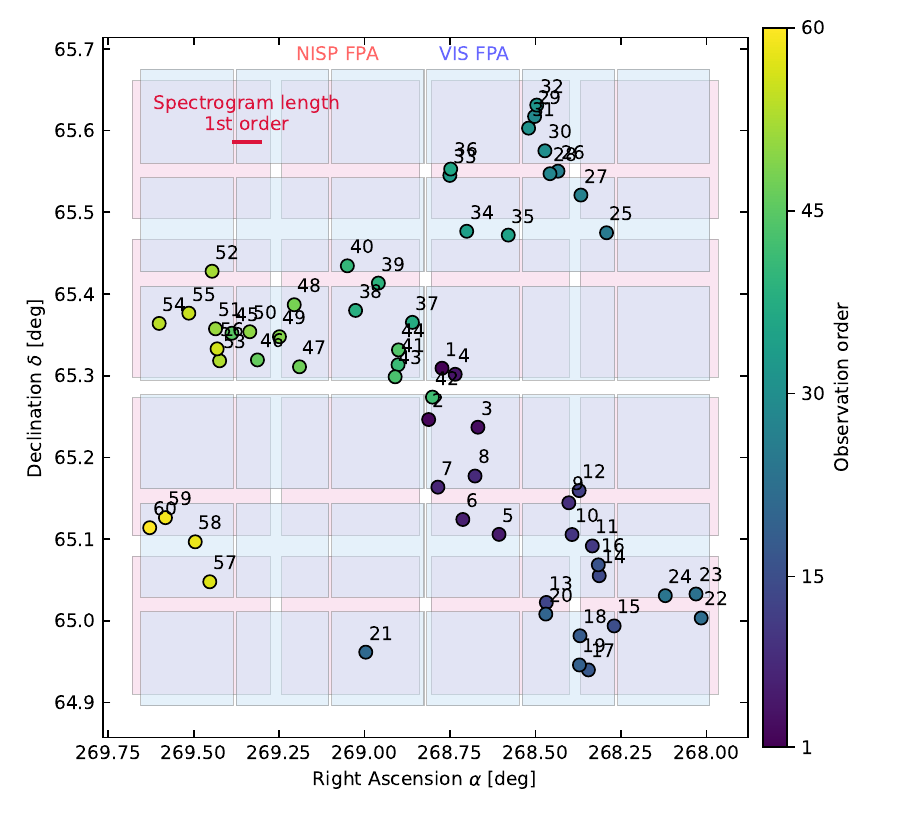}
    \caption{Self-calibration dither pattern, fixed on the sky. The shaded squares show the sizes of the \acp{FPA} relative to the pattern. The \acp{FPA} are centred on each dither point, with the position angle rotating by \ang{360;;} throughout the year. Dither offsets are encoded in `eigenslews' in the spacecraft frame, that is, the angle between a field quaternion and the next. Therefore, an eigenslew for a specific dither step becomes time-dependent, because the spacecraft changes its orientation with respect to the sky due to orbit progression. Eigenslews must respect a minimum length of \ang{;;73}. The colours and numbers of the dither positions show the order in which \Euclid can step through this pattern at any time without violating the minimum eigenslew length. This demonstrates the complexity in the planning and construction of some calibration observations. The red scale bar in the upper left corner shows the length of a first-order red-grism spectrogram. Details are given in \cref{sec:selfcal_ditherpattern}.}  
    \label{fig:selfcal_ditherpattern}
\end{SCfigure*}

\subsection{Location of the self-calibration field\label{sec:selfcal-location}}
The self-calibration field must have perennial visibility. \Euclid's \ac{CVZ} is set by thermal considerations: the requirement ${\rm SAA}\geq\ang{87}$ -- applicable to any time in the year -- restricts the maximum angular distance from an ecliptic pole to \ang{3;;}. Accounting for the dither-pattern size (\cref{sec:selfcal_ditherpattern}), \Euclid's \ac{FoV}, and perspective changes during the wide L2 halo orbit \citep[section 2.2 in][]{DR1-TP026b}, the centre of the self-calibration field must fall within \ang{2.3} of a pole.

\subsubsection{Exclusion of the southern continuous viewing zone\label{sec:selfcal_SEPexclusion}}
To avoid method-dependent biases, calibration fields are processed and analysed with the same tools as the nominal, sparse survey fields. This rules out the area around the \ac{SEP} near the \ac{LMC}, where heavy crowding would make calibration products unrepresentative of the science observations. In addition, the \ac{LMC}'s stellar population is considerably bluer than the Galactic halo population seen in the \ac{EWS} (\cref{fig:cmd_nep_sep_histo}), which would introduce undesired biases in astrometric and photometric calibration products \citep[see \cref{sec:DR2_lsf_chromaticity}, and sections 4.2.3][]{DR1-TP002}.

There is a risk that the \ac{LSF} and other calibration products obtained from this line of sight are not representative of \ac{EWS} conditions, for example due to the chromaticity of the VIS \ac{PSF} \citep[see sections 4.2.3 and 4.6 in][]{DR1-TP002}.

\subsubsection{Placement in the northern continuous viewing zone\label{sec:selfcal_northernplacement}}
The stellar density near the \ac{NEP} is several times higher than the average across the \ac{EWS}, providing good statistical \ac{SNR}, yet it is low enough for standard pipeline tools. The halo stars are also redder and spectrally closer to both typical survey stars and field galaxies (\cref{fig:cmd_nep_sep_histo}). The self-calibration dither pattern (\cref{sec:selfcal_ditherpattern}) requires 2.0\,deg$^2$, which fit comfortably within the 16.5\,deg$^2$ area accessible within \ang{2.3;;} of the pole. 

The final centring at $(\alpha,\delta) = (\ang{268.8127}, +\ang{65.286})$ was determined pre-launch, accounting for several constraints. Increased reddening toward the Galactic plane to the north-east, together with the presence of NGC~6543 -- a bright \ac{PN} of \ang{;6;} diameter -- motivated a shift toward the south-west. There, we established three faint hot \acp{WD} as absolute flux calibrators with the \ac{HST}\footnote{Programme IDs \href{https://archive.stsci.edu/proposal_search.php?mission=hst&id=16702}{16702 (PI: Appleton)} and \href{https://archive.stsci.edu/proposal_search.php?mission=hst&id=17442}{17442 (PI: Deustua)}.}, and at least one was required to fall within the self-calibration field. A final step ensured no stars brighter than $7.0$\,AB\,mag in any of the $g,r,i,J,H$, and $K_{\rm s}$ bands appeared in the field, based on the Asteroid Terrestrial-impact Last Alert System (ATLAS) catalogue of \cite{tonry2018}.

\subsection{Dither pattern\label{sec:selfcal_ditherpattern}}
The dither pattern used to build the \ac{LSF} must:
\begin{itemize}
    \item{sample all relevant spatial scales, from the diagonal diameter of the \ac{FPA} down to the smallest scale of 
    \ang{;;10} imposed by the scientific requirements;}
    \item{be fully contained within the \ac{CVZ};}
    \item{be fixed on the sky rather than rotating with \Euclid's orbit around the Sun, for operational reasons;}
    \item{fragment into clusters of four pseudo-random dithers with separations 
    $\leq\ang{;5;}$, to support early versions of the spectra-order decontamination pipeline \citep{Q1-TP004}.}
\end{itemize}
\begin{figure}[t]
\centering
\includegraphics[width=.99\linewidth]{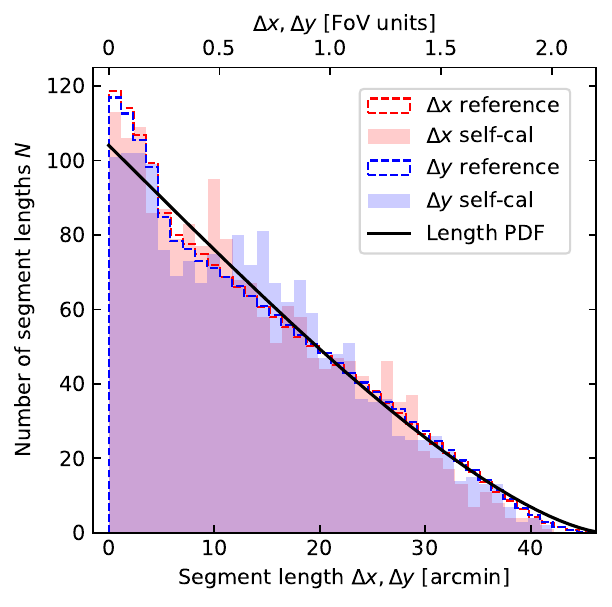}
\caption{Distributions of line-segment lengths connecting the dither points for the self-calibration dither pattern, along the horizontal and vertical axes (shaded histograms). Among the 1000 realisations probed, it has the smallest symmetric absolute difference with respect to the mean distribution of these realisations (reference, dashed lines). The solid black line shows the theoretical length distribution for random points in a circle. The excess at scales smaller than \ang{;5;} is due to the enforced sub-grouping for spectroscopy.}
\label{fig:selfcal_dither_chords_pdf}
\end{figure}
After correction by the \ac{LSF}, the relative photometric accuracies for VIS, NISP-P, and NISP-S must be 1.0\%, 1.5\%, and 0.7\%, respectively. \cite{holmes2012} find that random dither patterns outperform regular grids in terms of accuracy, and our simulations show they also converge faster. Accordingly, 60 dither points are required to ensure the relative photometric requirements for the VIS and NISP-P imaging channels are met, and by extrapolation also that of the NISP-S dispersed images. 

To find the optimal dither pattern, we generated 1000 pseudo-random and compliant realisations of 60 dithers. Then, for each pattern, we computed the projected lengths onto the horizontal and vertical axes of all 1770 possible line segments between two dither points (regardless of the instrument). The mean length frequency decreases with scale (\cref{fig:selfcal_dither_chords_pdf}), with an excess below \ang{;5;} arising from the sub-grouping imposed by the spectroscopic decontamination pipeline. The distribution shape naturally supports \ac{LSF} reconstruction: the many short-separation segments sample the high-frequency modes, while the fewer low-frequency modes that fit in the \ac{FoV} require correspondingly fewer long segments. To identify the best pattern among the 1000 realisations, we minimised the summed absolute deviations from both mean distributions simultaneously. This ensures that no angular scale is over- or under-represented for the computation of the \ac{LSF}; this is verified by the theoretical length distribution for line segments randomly placed in a circle  (black line in \cref{fig:selfcal_dither_chords_pdf}), which the chosen  dither pattern follows closely.

The self-calibration dither pattern is shown in \cref{fig:selfcal_ditherpattern}. It has performed well for all three instrument modes, delivering \acp{LSF} that absorb residual systematic zero-point variations on all relevant angular scales after the preceding calibration steps \citep{DR1-TP002,DR1-TP003,DR1-TP024}. For VIS, the minimum \ac{SNR} threshold of 80 on individual stellar measurements -- imposed to reduce sensitivity to faint cosmic rays -- proved unproblematic, despite yielding a lower usable source density than anticipated. The stable VIS throughput allowed us to combine the data of five self-calibration visits for DR1, benefiting both from the larger statistical sample and from the additional sampling baselines resulting from the approximately $\ang{30;;}$ field rotation between monthly visits. For NISP, we use the individual monthly \acp{LSF} for DR1 because of the ongoing detector contamination \citep[section 6 in][]{DR1-TP026b}.

The chosen dither pattern will remain unchanged for the rest of the mission for several reasons: (i) the pattern results in excellent calibration products; (ii) the resources of finding a more performant pattern are considerable and not available in the \ac{SGS}; (iii) the gain would be unclear; (iv) maintaining the same dither pattern ensures that any statistically significant changes in calibration products are due to actual changes in the \ac{PLM} and not due to a discontinuity in our calibration scheme. 

\subsection{Observations at the dither positions\label{sec:selfcal_block}}
At each of the 60 dithers we execute the ROS\_D1 that provides a nominal and short VIS exposure, as well as one image for each NISP filter, and therefore the imaging \acp{LSF} can be built after every visit. The spectroscopic \acp{LSF} for all five grism configurations -- four red, one blue -- require more visits because the ROS\_D1 includes only one spectroscopic exposure. We cycle through four self-calibration variants, in which the imaging configuration is identical; the spectroscopic configuration interleaves all red-grism configurations for all dither positions, thus requiring four visits to complete the spectroscopic data set. 

The blue grism was integrated differently, because its single dispersion direction does not provide sufficient angles for order decontamination. In each of the four self-calibration variants, the 60 dithers are therefore bracketed by eight blue-grism ROS\_D1 at each end, repeating 16 of the 60 dithers. The \ac{AA} is offset by $\pm\ang{1.4;;}$ to $\pm\ang{2.0;;}$ for the two blue-grism brackets to provide the necessary angles for order decontamination; the grism wheel itself must be kept fixed, since rotating it would introduce vignetting in the blue-grism data. These small \ac{AA} rotations do not disturb the thermal equilibrium, which is more susceptible to potentially large \ac{SAA} changes due to the return trip from a survey field to the \ac{NEP} near ${\rm SAA}=\ang{90}$ \citep[section 4 in][]{DR1-TP026b}. After four self-calibration visits, the blue grism has stepped through all 60 dither positions at least once, and the 16 additional ROS\_D1 runs provide extra data for the monthly imaging \acp{LSF}.

\begin{table}[t]
\caption{\Ac{ROS} data per field (or day)} 
\label{table:ROS}
\centering
\begin{tabular}{ll}
\hline\hline
\noalign{\vskip 1pt}
\multicolumn{2}{c}{VIS}\\
4 \IE-band nominal exposures & 566\,s each\\
2 \IE-band short exposures & 95\,s each\\
{\bf Bias} & 2 per day \\
{\bf Dark} & 4 per day \\
{\bf Flat} & 6 per day \\
{\bf Trap pumping} & 6 per day \\
{\bf Charge injection} & 8 per day \\
\hline
\noalign{\vskip 1pt}
\multicolumn{2}{c}{NISP}\\
4 red-grism spectro exposures & 574\,s each \\
4 \YE-band exposures & 112\,s each \\
4 \JE-band exposures & 112\,s each \\
4 \HE-band exposures & 112\,s each \\
{\bf 1 slew dark} & 112\,s \\
\hline
\end{tabular}
 \footnotesize
 \tablefoot{The \ac{ROS} is executed around 20 times per day, cycling through different configurations to provide a range of calibrations (marked with bold face characters). \ac{ROS} configurations using the blue grism are used only for the \ac{EDS} and \acp{EAF}. The NISP slew darks offer a great resource to monitor detector health, effects of space weather, and charge persistence \citep{Gabarra2026}. Most NISP calibrations are taken outside the \ac{ROS} with dedicated calibration blocks.}
\end{table}

\section{Calibration plan\label{sec:InflightCalPlan}}
\subsection{Performance-verification phase\label{sec:PVcalibs}}
Following spacecraft commissioning \citep[see e.g.][for NISP]{cogato2024}, 60 days were allocated for \acf{PV} to characterise the \ac{PLM} under in-flight conditions, and to update ground-calibration data \citep[see also][]{EuclidSkyOverview,EuclidSkyNISP}. A Gantt chart of the \ac{PV} schedule is shown in \cref{fig:PV_schedule}, and technical details are presented in section 5 of \cite{DR1-TP027}.

Whenever feasible, we used the \ac{ROS} for calibration observations (\cref{sec:ROS}). Its use has the added benefit of providing astrophysical  data that would normally not be taken, such as deep exposures toward the Magellanic Clouds for wavelength and distortion calibration \citep{Paterson-EP32,DR1-TP024}, and observations of the Orion star-forming regions \citep{Q1-TP001}. Approximately 40\% of \ac{PV} activities utilised the \ac{ROS}. Apart from numerous technical characterisations and calibrations, we validated the zodiacal background and Galactic stray-light models used for survey \ac{SNR} estimates, and obtained survey-like data from passes over the \ac{EDS} areas and the COSMOS field (\cref{tab:characterisations}).

Determination of the \ac{PLM}'s thermal timescale was a central task in \ac{PV}. Changes in spacecraft attitude with respect to the Sun result in a thermo-optical response as the \ac{PLM} equilibrates. The duration of this response must be known to ensure that VIS data are suitable for \ac{PSF} calibration and shear measurements. We determined an e-folding timescale of 21\,h by first allowing the spacecraft to equilibrate at $\ac{AA}=+\ang{3.5}$ and then reorienting it by a deliberately large amount, to $\ac{AA}=-\ang{7.5}$, to create a considerable thermal disturbance not encountered in nominal survey operations. We then monitored the \ac{PSF} continuously for several days until stability was restored. In \cite{DR1-TP026b} we analyse the principal thermal characteristics from two years of routine survey operations, showing that changes in \ac{SAA} have a much more profound effect than those in \ac{AA} \citep[see also][]{EU-Anselmi}.

\begin{figure*}[t]
\centering
\includegraphics[width=1.0\hsize]{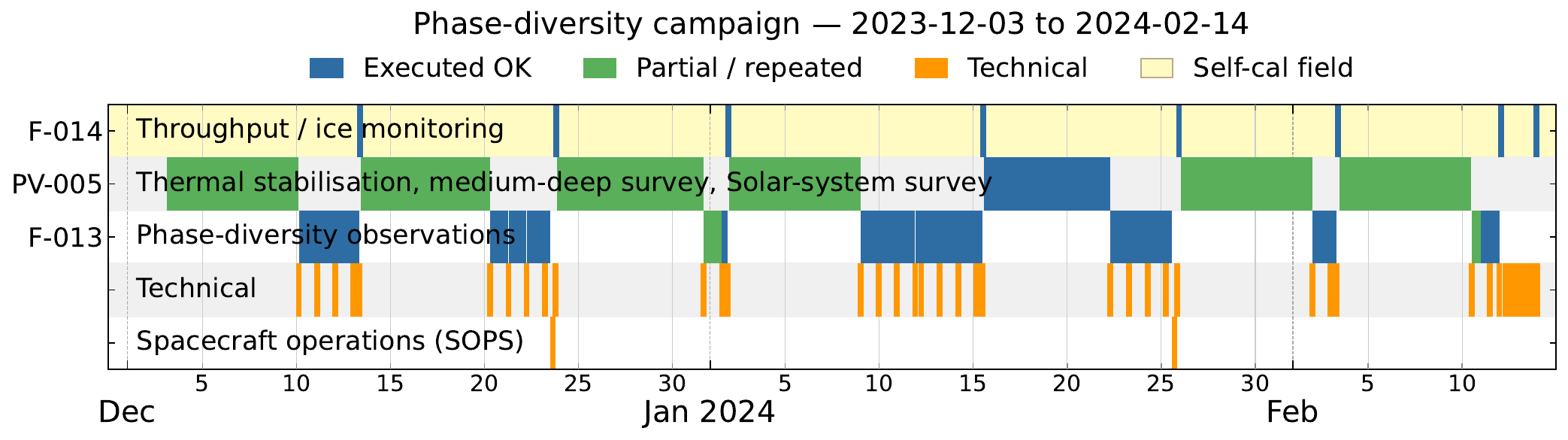}
\caption{Executed \ac{PDC} schedule. F-013 represents the in-focus, intra- and extra-focal phase-diversity observations, associated with numerous refocusing steps shown in orange. The \ac{PDC} requires maximum thermal stability with instruments operating in their nominal fashion, achieved by specially tailored filler surveys (part of PV-005, see also see \cref{sec:pdc}). Regular throughput monitoring (F-014) was also required to monitor ice contamination.}
\label{fig:PDC_schedule}
\end{figure*}

\subsection{Survey phase\label{sec:surveycalibs}}
The 18 different \CALBLOCK s scheduled in the survey are summarised in \cref{sec:apdx_calblocks}. They span the full range from hardware calibrations to target characterisations. Data products derived from them -- and from \ac{PV} -- are summarised in \cref{sec:nisp_common_calproducts,sec:nisp-p-calproducts,sec:nisp-s-calproducts,sec:vis-calproducts}; processing details are given in the companion papers \citep[see][]{DR1-TP001}.

\begin{table}[t]
\footnotesize
\caption{Euclid Auxiliary Fields (EAFs).}
\label{table:EAF}
\begin{tabular}{lccccc}
\hline\hline
\noalign{\vskip 1pt}
Field & RA & Dec & Area & DR3 & DR1\\
 &  &  & deg$^2$ & Depth & \\
\hline
\noalign{\vskip 1pt}
{\bf AEGIS} & \ang{214.827} & $+$\ang{52.82} & 1.0 & $4\times$& 100\%\\
{\bf CDFS} & \phantom{0}\ang{53.117} & $-$\ang{27.81} & 0.5 & $5\times$& 100\%\\
{\bf EUDF} & \ang{268.813} & $+$\ang{65.29} & 2.0 & $8\times$ & $50$\%\\
{\bf COSMOS} & \ang{150.119} & $+$\ang{02.21} & 2.0 & $5\times$& 61\%\\
GOODS-N & \ang{189.250} & $+$\ang{62.25} & 0.5 & $4\times$& 0\%\\
VVDS & \phantom{0}\ang{36.500} & $-$\ang{04.50} & 0.5 & $5\times$& 0\%\\
SXDS & \phantom{0}\ang{34.500} & $-$\ang{05.00} & 2.0 & $5\times$& 0\%\\
\hline
\end{tabular}
 \tablefoot{Location, area, and planned DR3 depth compared to the \ac{EWS}. Fields printed in bold-face are included in DR1, with their completeness in terms of exposure time shown in the last column. The COSMOS field consists of a $2\times2$ mosaic, of which in DR1 50\% of the area has 30\% depth, and a central pointing from \ac{PV} that has 100\% depth.}
\end{table}

Among the two with the most added astrophysical value are: first, the observations of several hundred thousand galaxies to $6\times$ the \ac{SNR} of the \ac{EWS}, to calibrate the weak lensing noise bias \citep{refregier2012,hoekstra2017} and to characterise the morphologies of galaxies; and second, for the galaxy clustering science a large spectroscopic sample of galaxies is needed with the aim of 99\% completeness and purity \citep{EP-Monaco2}, taken with the red and blue grisms. These two cases were merged with the separate scientific demand for a deep survey of at least 40\,deg$^2$, into the \ac{EDS}, covering 53\,deg$^2$ \citep{DR1-TP027}. These deep data facilitate the fine-tuning of pipeline parameters, such as \ac{SNR}-dependent thresholds for masking the VIS dichroic and crosstalk ghosts. The \ac{EUDF} is approximately $1.5$\,mag deeper still over 2\,deg$^2$, serving numerous technical, scientific, and cross-calibration purposes with other surveys (\cref{sec:roman_crosscalib}).

Additional deep fields (\cref{table:EAF}) are used to calibrate photometric redshifts for \Euclid's passbands \citep{DR1-TP011}, and the effect of colour gradients in galaxies on shape measurement in the presence of a chromatic \ac{PSF} \citep{Semboloni+13,Er+18,EP-Jansen}. The \acfp{EAF} were selected to have a large number of galaxies with known spectroscopic redshifts, as well as deep coverage with \ac{HST} and ground-based observatories. We observe them repeatedly with the \ac{ROS}, with every visit offset along a Vogel spiral \citep{Vogel1979} for homogeneous total exposure-time maps.

\subsection{Phase-diversity campaign\label{sec:pdc}}
Our strategy to build and apply a \ac{PSF} model consists of two complementary components. The first component is the \acf{PDC}, used to construct a physics-based wavefront model that relates spacecraft attitude and thermal state to the \ac{PSF} across the \ac{FPA}. It was executed once, before the start of survey operations, and is presumed to remain valid throughout \Euclid's lifetime because of the SiC design. A Gantt chart of the \ac{PDC} is shown in \cref{fig:PDC_schedule}. The second set of components are the visits of \ac{PSF} calibration fields (\cref{sec:psfmonitoring}) that act as frequent empirical checkpoints, validating this model against observed stellar fields at the actual attitudes and epochs of the survey. Implementation details are given in section 6 in \cite{DR1-TP027}.

Phase diversity is a wavefront-sensing technique in which the same stellar field is imaged at the nominal focus and at two defocused positions, one intra- and one extra-focal. To defocus, the \ac{M2M} is moved by $\pm18$\,\micron\ along the optical axis, shifting the optical focal plane by 113\,\micron\ with respect to the detector plane. This broadens the nominal \ac{PSF} -- with a typical \ac{FWHM} of 1.4\,pixels \citep{DR1-TP002} -- into a non-uniform disc of 8\,pixels diameter. An excellent \ac{SNR} is required so that photon noise is not mistaken for true surface-brightness variations in the discs. The differences between the intra- and extra-focal disc images encode the wavefront aberrations, from which the full pupil-plane \ac{WFE} is reconstructed. The \ac{WFE} is chromatic because the 200 dielectric coating layers on the dichroic beamsplitter are non-uniform; their thickness variations were determined on the ground \citep{baron2022} and are included in the \ac{PSF} model. The model is discussed in detail in Miller et al. (in prep.).

Utmost thermal stability is required, achieved by maintaining a fixed spacecraft attitude for seven days \citep[see][and \cref{fig:PDC_schedule}]{DR1-TP026b} before the \ac{PDC} observations at that attitude. This condition could not be met by any subset of the \ac{EWS}, due to frequent attitude changes. Separate filler surveys were designed for each \ac{PDC} attitude, using the \ac{ROS} to maintain the typical heat signatures of the instruments injected into the \ac{PLM} \citep[section 4 in][]{DR1-TP026b}. These surveys have unique tiling patterns and comprise an open-cluster field, the Euclid Medium Deep Survey, and the Ecliptic Survey that already produced spin periods for nearly 900 asteroids \citep{Irureta26}; more details are given in sections 6.2 to 6.4 in \cite{DR1-TP027}.

In total, ten distinct $[{\rm SAA}, {\rm AA}]$ configurations were chosen to sample the range of thermal states encountered during the survey. Three configurations included both defocused and in-focus observations, the remaining seven comprised only in-focus observations. This mixed approach reduced the risk of vacuum-induced failures when actuating a mechanism \citep{kaiser2026}. It also limited the fraction of the \ac{M2M}'s lifetime travel-distance budget consumed by \ac{PDC} activities early in the mission. The entire \ac{PDC} was scheduled as a single uninterrupted block of 2.5 months duration (\cref{fig:PDC_schedule}) after \ac{PV} and before the start of survey operations, to have low and similar levels of molecular contamination across the data set \citep[sections 5 and 6 in][]{DR1-TP026b}.

Each of the three focus positions required 150 exposures of 289\,s, dithered over a \ang{;;100}-diameter circle. While 100 exposures would have delivered sufficient \ac{SNR}, we were near the maximum of the 25th solar cycle, with powerful flares rendering the data from several VIS detectors unusable for \ac{PDC} for several hours at a time. Given \Euclid's planning timescale of five weeks, a large concurrent flare could result in data loss, requiring a full repeat including thermal stabilisation. Based on the 2023 flare statistics and the associated data losses since launch, we increased the number of exposures per focus position to 150, lasting 18\,h. This lowered the statistical risk of unrecoverable data loss to a few per cent, with the benefit of higher \ac{SNR} in the absence of a flare.

\begin{figure*}[t]
    \includegraphics[width=1\linewidth]{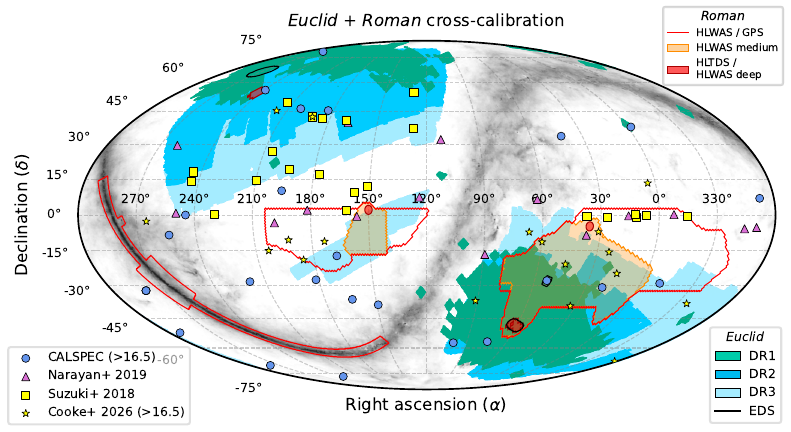}
    \caption{Euclid DR1 to DR3 footprints as in \ac{RSD} 2026A. Overlaid are the \textit{Roman} wide and deep surveys, showing the overlap area available for cross-calibrating the spectrophotometry of both missions. We also show the  network of absolute flux-calibration standards that has considerable overlap with DR2 and DR3. Apart from the \ac{HST} CALSPEC standards, the purple triangles show the faint \ac{WD} standards of \cite{narayan2019}, and yellow symbols show the blackbody-like standards of \cite{suzuki2018} and \cite{cooke2026}.}
    \label{fig:euclid_footprints_equatorial}
\end{figure*}

\subsection{PSF calibration fields\label{sec:psfmonitoring}}
The \ac{PSF} model predicts the \ac{PSF} for every survey field, and it must be validated and monitored as the survey progresses. This is done via 23 dedicated \ac{PSF} calibration fields, distributed across a range of ecliptic longitudes around both ecliptic poles, for maximum visibility at typical spacecraft attitudes \citep{DR1-TP027}. Ground-based spectral energy distributions and \textit{Gaia} photometry are required to characterise the stellar population in the magnitude range relevant for the \ac{EWS} ($17.5 < \IE < 21$). 

Each field visit consists of 100 exposures of 289\,s each, dithered along a Vogel spiral \citep{Vogel1979} to avoid clustering of dithers while at the same time achieving uniform sampling. The required \ac{SNR} is five times higher than that of the \ac{EWS}. The spacecraft's attitude during a visit is matched to the mean attitude of the contemporary survey patch, so that the obtained \ac{PSF} is representative. Observations are scheduled every 2--6 weeks, inserted into the observations of a contemporary patch at least three days after that patch began to allow for thermal equilibrium. Not every patch has an associated \ac{PSF}-calibration visit \citep[see also section 2.4 in][]{DR1-TP027}. If a visit fails, it cannot simply be rescheduled, since the patch and its associated thermal state will have moved on. In that case, a \ac{PSF} calibration field observed at a similar thermal state is used as a substitute, or the \ac{PSF} model itself is deemed sufficiently accurate for that state. 

\section{Limitations of DR1 and DR2 improvements\label{sec:DR2}}
\subsection{Continuous zero-point modelling\label{sec:DR2_zp}}
For DR1, we have implemented a piecewise linear fit of the zero point for VIS before the first decontamination, and a constant zero point following the second decontamination \citep[section 2.2][]{DR1-TP002}. For NISP-P, we adopted piece-wise constant zero points over the entire period \citep[figure 7 in][]{DR1-TP003}. For DR2, we have sufficiently long baselines and a large volume of data processed with the same pipeline configuration. This will enable us to establish a finer and continuous interpolation of the zero points at any epoch in the survey.

\subsection{Network of absolute flux standards\label{sec:DR2_absflux}}
The DR1 absolute flux calibration of VIS is indirectly tied to \ac{HST} CALSPEC by calibration against \textit{Gaia} DR3 $G$-band magnitudes. The \IE and $G$ passbands largely overlap, resulting in a well-defined transformation for stellar photometry as a function of \textit{Gaia} $G_{\rm BP}-G_{\rm RP}$ colour \citep[see figure 11 in][]{DR1-TP009}. For NISP-P, the absolute flux scale is tied to a hot \ac{WD} secondary \ac{HST} standard (J175318+644502, \cref{sec:selfcal_northernplacement}) in the self-calibration field, and three other CALSPEC standards. 

\Cref{fig:euclid_footprints_equatorial} shows that within the DR2 area, a considerable number of principal flux-calibration standards will become available. Among them are suitably faint CALSPEC standards \citep{bohlin2020,bohlin2025}, the faint \acp{WD} of \cite{narayan2019}, and blackbody-like helium \acp{WD} \citep{suzuki2018,cooke2026}. Anchored to this network of standards, we expect \Euclid's absolute flux calibration to become traceable via an unbroken calibration chain linked to standards maintained by the National Institute of Standards and Technology \citep[NIST;][]{peretz2024calibration,deustua2025,plavchan2025}, at the sub-per-cent level, and without requiring dedicated observations on our side.

On the spectroscopic side, the situation is more difficult due to overlap of first-order spectrograms. Only a few \ac{HST} CALSPEC standards are available with low enough order contamination from field crowding, and only at certain times of the year when the combination of position angle and dispersion directions are favourable for observing with a specific grism. Absolute spectrophotometric calibration of all grisms was obtained during \ac{PV} using CALSPEC standard GRW+70\,5824, and presumed to be temporally stable \citep[AFX solution, section 4.3 in][]{DR1-TP024}. For DR2 we need to account for time variability from detector contamination, in particular for the blue grism (\cref{sec:DR2_nisp_ice_photometry}). This can be obtained from secondary calibrators in the self-calibration data; spot-checks for the red-grism data can be done with brighter CALSPEC and blackbody standards \citep{suzuki2018,cooke2026} in the DR2 and DR3 area. 

\subsection{Large-scale flat\label{sec:DR2_lsf}}

\subsubsection{Time dependence\label{sec:DR2_lsf_timevar}}
The \acf{LSF} is time-dependent due to variable molecular contamination of optical surfaces. For DR1 and VIS, a high-confidence, single \ac{LSF} was used, constructed from five visits of the self-calibration field. For NISP, a monthly-built \ac{LSF} was applied piece-wise without interpolation in time. This approach achieves relative photometric accuracies of 0.08\%--0.14\% at the VIS quadrant level, and 0.9\% globally for NISP (\cref{table:calcda}), better than required.

For DR2, we will have a large data set processed with the same pipeline configuration, enabling the detection of more subtle effects. The VIS \ac{LSF} exhibits spatially variable transmission changes on a monthly basis, with low spatial frequency and an amplitude of around 0.4\%, presumably from slight contamination activity on \ac{FoM3}. Using monthly \ac{LSF} builds, or an interpolating function to any epoch in the survey, will improve the relative photometric accuracy in DR2. 

NISP is located upstream of \ac{FoM3} and is thus unaffected by its contamination. However, ice contamination is active on the NISP detectors, see \cref{sec:DR2_nisp_ice_photometry} and section 5.4.2 in \cite{DR1-TP026b}, and thus affects the \ac{LSF}, together with charge-persistence effects (\cref{sec:DR2_persistence}). Whether both effects can be isolated and a purely optical \ac{LSF} can be constructed before the DR2 software freeze is currently unclear.

\subsubsection{Chromaticity\label{sec:DR2_lsf_chromaticity}}
The \ac{LSF} is chromatic due to oblique angles of incidence on FoM1 and FoM2 with their hybrid coatings, and in particular on the dichroic beam splitter. The VIS \ac{LSF} is built from \textit{Gaia} stars, restricted in \textit{Gaia} colour to $G_{\rm BP}-G_{\rm RP}=0.2$--2.5 and to $\IE=18.5$--19.8. Since typical \ac{EWS} galaxies are redder than Galactic stars (\cref{fig:cmd_nep_sep_histo}), the `stellar' \ac{LSF} introduces a field-dependent bias in photometry. For a preliminary analysis, we divided the stellar sample with \textit{Gaia} colours $G_{\rm BP}-G_{\rm RP}=0.44$--2.84 into a blue, green, and red bin with equal number of sources, and recomputed the corresponding \ac{LSF}. There is a maximum deviation of approximately 1.0\% to 1.5\,\% between the blue and red \acp{LSF} near the right edge of the VIS \ac{FPA}, where we also observe a higher chromatic shift in the astrometric residuals between blue and red sources \citep[see figure 7 in][]{DR1-TP002}. Since the \ac{LSF} is an achromatic correction at the pixel level, we consider a fine-tuning of the allowed $G_{\rm BP}-G_{\rm RP}$ range that minimises the photometric bias for galaxies.

The NISP \ac{LSF} is also subject to optical chromaticity. However, due to the well-defined bandpass filtering \citep{Schirmer-EP18} and the longer wavelengths, we presume that this effect is considerably smaller than for VIS, and a correction might not be warranted for DR2.

\begin{figure}[t]
    \includegraphics[width=\linewidth]{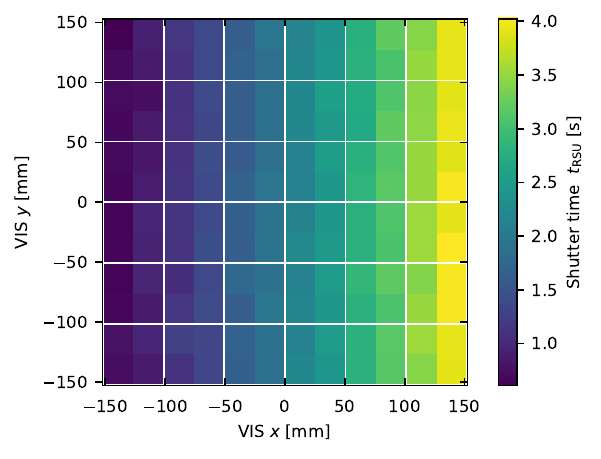}
    \caption{VIS shutter map or exposure-time difference, obtained from pairs of images with different exposure times. In DR2, the shutter effect will be explicitly removed, separating mechanical from optical effects.}
    \label{fig:visshuttermap}
\end{figure}

\subsection{VIS shutter correction\label{sec:dr2_visshutter}}
VIS employs a blade-type shutter mounted in the converging beam path ahead of the \ac{FPA} \citep{EuclidSkyVIS}. The blade rotates about a hinge positioned near the \ac{FPA}, with its leading edge sweeping through the convergent beam during closure. Due to the beam geometry, the effective exposure time exhibits a considerable nonlinear gradient along the \ac{FPA}'s horizontal axis (\cref{fig:visshuttermap}). 

The shutter operation introduces an exposure-time gradient of approximately 4.0\,s across the \ac{FPA} (\cref{fig:visshuttermap}), causing a zero-point change of 0.8\% for nominal exposures ($560.5$\,s) and 4.7\% for short exposures ($89.5$\,s). The optical component of the \ac{LSF} (\cref{sec:DR2_lsf}) varies by around 1\%; this is similar to the shutter effect in the nominal exposures, and dominated by the shutter in the short exposures. 

The shutter effect is purely geometric, achromatic, very stable in time, and unaffected by ice contamination. In DR2 it will be removed prior to the computation of the \ac{LSF}, thus separating mechanical from optical components and providing consistent treatment of exposures with different exposure times. 

\subsection{NISP persistence correction\label{sec:DR2_persistence}}

The persistence behaviour from charge traps in the NISP detectors depends on the trap population in a pixel, its illumination and saturation history, the brightness of the local stimulus, and the ionisation mechanism. The cumulative persistence signal from the various traps in a pixel has an approximate power-law decay in time, with a nonlinear dependence on the factors mentioned above \citep{EU-Kubik}. We find that overall the persistence response has decreased since launch, possibly because of repeated (super-)saturation by bright sources, which most pixels experience approximately every 2--3 months of survey operations. It appears that this passivates some charge traps on time-scales of months, an inoculation effect that we already saw in ground tests before launch; it has also been described for other detectors of similar architecture \citep{mcleod2016}. 

Before launch, we did consider to super-saturate the detectors with the calibration lamp \citep{EuclidSkyNISPCU}. Done on a monthly basis in flight, this could have resulted in a reduced, controlled, and approximately stable persistence state, while also guarding against the spotty and unpredictable natural inoculation caused by bright sources. However, it is currently unclear what this inoculation actually does at the diode level in the detector. We decided to not use inoculation, because mitigating a little understood effect (persistence) with a procedure that is even less understood carried too many risks for the detectors and their overall calibration.

Because of this complexity, while a persistence correction is planned in the long run, DR1 relies on masking pixels affected by persistence. This masking uses an empirical power-law model obtained from \ac{PV} data, accounting for up to five hours of illumination history. For NISP photometry images all spectroscopic observations from this past time interval are used for masking, with an additional Hough transform-based step to further detect and improve spectroscopy persistence streaks. For spectroscopic data both past photometric and spectroscopic observations are used for masking. The masking is conservative, resulting in a fairly high masked pixel fraction of approximately 5\% across the \ac{FPA}. Details about persistence characterisation and masking are given in \cite{EU-Kubik}.

For DR2, we are investigating several improvements, such as parameter optimisation to lower the masked pixel fraction to possibly 2\%. For actual persistence correction, a more accurate model is needed that includes signal dependence, time evolution, and the different response to cosmic-ray hits.

\begin{figure}[t]
    \centering
    \includegraphics[width=1\hsize]{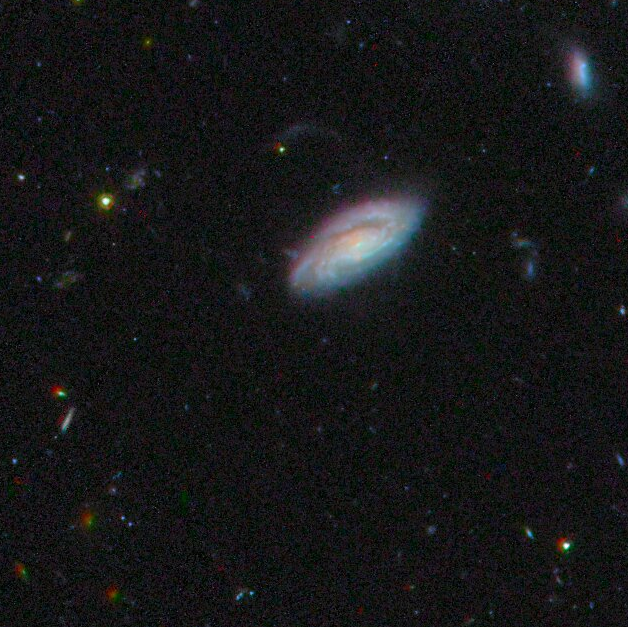}
    \caption{Insufficient NISP distortion correction, showing as a residual colour gradient across sources. This is an extreme example with an offset of 0.8\,pixel, or \ang{;;0.24}. Overall, less than 1\% of the DR1 area is affected by this, confined to areas of a few arcminutes diameter. }
    \label{fig:nispdistortion}
\end{figure}

\subsection{NISP distortion models\label{sec:DR2_nisp_distortion}}
The NISP filters have optical power \citep{EuclidSkyNISP} and thus contribute to the total optical distortion seen by the NISP detectors. The filter-wheel assembly does not reproduce filter positions perfectly, resulting in slight but measurable changes in the optical distortion between exposures. To account for this, the NIR-PF \citep{Q1-TP003} determines the camera distortion for every exposure individually. 

\begin{figure*}[t]
    \centering
    \includegraphics[width=1\hsize]{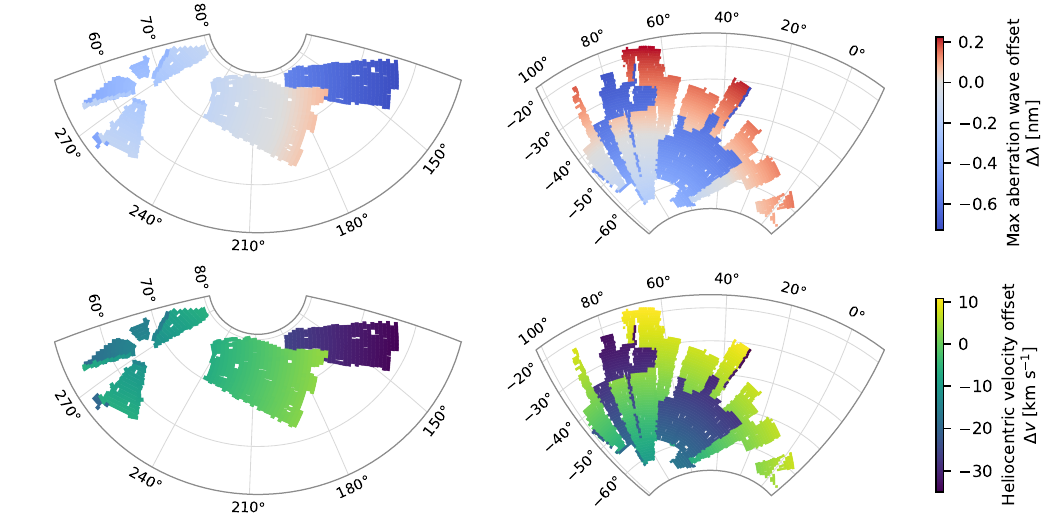}
    \caption{Wavelength systematic errors in DR1 spectrograms. \textit{Top}: Maximum wavelength zero-point error a red-grism spectrum can have in a single exposure for individual sources. The error is caused by uncorrected differential aberration, varies across the field, and changes sign depending on whether observations were taken in the leading or trailing direction with respect to the spacecraft's apex. The effect is 10\% smaller for the blue grism. \textit{Bottom}: Due to absent heliocentric conversions in calibration and science data, a field-dependent velocity offset arises. The offset is constant across 
    \ac{FoV} and the same for all grisms. Details are given in \cref{sec:aberration}.}
    \label{fig:aberration}
\end{figure*}

In survey areas with very low source density, aggressive persistence masking can result in a critically low source density in two detectors that have higher persistence than the others. As a result, area fractions of 0.70\%, 0.23\%, and 0.14\% of DR1 have an astrometric residual exceeding \ang{;;0.05}, \ang{;;0.08}, and \ang{;;0.1}, respectively. Formally, our requirement is that after all calibrations the total $1\,\sigma$-error in the measured position of an object must be smaller than 0.3\,pixels or \ang{;;0.09}. The affected areas are typically \ang{;1;} to \ang{;3;} in size. Complementing the nominal \ac{SGS} data products, we provide {\tt HEALPix} maps \citep{Gorski2005} with the astrometric error as a function of sky position. A zoom-in into one affected tile is shown in \cref{fig:nispdistortion}. Details are discussed in sections 3.6 and 4.1 of \cite{DR1-TP004}.

This limitation does not affect the accuracy of photometric redshifts \citep{Q1-TP005}, because the astrometric residuals are small compared to the \ac{MER} aperture radii \citep{Q1-TP004} used for multi-wavelength photometry. In addition, the photo-$z$ pipeline allows for a few per cent of systematic flux uncertainty to accommodate any small patch-to-patch zero-point shifts that may exist, and therefore we expect a negligible impact on \Euclid core science. However, photometry of individual compact sources that fall into affected areas -- identifiable in the HEALpix maps -- need to account for this effect in DR1. We have verified that an adjustment to the persistence-masking threshold for these two detectors resolves the issue, and therefore it will not be present in DR2.

\subsection{NISP-S wavelength-calibration systematic errors\label{sec:aberration}}
The DR1 spectra carry two small systematic wavelength errors of similar magnitude. They are at or below our requirements, but detectable in comparisons with reference spectra. Corrections are planned for DR2.

The first systematic effect is an uncorrected plate-scale change caused by
differential velocity aberration. The bulk aberration offsets are absorbed when registering the dispersed images on the sky, by comparing the aberrated positions of the 0th-order spectrograms against aberration-free coordinates in the barycentric \ac{ICRS}, accounting for shift and rotation but not for scale. Therefore, the plate-scale change remained, introducing a field-dependent error in the predicted position of the first order along the dispersion axis; this error varies with epoch and ecliptic coordinates. The same systematic entered the dispersion calibration, based on \ac{PV} observations of a \ac{PN} in October
2023, and was propagated to all DR1 fields. The combined effect is a zero-point error that shifts all wavelengths in the first-order spectrum of a source by the same amount $\Delta\lambda$. The shift varies linearly with the source position along the dispersion direction: it vanishes on the line 
through the field centre perpendicular to the dispersion direction and reaches
its maximum at the field edges, around 0.5\,pixel, that is 0.7\,nm for the red
grisms and 0.6\,nm for the blue grism. In velocity, this corresponds to
110--175\,km\,s$^{-1}$ for the red grisms and 140--205\,km\,s$^{-1}$ for the
blue grism, decreasing with wavelength across the first order. The maximum
shift per field in DR1 is shown in the upper row of \cref{fig:aberration}.
Because the shift is antisymmetric about the field centre, sources at opposite
edges of the same field and exposure differ by up to twice this value. Grisms with opposite dispersion directions have shifts of opposite sign; in combined spectrograms the shifts therefore cancel only partially, owing to the dither offsets and masking, and spectral features are broadened.

The second systematic effect arises because the \ac{PN} emission-line wavelengths determined by \cite{Paterson-EP32} were not Doppler-corrected from the barycentric frame to the spacecraft frame at the epoch of the \ac{PN} observations, and because DR1 spectra are projected onto a wavelength grid in the spacecraft frame instead of the barycentric frame. Together, these scale 
all wavelengths by a common factor, so measured wavelengths must be corrected as $\lambda_{\rm corr}=(1+\delta)\,\lambda_{\rm meas}$, with $|\delta|\lesssim1.5\times10^{-4}$. The factor is constant per field and the same for all grisms. The resulting velocity offset, $-c\,\delta$, where $c$ is the speed of light, is shown in the lower row of \cref{fig:aberration}.

\begin{figure}[t]
    \centering
    \includegraphics[width=1\linewidth]{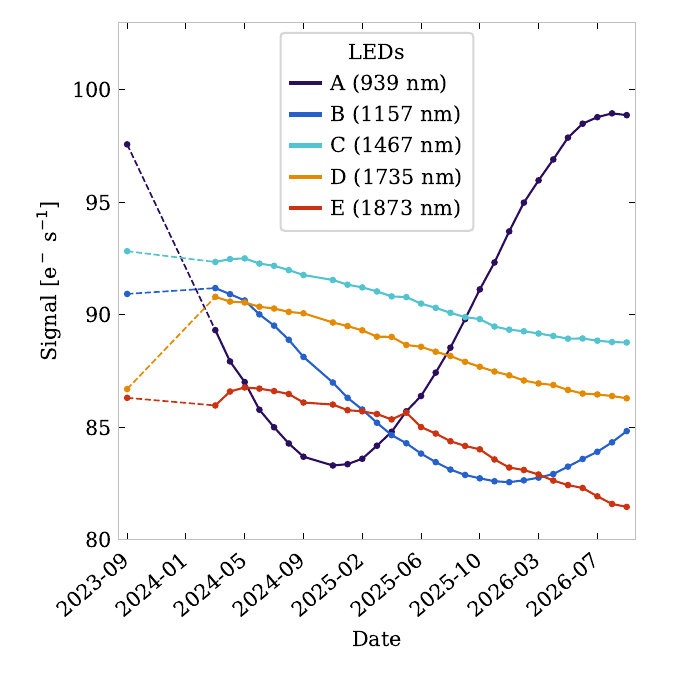}
    \caption{Evolution of flat-field brightness for one NISP detector, modulated by optical interference from the growing ice film on the detector surface. The DR1 period extends up until May 2025.}
    \label{fig:det13_signal}
\end{figure}

\subsection{NISP contamination effects and correction\label{sec:DR2_nisp_ice_photometry}}
Growing water ice on the detectors' \ac{AR} coating stacks introduces a time-variable component to the detector \ac{QE} by means of optical interference. This affects predominantly the \YE band and blue-grism data (\cref{fig:det13_signal}).

In DR1, photometric effects in the imaging data were corrected for -- implicitly to first order -- by the monthly relative flux calibration \citep[section 3.2.2 in][]{Q1-TP003}. The residual colour terms remained small enough to not violate the relative photometry calibration requirements \citep[see \cref{table:calcda}, and section 4.3 in][]{DR1-TP003}. On the spectroscopic side, the relative and absolute flux calibration of the blue-grism data are based on \ac{PV} observations and were not updated monthly, and thus the slow transmission modulation is fully present in the DR1 data. The red-grism data are susceptible to ice mostly at the blue end (1250 to 1350\,nm) of their sensitivity range that extends to 1850\,nm \citep[see also section 4.3 in][]{DR1-TP024}; we consider them weakly affected. The completeness and purity estimates for the spectroscopic reference sample (\cref{sec:redshift_requirements_unmet}) are based on the detectability of emission lines, which degrades or improves depending on the phase of the interference fringe. This could cause notable differences when comparing samples selected in the \ac{EDF-N} with those from the \ac{EDF-S}, as the blue-grism observations of these fields are not scheduled at the same epochs.

The DR2 period will have maximum effects from interference. We estimate the following temporal and spatial flux variations in the uncalibrated LE1 data, that is before any processing:  
\begin{itemize}
    \item{blue-grism data, 15 to 20\% in the 930--1350\,nm range,}
    \item{red-grism data, 3 to 7\% in the 1250--1850\,nm range,}
    \item{\YE band, 0.08 to 0.12\,mag, dependent on \ac{SED},}
    \item{\JE band, 0.03 to 0.06\,mag,}
    \item{\HE band, 0.01 to 0.03\,mag.}
\end{itemize}

These effects are too large to be corrected to within CalCD-A requirements (\cref{table:calcda}) with the DR1 pipeline. The mitigation scheme that is currently developed is as follows:
Using the complex refractive index of the H2RG substrate, we are building four-layer \ac{AR} coating-stack models that approximate the measured \ac{QE} for the uncontaminated detectors. Using the known refractive index of amorphous ice, we then infer the ice thickness and contamination rate using the monthly flat measurements at five wavelengths. Together, these yield models for the detectors' effective \acp{QE} at any time in the survey. These can be used to recompute effective central wavelengths for the imaging passbands needed for photo-$z$ estimates \citep{DR1-TP011}, and explicit wavelength-dependent corrections for all grism data. This is work in progress, with a DR2 or DR3 perspective.

\section{Photometric transformations between \Euclid and each of \textit{Roman}, JWST, and CSST \label{sec:roman_crosscalib}}
The large overlap between the \Euclid and \textit{Roman} surveys (\cref{fig:euclid_footprints_equatorial}) will enable photometric consistency checks over an area of approximately 1000\,deg$^2$. Using \ac{JWST} data over smaller areas, the relative photometric accuracy of the nonlinearity corrections implemented by \Euclid and \ac{JWST} can be assessed over a wider dynamic range, because sources that appear faint for NISP are bright for \ac{JWST}. To prepare for these tests, we need accurate reference transformations between these photometric systems and high-\ac{SNR} data, that is, either brighter stars selected from a wide area, or data from the deep fields. Ideally suited are warm F-, G-, and K-type dwarfs, which form quasi-linear and well-defined sequences in broadband colour--colour space. Cooler stars have strong molecular absorption bands causing photometric scatter, and in hotter stars the increasing ionisation of hydrogen changes the atmospheric opacity, making colour--colour relations nonlinear.

For these cross-comparisons we need to address two problems: deriving reference transformations from well-defined sequences of warm stars in synthetic colour--colour space; and identifying their stellar counterparts in the survey data to enable a direct photometric comparison across surveys. To quantify the effect of extinction, we considered two lines of sight: the \ac{EDF-S} [Galactic coordinates $\ell = 256\degr$, $b = -47\degr$, $E(B-V)=0.01$--$0.02$], which will be fully covered by both the medium and deep layers of the \textit{Roman} High Latitude Wide Area Survey (HLWAS); and the \ac{EUDF} [$\ell = 95\degr$, $b = +30\degr$, $E(B-V)=0.04$ to $0.05$], which contains the smaller \ac{JWST} \ac{NEXUS} field \citep{shen2024}. 

\subsection{Stellar population synthesis}
We simulated the Galactic stellar populations along the two lines of sight using the Besan\c{c}on Galactic model\footnote{\url{https://model.obs-besancon.fr/}} \citep{Robin2003}, which describes the spatial density profiles of the thin and thick stellar discs, the bulge, and the halo. For each star the model provides the fundamental atmospheric parameters -- effective temperature, surface gravity, metallicity, and stellar radius ($T_{\rm eff}$, $\log g$, $[{\rm M/H}]$, $R_\ast$) -- the heliocentric distance $d$, and the extinction $A_V$ integrated along the line of sight to the stellar distance using the \citet{Marshall2006} three-dimensional dust map. For each line of sight we performed a cone search with circular area of 1\,deg$^2$ to a distance of 100\,kpc, obtaining statistically representative stellar samples. 

Our tests show that the line-of-sight extinction dependence  introduces effects on the order of 1\,mmag only. Therefore, we concatenated the two catalogues and conclude that the transformations shown further below are valid for the entire \ac{EWS} area. The joint catalogue was filtered with $16.5 \leq\IE\leq28.0$ to match the approximate brightness limits of deep \Euclid data, retaining 30\,700 of 40\,367 stars.

\subsection{Broadband photometry from synthetic spectra}
We compute synthetic photometry by matching each simulated star to the nearest node of the Phoenix synthetic stellar atmosphere grid\footnote{\url{https://ssb.stsci.edu/trds/tarfiles/synphot5.tar.gz}} \citep{allard2012,Husser2013} in the 3-dimensional parameter space ($T_{\rm eff}/1000\,{\rm K}$, $\log g$, $[{\rm M/H}]/0.5$). The Phoenix models provide the flux density $F_\lambda$ emitted by a stellar surface element with unit area 1\,cm$^2$ into the hemisphere above, and the star's luminosity density is then
\begin{equation}
   L_\lambda = 4\pi\,R_\ast^{\,2}\,F_\lambda\;.
\end{equation}
At a distance $d$, the observer receives the extincted flux density
\begin{equation}
   f_\lambda = \frac{L_\lambda}{4\pi d^2}\,10^{\scalebox{0.85}{$-0.4\,A_V\,k(\lambda)$}}\,= F_\lambda\,\left(\frac{R_\ast}{d}\right)^2\,10^{\scalebox{0.85}{$-0.4\,A_V\,k(\lambda)$}}\,,
\end{equation}
where $k(\lambda)$ is the extinction curve evaluated with {\tt dust\_extinction} \citep{gordon2024} using the \cite{Gordon2023} law with standard extinction parameter $R_V = 3.1$. 

The apparent AB magnitude in a passband with total system throughput $T(\lambda)$ is
\begin{equation}
  m_{\rm AB} = -2.5\log_{10}\left[
    \frac{\int f_\lambda\,
          T(\lambda)\,\mathrm{d}\lambda}
    {3631\,{\rm Jy}\,c\int T(\lambda)\,\lambda^{-2}\,\mathrm{d}\lambda}\right]\,.
    \label{eq:synphot}
\end{equation}
For $T(\lambda)$ we use \cite{Schirmer-EP18} for NISP, the mission data base for VIS, and {\tt sedpy}\footnote{\url{https://github.com/bd-j/sedpy}} for \textit{Roman} and \ac{JWST}; updated throughput curves for \ac{CSST} were kindly provided by \cite{Liu2026privcomm}, including the wide ${\rm WV_{CSST}}$ and ${\rm WI_{\rm CSST}}$ filters that replaced $y_{\rm CSST}$. For respective details about \ac{JWST}, \textit{Roman}, and \ac{CSST} passbands we refer to \cite{rieke2005}, \cite{switzer2025}, and \cite{luo2026}.

\subsection{Rejection of cool stars}
There is a great abundance of cool red dwarfs with molecular absorption bands from \ce{TiO} and \ce{H2O}, producing considerable intrinsic photometric scatter, which is problematic for photometric comparison purposes. We used the spectral signature of these molecules to select against cool stars in real survey data. \ce{TiO} forms in stellar photospheres below $4000$\,K \citep[spectral type K7V and later, see][]{pecaut2013} and strongly absorbs across the \IE band. Absorption by \ce{H2O} becomes prominent below $3000$\,K (spectral type M5V and later), with its strongest features appearing above 1.3\,\micron\ and below that wavelength only in much colder L-type dwarfs. The \YE-band flux is therefore mostly unaffected by both molecules, and the \ce{TiO} absorption in the \IE band drives the $\IE - \YE$ colour to large positive values for cool red dwarfs.

Using the synthetic photometry from \cref{eq:synphot}, we find that a single threshold $\IE - \YE < 0.69$ provides a clean selection against stars cooler than 4000\,K: applied to our noise-free, synthetic catalogue it rejects 92\,\% of stars with $T_{\rm eff} < 4000$\,K while retaining 97\,\% of warmer FGK-types. Adding the redder \JE and \HE bands to form a multi-colour cut does not improve on this. A more conservative threshold of $\IE - \YE < 0.50$ rejects 97.4\% of stars colder than 4500\,K, and retains 91.5\% of stars warmer than that; we did not use this setting because of its lower completeness.

\begin{figure}[t]
    \includegraphics[width=\hsize]{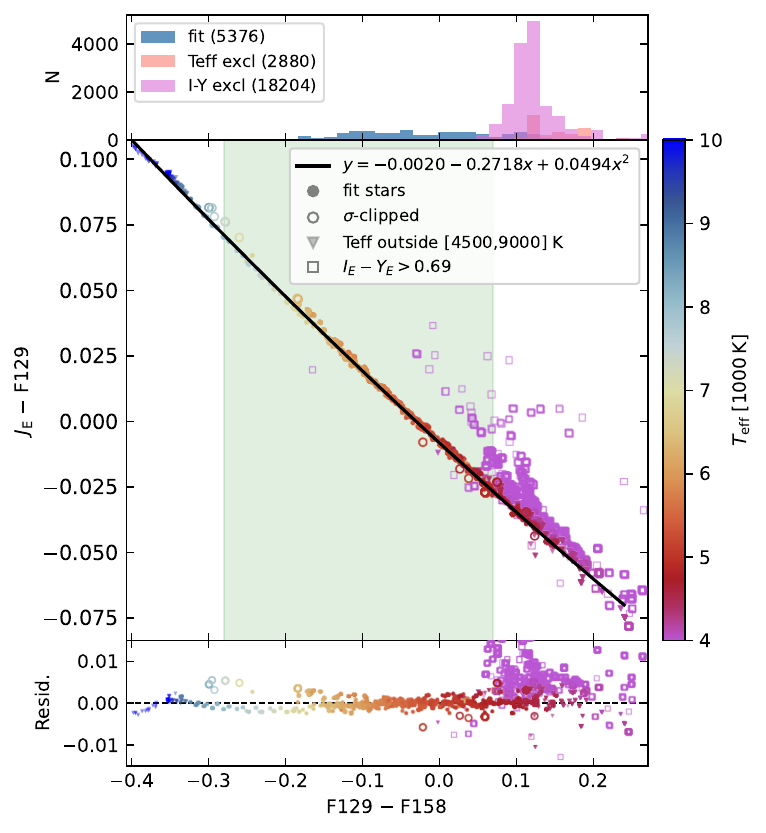}
    \caption{Synthetic transformation between the \Euclid \JE band and the \textit{Roman} F129 and F158 bands. \textit{Top}: number of stars included in the fit, and excluded because of their effective temperature. \textit{Middle}: the shaded region shows the recommended \textit{Roman} colour range to select suitable sources. \textit{Bottom}: fit residuals.}
    \label{fig:transformation_euclid_roman}
\end{figure}

\subsection{Computing the reference transformations}
We fit second-order polynomials to the FGK-type sequence, unless a first-order fit provides an equally good description. Only stars that fall within $T_{\rm eff}=4500\,{\rm K}$ to $T_{\rm max}$ are included in the fit. The upper limit, $T_{\rm max}$, was fine-tuned for every transformation because -- depending on the curvature of the FGK sequence -- including hotter stars resulted in a better fit, or hotter stars simply extended the linear FGK sequence and there was no reason to exclude them. The fit itself was performed iteratively using a $2.5\,\sigma$ outlier rejection until convergence was achieved.

An example plot is shown in \cref{fig:transformation_euclid_roman}, all other plots in \cref{fig:e2r,fig:r2e,fig:e2j,fig:j2e,fig:e2c,fig:c2e}, and \cref{tab:colour_terms} has all transformation equations. We validated our code by comparing selected \Euclid-\ac{JWST} transformations against those published by \cite{durbin2025} in the Vega system. The slopes of the fits agree to within a few per cent. The absolute values could not be directly compared, because \cite{durbin2025} used a non-standard AB renormalisation for Vega and JWST/NIRcam; after recomputing the standard AB conversion, the agreement on the absolute scale to 8\% or better. Recomputing the transformations for 2MASS \citep{cutri2003}, we do not find considerable qualitative or quantitative differences with our earlier work \citep[appendix C in][]{Schirmer-EP18}; there, we computed linear transformations for spectral types O to K to have an all-sky reference system for early commissioning work. Transformations to 2MASS have limited value due to the very small dynamic-range overlap with \Euclid, and therefore we do not reproduce them here.

\section{Summary and conclusions\label{sec:summary}}
We have presented the development, implementation, observations, and performance of the \Euclid mission calibration plan, starting out with the top-level calibration requirements derived from the scientific requirements. The calibration plan is centred on the self-calibration principle, is fully adequate, and required only a few minor modifications since launch.

Most top-level spectrophotometric requirements are already met by DR1. The assembly of a 99\% pure and complete spectroscopic redshift reference sample requires more observations, and building of the VIS \ac{PSF} model is still ongoing because of the considerable complexity of the optical system. Formally, all requirements must be met by DR3. 

We have summarised the most important upcoming changes for DR2 that will lead to considerable improvements, beginning with the VIS shutter correction, more accurate absolute flux calibration, better handling of persistence and contamination effects in NISP, and improved wavelength calibration. Lastly, we have computed photometric transformation equations between \Euclid on one side, and \textit{Roman}, \ac{JWST}, and \ac{CSST} on the other side. 

\begin{acknowledgements}
\AckEC
\AckDRone
Mischa Schirmer thanks the many people in the EC and at ESA who tirelessly contributed to developing and implementing the \Euclid calibration strategy over more than a decade. Working with you has been a privilege, shaped by your kindness, expertise, dedication, and rigorous scrutiny, all in pursuit of a single common goal.
\end{acknowledgements}

\bibliography{main,Euclid}
\begin{appendix}
\crefalias{section}{appendix}
  \onecolumn 

  \nolinenumbers

\section{MRD and CalCD-A requirements\label{sec:apdx_toplevelreqs}}

\begingroup
\small
\setlength{\extrarowheight}{2pt}

\begin{longtable}{|>{\columncolor{lightblue!20}}p{0.11\textwidth} |p{0.52\textwidth}| p{0.29\textwidth}|}
\caption{Calibration- and performance-related MRD requirements. Colour coding: green, the requirement is met; yellow, partially met or status not yet available; blue, observations still ongoing.}\\
\hline
\textbf{ID} & \textbf{MRD requirement} & \textbf{DR1 status} \\
\hline
\endfirsthead
\caption{continued}\\
\hline
\textbf{ID} & \textbf{MRD requirement} & \textbf{DR1 status} \\
\hline
\endhead
\hline
\endfoot

\multicolumn{3}{|l|}{\rule{0pt}{12pt}\textit{Weak lensing}\rule[-6pt]{0pt}{6pt}} \\
\hline

{\bf MRD-WL-001} & {\bf VIS off-band throughput}

The VIS channel off-band throughput shall be $<10^{-3}$ on average in the range 300--525\,nm with the transition between on- and off-band $<25$\,nm. & \cellcolor{lightgreen} Mean throughput 
$\approx2\times10^{-3}$, dielectric coating on \ac{FoM3} was replaced with silver coating. Transition width $\approx10$\,nm. \\
\hline

{\bf MRD-WL-002} & {\bf VIS \ac{PSF} \ac{FWHM}}

The \ac{FWHM} of the VIS channel \ac{PSF} at 800\,nm shall be $\leq \ang{;;0.18}$ over one exposure. This requirement does not include the effects of pixelisation. & \cellcolor{lightgreen} Median ${\rm FWHM}\approx\ang{;;0.140}$, figure 9 in \cite{DR1-TP002}.\\
\hline

{\bf MRD-WL-003} & {\bf VIS pixel scale}

The pixel scale of the VIS channel shall be $\ang{;;0.10}\pm\ang{;;0.01}$. & \cellcolor{lightgreen} \ang{;;0.0991}--\ang{;;0.1007} across the \ac{FoV}.\\
\hline

{\bf MRD-WL-004} & {\bf VIS \ac{PSF} ellipticity}

The ellipticity of the VIS channel \ac{PSF}, measured from an image that has been exposed for the nominal integration time, shall be $<0.15$. & \cellcolor{lightgreen} $\lesssim0.025$; Miller et al. (in prep.), and Whittam et al. (in prep.).\\
\hline

{\bf MRD-WL-005} & {\bf VIS \ac{PSF} profile}

The profile of the VIS channel \ac{PSF} over one exposure at 800\,nm shall be such that $(R_\mathrm{PSF}/R_\mathrm{ref})^2 < 4$.
\newline $R_\mathrm{ref}$ is the value for a Gaussian profile with an \ac{FWHM} of \ang{;;0.2}. As a goal, a value of $(R_\mathrm{PSF}/R_\mathrm{ref})^2 < 3$ should be pursued. & \cellcolor{lightgreen} Approximately 1.9 (Miller et al., in prep.).\\
\hline

{\bf MRD-WL-006} & {\bf VIS \ac{PSF} model (ellipticity)}

The spatio-temporal variation of the VIS channel \ac{PSF} shall allow a model of the \ac{PSF} as a function of spatial position and time to be reconstructed at any point with a residual model uncertainty of $\sigma(e_i) < 2\times10^{-4}$ per exposure and per ellipticity component with respect to the system \ac{PSF}. The model should be constructed from stellar observations of as many exposures as required.& \cellcolor{lightyellow} See \cref{sec:psf_requirements_unmet}.\\
\hline

{\bf MRD-WL-007} & {\bf VIS \ac{PSF} model (profile)}

The spatio-temporal variation of the VIS \ac{PSF} shall allow a model of the \ac{PSF} as a function of spatial position and time to be reconstructed at any point with a residual model uncertainty of $\sigma(R^2_\mathrm{res})/\langle R^2 \rangle < 10^{-3}$ with respect to the system \ac{PSF}. The model should be constructed from stellar observations of as many exposures as required.& \cellcolor{lightyellow} See \cref{sec:psf_requirements_unmet}.\\
\hline

{\bf MRD-WL-009} & {\bf VIS geometric distortion}

The \ac{RMS} post-calibration geometric distortion in the VIS channel averaged within a circular aperture of radius \ang{;4;} shall be $<0.003\%$. As a goal, a value $<0.001\%$ should be pursued. & \cellcolor{lightgreen} RMS distortion gradient over \ang{;4;} of $7\times10^{-5}$\%, maximum $2\times10^{-3}$\% in any single exposure; edge ellipticity exceeds $3\times10^{-5}$ below \ang{;;22}; section 4.2.1, figure 4 in \cite{DR1-TP002}.\\
\hline

{\bf MRD-WL-010} & {\bf NISP-P off-band throughput}

The NISP-P off-band throughput shall be $<10^{-3}$. & \cellcolor{lightgreen} $<10^{-4}$, figure 13 in \cite{Schirmer-EP18}.\\
\hline

{\bf MRD-WL-011} & {\bf NISP-P photometric bands}

The NISP-P channel shall have three or more wavelength bands. & \cellcolor{lightgreen} \YE, \JE, \HE bands.\\
\hline

{\bf MRD-WL-012} & {\bf NISP-P PSF profile and compactness (EE50)}

The NISP-P radii encircling 50\% of energy shall be $<\ang{;;0.4}$ for all wavelengths below 1486\,nm. & \cellcolor{lightgreen} \ang{;;0.28} or better in all bands, table 3 in \cite{DR1-TP005}.\\
\hline

{\bf MRD-WL-013} & {\bf NISP-P PSF profile and compactness (EE80)}

The NISP-P radii encircling 80\% of energy shall be $<\ang{;;0.7}$ for all wavelengths below 1486\,nm. & \cellcolor{lightgreen}  \ang{;;0.56} or better in all bands, table 3 in \cite{DR1-TP005}.\\
\hline

{\bf MRD-WL-014} & {\bf NISP-P pixel scale}

The NISP-P pixel scale shall be $\ang{;;0.30} \pm \ang{;;0.03}$. & \cellcolor{lightgreen} \ang{;;0.296}--\ang{;;0.299} across the \ac{FoV}.\\
\hline

{\bf MRD-WL-015} & {\bf NISP-P relative photometric error}

The NISP-P post-calibration relative photometric error shall be $<1.5\%$. & \cellcolor{lightgreen} 0.9\% from self-calibration visits \citep[section 4.3 and figure 8 in][]{DR1-TP003}, and $0.3\%$ from propagating the absolute calibration level through the DR1 period. \\
\hline

{\bf MRD-WL-016} & {\bf Shape-measurement multiplicative bias}

The multiplicative bias in shape measurement shall be known to an accuracy $\sigma(\mu) \leq 2\times10^{-3}$ under all observing conditions and for all types of objects used for weak lensing. As a goal, the multiplicative bias should be $|\mu| < 10^{-3}$ for monochromatic galaxies. & \cellcolor{lightyellow} See \cref{sec:psf_requirements_unmet}.\\
\hline

{\bf MRD-WL-017} & {\bf Shape-measurement additive bias}

The additive bias $c$ in shape measurement shall be known to an accuracy $\sigma(c) \leq 5\times10^{-4}$ under all observing conditions and for all types of objects used for weak lensing. & \cellcolor{lightyellow} See \cref{sec:psf_requirements_unmet}. \\
\hline

{\bf MRD-WL-018} & {\bf VIS \ac{PSF} model (ellipticity, non-convolutive contribution)}

The non-convolutive contribution to the residual model uncertainty of the ellipticity shall be $\sigma(e_\mathrm{NC}) < 1.1\times10^{-4}$ per exposure per ellipticity component with respect to the system \ac{PSF}. & \cellcolor{lightyellow} See \cref{sec:psf_requirements_unmet}.\\
\hline

{\bf MRD-WL-019} & {\bf VIS off-band throughput}

The VIS channel off-band throughput shall be $<10^{-3}$ on average in the range 935--1600\,nm with the transition between on- and off-band $<35$\,nm. & \cellcolor{lightgreen} Mean throughput $\lesssim8\times10^{-4}$, width $\approx23$\,nm.\\
\hline

{\bf MRD-WL-020} & {\bf VIS relative photometric error}

The VIS channel post-calibration relative photometric error shall be $<1.0\%$. & \cellcolor{lightgreen} 0.76\,mmag (internal) to 1.4\,mmag (external, with respect to \textit{Gaia}) at quadrant level \citep[see section 4.3 in][]{DR1-TP002}.\\
\hline

{\bf MRD-WL-021} & {\bf VIS absolute photometric error}

As a goal, the VIS channel post-calibration absolute photometric error should be $<5.0\%$. & \cellcolor{lightgreen} 3.8\,mmag mean offset w.r.t. \textit{Gaia} DR3; section 4.3 and table 1 in \cite{DR1-TP002}.\\
\hline

{\bf MRD-WL-022} & {\bf NISP-P absolute photometric error}

As a goal, the NISP-P channel post-calibration absolute photometric error should be $<5.0\%$. & \cellcolor{lightgreen} Tied to \ac{HST} standard to 0.5\% or better, figure 7 in \cite{DR1-TP003}; limited by the standards' intrinsic variability at the 1\% level.\\
\hline

{\bf MRD-WL-023} & {\bf VIS passband transition shape}

The VIS channel transition shape between on-band and off-band has to be known with an accuracy better than 1\%. & \cellcolor{lightgreen} Known at 1\,nm resolution with high \ac{SNR}. \ac{FoV}-dependence due to angle of incidence needs to be computed.\\
\hline

\multicolumn{3}{|l|}{\rule{0pt}{12pt}\textit{Galaxy clustering}\rule[-6pt]{0pt}{6pt}} \\
\hline

{\bf MRD-GC-001} & {\bf NISP-S line-flux limit at 1600\,nm}

The average effective H$\alpha$-line flux limit from a \ang{;;0.5} diameter source in the NISP spectroscopic channel shall be $\leq 2\times10^{-16}$\,erg\,cm$^{-2}$\,s$^{-1}$ at 1600\,nm.
\newline The flux limit is defined as the flux for which the \ac{SNR} is $>3.5$. & \cellcolor{lightgreen} $\approx2\times10^{-16}$\,erg\,cm$^{-2}$\,s$^{-1}$, section 5.2 in \cite{EuclidSkyNISP}.\\
\hline

\hypertarget{target:MRD-GC-002}{{\bf MRD-GC-002}} & {\bf NISP-S redshift completeness}

The completeness of the redshift measurements in the NISP spectroscopic channel shall be $>45\%$.
\newline Defined as the number of galaxies for which a redshift is measured at 50\% confidence level, divided by the total number of galaxies at the flux limit specified by MRD-GC-001. As a goal, a value $>65\%$ should be pursued. & \cellcolor{lightblue} See \cref{sec:redshift_requirements_unmet}.\\
\hline

{\bf MRD-GC-003} & {\bf NISP-S average flux limit}

The average flux limit for observations in the NISP spectroscopic channel at all wavelengths shall be $<1.2$ times the flux limit at 1600\,nm specified in MRD-GC-001. & \cellcolor{lightgreen} \\
\hline

{\bf MRD-GC-004} & {\bf NISP-S red-grism spectral resolution}

The NISP spectroscopic channel spectral resolution considering a reference \ang{;;0.5} diameter source shall be $\mathcal{R} = \lambda/\Delta\lambda > 380$ over the 1250--1850\,nm spectral range. Here, $\Delta\lambda$ is defined as the minimum wavelength separation at which two spectral lines produced by a \ang{;;0.5} object and with the same equivalent width can still be separated. & \cellcolor{lightgreen} $\mathcal{R}>480$, table 2 in \cite{EuclidSkyNISP}.\\
\hline

{\bf MRD-GC-005} & {\bf NISP-S position error of spectral features}

After calibration, the maximum error in the measured position of a spectral feature in the NISP-S red channel (1250--1850\,nm) shall be $<38\%$ of one resolution element. & \cellcolor{lightgreen} $0.35$\,pixel, or $18$\% total error (0th and 1st order), averaged over all grisms \cite[table 1 in][]{DR1-TP024}. \\
\hline

\hypertarget{target:MRD-GC-006}{{\bf MRD-GC-006}} & {\bf Spectroscopic purity subsample}

A purity $>99\%$ shall be achieved for a near-infrared spectroscopic subsample containing more than 120\,000 galaxies. The same selection criteria as for the spectroscopic galaxy sample chosen from the \ac{EWS} is to be applied. & \cellcolor{lightblue} See \cref{sec:redshift_requirements_unmet}.\\
\hline

\hypertarget{target:MRD-GC-007}{{\bf MRD-GC-007}} & {\bf Spectroscopic redshift purity}

The purity of the redshift measurements from NISP spectra shall be larger than 80\%.
\newline Purity is defined as the fraction of galaxies satisfying MRD-GC-001 and MRD-GC-002 for which the best redshift estimate has a standard deviation around their true value of $\sigma(z) < 0.001\,(1+z)$. & \cellcolor{lightblue} See \cref{sec:redshift_requirements_unmet} and \cite{DR1-TP025}.\\
\hline

{\bf MRD-GC-008} & {\bf NISP-S relative photometric error}

Within circles of diameter \ang{0.8} distributed over the whole survey, fluctuations in the zero point of the flux limit in the NISP-S channel shall be smaller than 0.7\% (\ac{RMS}). & \cellcolor{lightgreen} $\lesssim0.7$\% across DR1, section 5.5 in \cite{DR1-TP024}.\\
\hline

{\bf MRD-GC-009} & {\bf NISP-S absolute photometric error}

As a goal, the NISP spectroscopic channel post-calibration absolute photometric error should be $<5.0\%$ in the red band (1250--1850\,nm). & \cellcolor{lightyellow} Red grism: $\approx1\%$. Blue grism: up to 2\% at the short-wavelength end \citep[figure 5 in][]{DR1-TP024} only during \ac{PV}. Detector-contamination effects are still uncorrected, and the goal is missed for the blue-grism (\cref{sec:DR2_nisp_ice_photometry}).\\
\hline

\multicolumn{3}{|l|}{\rule{0pt}{12pt}\textit{Wide survey}\rule[-6pt]{0pt}{6pt}} \\
\hline

{\bf MRD-WS-001} & {\bf \ac{EWS} VIS depth}

Observations in the VIS channel in the \ac{EWS} shall achieve a depth of AB magnitude 24.5 for $10\,\sigma$ extended objects.
\newline Extended objects are defined as those with a \ac{SNR} greater than 10 and a size larger than 1.25 times the \ac{PSF} size. & \cellcolor{lightgreen} Turn-around point at $\approx25.5$, figure 22 in \cite{Q1-TP002}.\\
\hline

{\bf MRD-WS-002} & {\bf \ac{EWS} NISP-P depth}

Observations in each band of the NISP-P channel in the \ac{EWS} shall achieve a depth of AB magnitude 24 (in \YE, \JE, and \HE) for $5\,\sigma$ point sources. & \cellcolor{lightgreen} 25.0 in $2\times$ \ac{FWHM} aperture.\\
\hline

{\bf MRD-WS-003} & {\bf VIS number of dithers}

Observations in the VIS channel in the \ac{EWS} shall be performed with at least four dithers. & \cellcolor{lightgreen} Four dithers, section 4.1 in \cite{EuclidSkyOverview}. \\
\hline

{\bf MRD-WS-004} & {\bf NISP number of dithers}

Observations in the NISP-P and NISP-S channels in the \ac{EWS} shall be performed with at least four dithers in each band. & \cellcolor{lightgreen} Four dithers, section 4.1 in \cite{EuclidSkyOverview}. \\
\hline

{\bf MRD-WS-005} & {\bf NISP-S dispersion directions}

NISP-S observations in the \ac{EWS} shall be performed such that the spectra obtained have $\geq 3$ different orientations on the sky. & \cellcolor{lightgreen} Four angles, section 3.5.3 in \cite{EuclidSkyOverview}.\\
\hline

{\bf MRD-WS-006} & {\bf NISP-P images must complement NISP-S exposures}

NISP-P imaging of the fields covered by the NISP-S channel in the \ac{EWS} shall be acquired with depth fainter than 24 AB mag for $5\,\sigma$ point sources.
\newline The required depth can be achieved through combination of the different NISP-P bands bands. & \cellcolor{lightgreen} 25.0 in $2\times$ \ac{FWHM} aperture per band.\\
\hline

{\bf MRD-WS-007} & {\bf \ac{EWS} fraction of area useful for cosmological analysis}

At least 85\% of the \ac{EWS} covered area shall be useful for the core \Euclid science analysis. & \cellcolor{lightblue} Area loss estimated at 11.7\%  (88.3\% useful).\\
\hline

\multicolumn{3}{|l|}{\rule{0pt}{12pt}\textit{Deep survey}\rule[-6pt]{0pt}{6pt}} \\
\hline

{\bf MRD-DS-001} & {\bf \ac{EDS} imaging depth}

The stack of images obtained in the \ac{EDS} shall achieve a depth at least 2\,mag deeper than the \ac{EWS} depth in both the VIS and NISP-P imaging channels. & \cellcolor{lightblue} The \ac{EDS} does not yet cover all visits. Current depth-scaling measurements with the DR1 pipeline fall within 0.1\,mag of the requirement.\\
\hline

{\bf MRD-DS-002} & {\bf \ac{EDS} spectroscopic depth}

The stacked line flux obtained in the \ac{EDS} in the NISP-S channel shall achieve a depth of at least $5\times10^{-17}$\,erg\,s$^{-1}$\,cm$^{-2}$ ($3.5\,\sigma$) in the wavelength range common to the \ac{EWS}. & \cellcolor{lightblue}
The \ac{EDS} does not yet cover all visits.\\
\hline

{\bf MRD-DS-003} & {\bf \ac{EDS} blue-grism coverage}

The \ac{EDS} shall include observations with the blue band (935--1300\,nm) of the NISP-S channel over the same area as the red band. & \cellcolor{lightgreen} Implemented, with a blue:red exposure-time ratio of 5:3, section 4.2 in \cite{EuclidSkyOverview}.\\
\hline

{\bf MRD-DS-004} & {\bf \ac{EDS} blue-grism depth}

As a goal, the depth in the NISP-S blue channel (935--1300\,nm) for the stacked line flux should be better than $6\times10^{-17}$\,erg\,s$^{-1}$\,cm$^{-2}$ ($3.5\,\sigma$) at 1100\,nm. & \cellcolor{lightblue} The \ac{EDS} does not yet cover all visits. The blue:red exposure-time ratio of 5:3 was chosen to achieve comparable depth across both spectroscopic passbands.\\
\hline

{\bf MRD-DS-005} & {\bf NISP-S blue-grism spectral resolution}

The NISP-S blue channel spectral resolution considering a reference \ang{;;0.5} diameter source shall be $\mathcal{R} = \lambda/\Delta\lambda > 260$ in the 935--1300\,nm spectral range.
\newline The resolution element ($\Delta\lambda$) is defined as the minimum wavelength separation at which two spectral lines produced by a \ang{;;0.5} object and with the same equivalent width can still be separated. & \cellcolor{lightgreen} $\mathcal{R}>400$, table 2 in \cite{EuclidSkyNISP}.

\label{table:MRD}
\end{longtable}
\endgroup


\begingroup
\small
\setlength\extrarowheight{2pt}
\begin{longtable}{|>{\columncolor{lightgreen!20}}p{0.15\textwidth}|p{0.50\textwidth}|p{0.29\textwidth}|}
\caption{CalCD-A requirements on the calibration plan. They must be met at the end of the mission in DR3. Colour coding: green, the requirement is met; yellow, partially met or status not yet available; blue, observations still ongoing.}\label{table:calcda}\\
\hline
\textbf{ID} & \textbf{CalCD-A requirement} & \textbf{DR1 status} \\
\hline
\endfirsthead
\multicolumn{3}{c}{\tablename\ \thetable{} -- continued from previous page} \\
\hline
\textbf{ID} & \textbf{CalCD-A requirement} & \textbf{DR1 status} \\
\hline
\endhead
\hline
\multicolumn{3}{r}{\textit{Continued on next page}} \\
\endfoot
\hline
\endlastfoot
{\bf R-CAL-A-VS-1000} & {\bf VIS \ac{PSF} ellipticity post-calibration error}

Assuming a perfect extraction, the data should be calibrated such that the intrinsic contribution to the error in the determination of the ellipticity components of the \ac{PSF} model is $\sigma(\varepsilon) \leq 7.0 \times 10^{-5}$. &
\cellcolor{lightyellow}See \cref{sec:psf_requirements_unmet}.
\\
\hline
{\bf R-CAL-A-VS-2000} & {\bf VIS \ac{PSF} $R^2$ post-calibration error}

Assuming a perfect extraction, the data shall be calibrated such that the intrinsic contribution to the error in the determination of the relative $R^2$ of the \ac{PSF} model is $\sigma(R^2)/R^2 \leq 2.5 \times 10^{-4}$. &
\cellcolor{lightyellow}See \cref{sec:psf_requirements_unmet}.
\\
\hline
{\bf R-CAL-A-VS-3000} &  {\bf VIS distortion correction}

The \ac{RMS} of the post-calibration geometric distortion in VIS measured on a scale of \ang{;4;} shall be $< 0.003\%$. & \cellcolor{lightgreen} 0.08\,mas \ac{RMS}, or $8\times10^{-5}\,\%$,  section 4.2 in \cite{DR1-TP002}.\\
\hline
{\bf R-CAL-A-VS-4000} &  {\bf VIS relative photometric error}

After all calibrations, the relative photometric error of the VIS instrument shall be $<1.0\%$. &
\cellcolor{lightgreen} 0.76\,mmag (internal), 1.4\,mmag (external) at quadrant level; section 4.3 and table 1 \cite{DR1-TP002}.\\
\hline

{\bf R-CAL-A-VS-5000} &  {\bf VIS absolute photometric error}

After all calibrations, the absolute photometric error of the VIS instrument shall be
$< 5.0\%$. &
\cellcolor{lightgreen} 3.8\,mmag mean offset w.r.t. \textit{Gaia} DR3; section 4.3 and table 1 in \cite{DR1-TP002}.\\
\hline
{\bf R-CAL-A-VS-6000} &  {\bf VIS colour gradient}

The following fields shall be observed with VIS to an \ac{SNR} 4 times better than the \ac{EWS}: AEGIS, SXDS, COSMOS, CDFS, GOODS-NORTH. &
\cellcolor{lightblue}AEGIS and CDFS completed, observations ongoing.\\
\hline
{\bf R-CAL-A-VS-7000} &  {\bf VIS noise bias fields}

40\,deg$^2$ shall be observed with VIS to an \ac{SNR} $>6$ times that of the \ac{EWS}. The fields must also be observed by EXT, at least as deep as the \ac{EWS}.

&

\cellcolor{lightblue}Observations ongoing until completed in DR3.
\\
\hline
{\bf R-CAL-A-NP-1000} &  {\bf NISP-P relative photometric error}

After all calibrations, the relative photometric error of NISP-P shall be $<1.5\%$. & \cellcolor{lightgreen} 0.9\% from self-calibration visits \cite[section 4.3 and figure 8 in][]{DR1-TP003}, and $0.3\%$ from propagating the absolute calibration level through the DR1 period.
\\
\hline

{\bf R-CAL-A-NP-3000} &  {\bf NISP-P absolute photometric error}

After all calibrations, the absolute photometric error of NISP-P shall be $<5.0\%$. & \cellcolor{lightgreen} Tied to \ac{HST} standard to 0.5\% or better \citep[figure 7 in][]{DR1-TP003}; limited by the standards' intrinsic variability at the 1\% level.
\\
\hline
{\bf R-CAL-A-NP-4000} &  {\bf NISP-P colour-gradient fields}

The following fields shall be observed in all NISP-P bands to an \ac{SNR} of four times better than the \ac{EWS}: AEGIS, SXDS, COSMOS, CDFS, GOODS-NORTH. &
\cellcolor{lightblue}AEGIS and CDFS completed, observations ongoing.
\\
\hline
{\bf R-CAL-A-NP-5000} &  {\bf NISP-P photo-z calibration}

A sample of 20\,000 galaxies representative of the \ac{EWS}, which have adequate external visible and near-infrared spectroscopic observations by launch, shall be observed in all NISP-P bands and VIS to an \ac{SNR} 5 times better than the \ac{EWS}. At least four different lines-of-sight are required to account for cosmic variance, and the fields must be observed by EXT. &
\cellcolor{lightblue}COSMOS, VVDS, CDFS, and SXDS fields; observations ongoing.
\\
\hline
{\bf R-CAL-A-NS-1000} &  {\bf NISP-S wavelength error}

The error in the determination of the wavelength of a detected emission line on one spectral element -- assumed to be four pixels -- in the NISP-S images shall be $<0.8$ pixel. &
\cellcolor{lightgreen} $\lesssim0.5$ pixel combined uncertainty of the full calibration chain \citep[section 5 in][]{DR1-TP024}, with residual systematics (see \cref{sec:aberration}).
\\
\hline
{\bf R-CAL-A-NS-3000} &  {\bf NISP-S absolute spectrophotometric error}

After all calibrations, the absolute spectrophotometric error of  NISP-S shall be $<5.0\%$. & \cellcolor{lightyellow} Met with caveats (see MRD-GC-009 in \cref{table:MRD}).
\\
\hline

{\bf R-CAL-A-NS-4000} &  {\bf Purity spectroscopic sub-sample}

A NISP spectroscopic subsample, with the same selection criteria as that of the spectroscopic galaxy sample chosen from the \ac{EWS}, containing more than 120\,000 galaxies shall be provided with purity $>99\%$. &
\cellcolor{lightblue}Observations ongoing until completed in DR3.
\\
\hline

{\bf R-CAL-A-NS-5000} &  {\bf NISP-S spatial relative spectrophotometric error}

Within the \ac{EWS}, a spatially varying solution for the relative flux calibration shall be provided. The solution shall be such that the relative flux at the flux limit of the sample, averaged over circles of diameter \ang{0.8;;}, has a distribution with width less than $0.7\%$ (\ac{RMS}). & \cellcolor{lightyellow} $\lesssim0.7$\% across DR1 for the red grism \citep[section 5.5 in][]{DR1-TP024}. Blue grism still affected by contamination (see \cref{sec:DR2_nisp_ice_photometry}).
\\
\hline
\end{longtable}
\footnotesize
\label{tab:calcd-a}
\endgroup

\clearpage

\section{Tables and figures\label{sec:apdx_figs}}

\begin{figure}[h]
    \centering
    \includegraphics[height=0.50\textheight,angle=90]{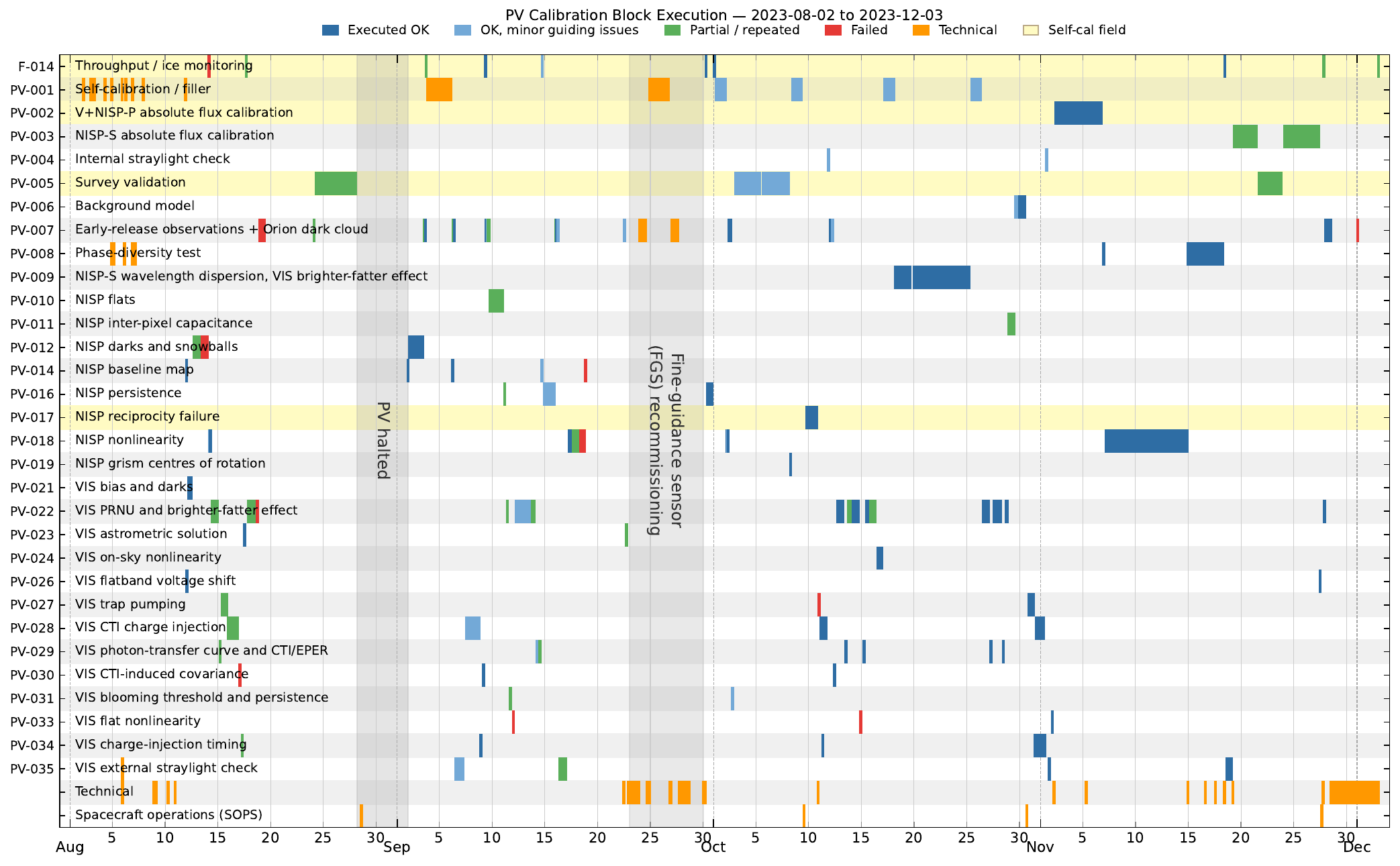}
    \caption{\ac{PV} schedule. The $y$-axis shows the \CALBLOCK\ IDs. Finer details are suppressed to maintain readability, and a minimum length was assigned to very short activities to keep them visible. Unplanned activities are shown in orange and were either rolled into matching \CALBLOCK s or into the `Technical' group. A yellow background indicates that some or all data were taken in the self-calibration field, maximising spacecraft monitoring and the depth of the \ac{EUDF}. Due to issues with the \ac{FGS}, \ac{PV} was halted at the end of August 2023, and resumed a few days later with tasks that did not require high pointing stability. At the end of September 2023, the \ac{FGS} was recommissioned and \ac{PV} activities resumed.}  
    \label{fig:PV_schedule}
\end{figure}

\begin{table*}[h]
\centering
\caption{Transformation equations from and to the \Euclid photometric system, for stellar systems. }
\label{tab:colour_terms}
\begin{tabular}{llrrll}
\hline\noalign{\smallskip}
Transformation & $x$ & RMS & \multicolumn{2}{l}{Required preselection to reject} \\
  & & & \multicolumn{2}{l}{hot and cool stars} \\
\hline\noalign{\vskip 1pt}
\multicolumn{5}{l}{\textit{Euclid$\to$Roman}} \\
\hline\noalign{\vskip 1pt}
    $\mathrm{F062} = \IE +0.0451 +0.3712\,x +0.0612\,x^2$ & $x=\IE-\YE$ & $0.0090$ & $\IE-\YE < 0.69$ & \multicolumn{1}{l}{$x \in [-0.20,\,0.50]$} \\
    $\mathrm{F087} = \IE -0.0471 -0.5742\,x +0.0462\,x^2$ & $x=\IE-\YE$ & $0.0035$ & $\IE-\YE < 0.69$ & \multicolumn{1}{l}{$x \in [-0.20,\,0.50]$} \\
    \boldmath $\mathrm{F106} = \YE -0.0062 +0.0313\,x -0.0046\,x^2$ & $x=\IE-\YE$ & $0.0006$ & $\IE-\YE < 0.69$ & \multicolumn{1}{l}{$x \in [-0.25,\,0.60]$} \\
    $\mathrm{F106} = \YE +0.0001 +0.0588\,x -0.0226\,x^2$ & $x=\YE-\JE$ & $0.0007$ & $\IE-\YE < 0.69$ & \multicolumn{1}{l}{$x \in [-0.33,\,0.12]$} \\
    \boldmath $\mathrm{F129} = \JE -0.0021 +0.2839\,x +0.0360\,x^2$ & $x=\YE-\JE$ & $0.0011$ & $\IE-\YE < 0.69$ & \multicolumn{1}{l}{$x \in [-0.30,\,0.12]$} \\
    $\mathrm{F129} = \JE +0.0333 +0.2528\,x -0.0450\,x^2$ & $x=\JE-\HE$ & $0.0013$ & $\IE-\YE < 0.69$ & \multicolumn{1}{l}{$x \in [-0.35,\,-0.05]$} \\
    $\mathrm{F158} = \HE -0.0586 +0.3104\,x$ & $x=\JE-\HE$ & $0.0013$ & $\IE-\YE < 0.69$ & \multicolumn{1}{l}{$x \in [-0.35,\,-0.02]$} \\
    $\mathrm{F184} = \HE +0.0565 -0.0842\,x -0.0518\,x^2$ & $x=\JE-\HE$ & $0.0021$ & $\IE-\YE < 0.69$ & \multicolumn{1}{l}{$x \in [-0.35,\,-0.03]$} \\
\hline\noalign{\vskip 1pt}
\multicolumn{5}{l}{\textit{Roman$\to$Euclid}} \\
\hline\noalign{\vskip 1pt}
    $\IE = \mathrm{F062} -0.0042 -0.3940\,x -0.0697\,x^2$ & $x=\mathrm{F062}-\mathrm{F087}$ & $0.0049$ & $\IE-\YE < 0.69$ & \multicolumn{1}{l}{$x \in [-0.23,\,0.60]$} \\
    \boldmath $\IE = \mathrm{F087} -0.0042 +0.6060\,x -0.0697\,x^2$ & $x=\mathrm{F062}-\mathrm{F087}$ & $0.0049$ & $\IE-\YE < 0.69$ & \multicolumn{1}{l}{$x \in [-0.23,\,0.60]$} \\
    \boldmath $\YE = \mathrm{F106} +0.0028 -0.0751\,x +0.0302\,x^2$ & $x=\mathrm{F087}-\mathrm{F106}$ & $0.0005$ & $\IE-\YE < 0.69$ & \multicolumn{1}{l}{$x \in [-0.08,\,0.23]$} \\
    \boldmath $\YE = \mathrm{F106} +0.0001 -0.0759\,x +0.0288\,x^2$ & $x=\mathrm{F106}-\mathrm{F129}$ & $0.0006$ & $\IE-\YE < 0.69$ & \multicolumn{1}{l}{$x \in [-0.30,\,0.08]$} \\
    \boldmath $\JE = \mathrm{F129} +0.0030 -0.3660\,x -0.0942\,x^2$ & $x=\mathrm{F106}-\mathrm{F129}$ & $0.0017$ & $\IE-\YE < 0.69$ & \multicolumn{1}{l}{$x \in [-0.25,\,0.12]$} \\
    \boldmath $\JE = \mathrm{F129} -0.0082 -0.2718\,x +0.0494\,x^2$ & $x=\mathrm{F129}-\mathrm{F158}$ & $0.0010$ & $\IE-\YE < 0.69$ & \multicolumn{1}{l}{$x \in [-0.28,\,0.07]$} \\
    $\HE = \mathrm{F158} +0.0889 -0.3285\,x -0.0346\,x^2$ & $x=\mathrm{F129}-\mathrm{F158}$ & $0.0022$ & $\IE-\YE < 0.69$ & \multicolumn{1}{l}{$x \in [-0.23,\,0.05]$} \\
    \boldmath $\HE = \mathrm{F158} -0.0392 -0.7789\,x +0.3488\,x^2$ & $x=\mathrm{F158}-\mathrm{F184}$ & $0.0014$ & $\IE-\YE < 0.69$ & \multicolumn{1}{l}{$x \in [-0.25,\,-0.12]$} \\
    \boldmath $\HE = \mathrm{F184} -0.0356 +0.2246\,x +0.2365\,x^2$ & $x=\mathrm{F158}-\mathrm{F184}$ & $0.0013$ & $\IE-\YE < 0.69$ & \multicolumn{1}{l}{$x \in [-0.23,\,-0.10]$} \\
\hline\noalign{\vskip 1pt}
\multicolumn{5}{l}{\textit{Euclid$\to$JWST}} \\
\hline\noalign{\vskip 1pt}
    \boldmath $\mathrm{F070W} = \IE -0.0274 -0.0634\,x +0.0455\,x^2$ & $x=\IE-\YE$ & $0.0047$ & $\IE-\YE < 0.69$ & \multicolumn{1}{l}{$x \in [-0.35,\,0.50]$} \\
    $\mathrm{F090W} = \YE -0.0417 +0.3531\,x$ & $x=\IE-\YE$ & $0.0038$ & $\IE-\YE < 0.69$ & \multicolumn{1}{l}{$x \in [-0.12,\,0.45]$} \\
    $\mathrm{F090W} = \YE +0.0309 +0.6553\,x$ & $x=\YE-\JE$ & $0.0044$ & $\IE-\YE < 0.69$ & \multicolumn{1}{l}{$x \in [-0.15,\,0.12]$} \\
    \boldmath $\mathrm{F115W} = \JE -0.0013 +0.7393\,x -0.0041\,x^2$ & $x=\YE-\JE$ & $0.0011$ & $\IE-\YE < 0.69$ & \multicolumn{1}{l}{$x \in [-0.33,\,0.13]$} \\
    $\mathrm{F115W} = \JE +0.0915 +0.6571\,x -0.1962\,x^2$ & $x=\JE-\HE$ & $0.0026$ & $\IE-\YE < 0.69$ & \multicolumn{1}{l}{$x \in [-0.35,\,-0.02]$} \\
    \boldmath $\mathrm{F150W} = \HE -0.0575 +0.5115\,x +0.0464\,x^2$ & $x=\JE-\HE$ & $0.0016$ & $\IE-\YE < 0.69$ & \multicolumn{1}{l}{$x \in [-0.37,\,0.00]$} \\
    \boldmath $\mathrm{F200W} = \HE +0.1667 -0.1063\,x$ & $x=\JE-\HE$ & $0.0028$ & $\IE-\YE < 0.69$ & \multicolumn{1}{l}{$x \in [-0.35,\,0.00]$} \\
\hline\noalign{\vskip 1pt}
\multicolumn{5}{l}{\textit{JWST$\to$Euclid}} \\
\hline\noalign{\vskip 1pt}
    $\IE = \mathrm{F070W} +0.0272 +0.1018\,x -0.1219\,x^2$ & $x=\mathrm{F070W}-\mathrm{F090W}$ & $0.0047$ & $\IE-\YE < 0.69$ & \multicolumn{1}{l}{$x \in [-0.18,\,0.35]$} \\
    $\YE = \mathrm{F090W} +0.0492 -0.5972\,x +0.0506\,x^2$ & $x=\mathrm{F070W}-\mathrm{F090W}$ & $0.0061$ & $\IE-\YE < 0.69$ & \multicolumn{1}{l}{$x \in [-0.05,\,0.30]$} \\
    \boldmath $\YE = \mathrm{F090W} -0.0078 -0.7143\,x -0.0186\,x^2$ & $x=\mathrm{F090W}-\mathrm{F115W}$ & $0.0007$ & $\IE-\YE < 0.69$ & \multicolumn{1}{l}{$x \in [-0.15,\,0.16]$} \\
    $\JE = \mathrm{F115W} +0.0278 -0.8044\,x$ & $x=\mathrm{F090W}-\mathrm{F115W}$ & $0.0063$ & $\IE-\YE < 0.69$ & \multicolumn{1}{l}{$x \in [-0.13,\,0.15]$} \\
    \boldmath $\JE = \mathrm{F115W} -0.0053 -0.5769\,x +0.0557\,x^2$ & $x=\mathrm{F115W}-\mathrm{F150W}$ & $0.0011$ & $\IE-\YE < 0.69$ & \multicolumn{1}{l}{$x \in [-0.40,\,0.15]$} \\
    $\HE = \mathrm{F150W} +0.1226 -0.4434\,x -0.1389\,x^2$ & $x=\mathrm{F115W}-\mathrm{F150W}$ & $0.0032$ & $\IE-\YE < 0.69$ & \multicolumn{1}{l}{$x \in [-0.33,\,0.10]$} \\
    \boldmath $\HE = \mathrm{F150W} -0.1370 -0.8255\,x +0.1066\,x^2$ & $x=\mathrm{F150W}-\mathrm{F200W}$ & $0.0023$ & $\IE-\YE < 0.69$ & \multicolumn{1}{l}{$x \in [-0.43,\,-0.23]$} \\
    $\HE = \mathrm{F200W} -0.1270 +0.1764\,x$ & $x=\mathrm{F150W}-\mathrm{F200W}$ & $0.0021$ & $\IE-\YE < 0.69$ & \multicolumn{1}{l}{$x \in [-0.44,\,-0.23]$} \\
\hline\noalign{\vskip 1pt}
\multicolumn{5}{l}{\textit{Euclid$\to$CSST}} \\
\hline\noalign{\vskip 1pt}
    $r_{\rm CSST} = \IE +0.0230 +0.3242\,x +0.0370\,x^2$ & $x=\IE-\YE$ & $0.0042$ & $\IE-\YE < 0.69$ & \multicolumn{1}{l}{$x \in [-0.35,\,0.60]$} \\
    $i_{\rm CSST} = \IE -0.0461 -0.2803\,x$ & $x=\IE-\YE$ & $0.0045$ & $\IE-\YE < 0.69$ & \multicolumn{1}{l}{$x \in [-0.22,\,0.50]$} \\
    ${\rm WV}_{\rm CSST} = \IE +0.2775 +1.3290\,x$ & $x=\IE-\YE$ & $0.0521$ & $\IE-\YE < 0.69$ & \multicolumn{1}{l}{$x \in [-0.02,\,0.50]$} \\
    ${\rm WI}_{\rm CSST} = \IE -0.0317 -0.2342\,x +0.0320\,x^2$ & $x=\IE-\YE$ & $0.0040$ & $\IE-\YE < 0.69$ & \multicolumn{1}{l}{$x \in [-0.20,\,0.50]$} \\
    $z_{\rm CSST} = \YE -0.0394 +0.3470\,x$ & $x=\IE-\YE$ & $0.0050$ & $\IE-\YE < 0.69$ & \multicolumn{1}{l}{$x \in [-0.08,\,0.50]$} \\
\hline\noalign{\vskip 1pt}
\multicolumn{5}{l}{\textit{CSST$\to$Euclid}} \\
\hline\noalign{\vskip 1pt}
    \boldmath $\IE = r_{\rm CSST} +0.0189 -0.5315\,x -0.1343\,x^2$ & $x=r_{\rm CSST}-i_{\rm CSST}$ & $0.0016$ & $\IE-\YE < 0.69$ & \multicolumn{1}{l}{$x \in [-0.25,\,0.50]$} \\
    \boldmath $\IE = {\rm WI}_{\rm CSST} +0.0118 +0.4212\,x -0.1454\,x^2$ & $x=r_{\rm CSST}-{\rm WI}_{\rm CSST}$ & $0.0018$ & $\IE-\YE < 0.69$ & \multicolumn{1}{l}{$x \in [-0.30,\,0.45]$} \\
    $\IE = z_{\rm CSST} +0.0546 +1.7254\,x$ & $x=i_{\rm CSST}-z_{\rm CSST}$ & $0.0060$ & $\IE-\YE < 0.69$ & \multicolumn{1}{l}{$x \in [-0.11,\,0.19]$} \\
    $\IE = z_{\rm CSST} +0.0269 +1.5353\,x$ & $x={\rm WI}_{\rm CSST}-z_{\rm CSST}$ & $0.0052$ & $\IE-\YE < 0.69$ & \multicolumn{1}{l}{$x \in [-0.13,\,0.25]$} \\
    $\IE = {\rm WV}_{\rm CSST} +0.0043 -0.8509\,x -0.0405\,x^2$ & $x={\rm WV}_{\rm CSST}-{\rm WI}_{\rm CSST}$ & $0.0054$ & $\IE-\YE < 0.69$ & \multicolumn{1}{l}{$x \in [-0.40,\,1.10]$} \\
\hline\noalign{\vskip 1pt}
\end{tabular}
\footnotesize
 \tablefoot{Transformations printed in bold have particularly low intrinsic scatter.}
\end{table*}


\begin{figure*}
\centering
\includegraphics[width=0.245\textwidth]{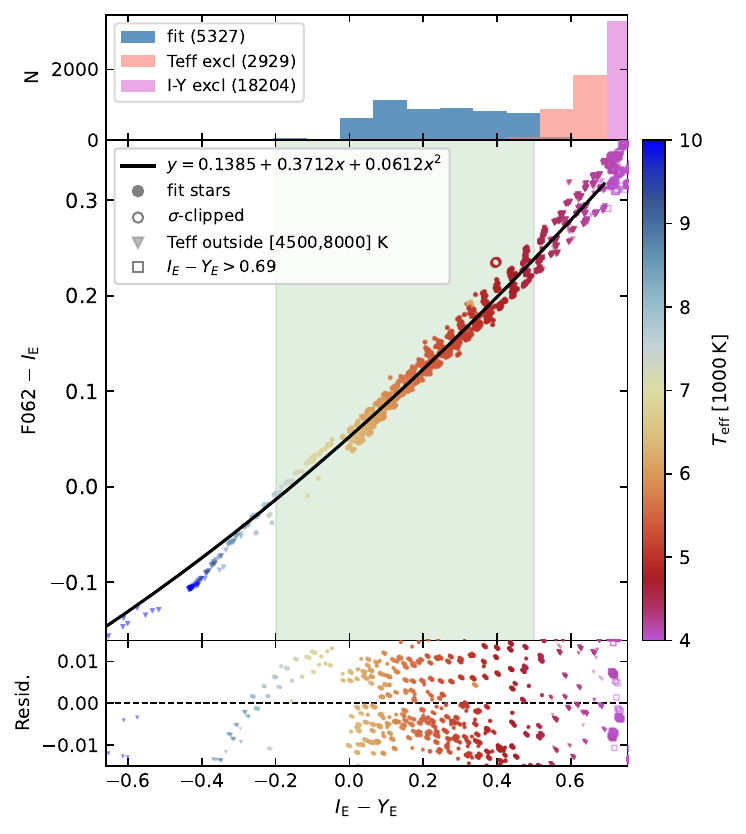}%
\includegraphics[width=0.245\textwidth]{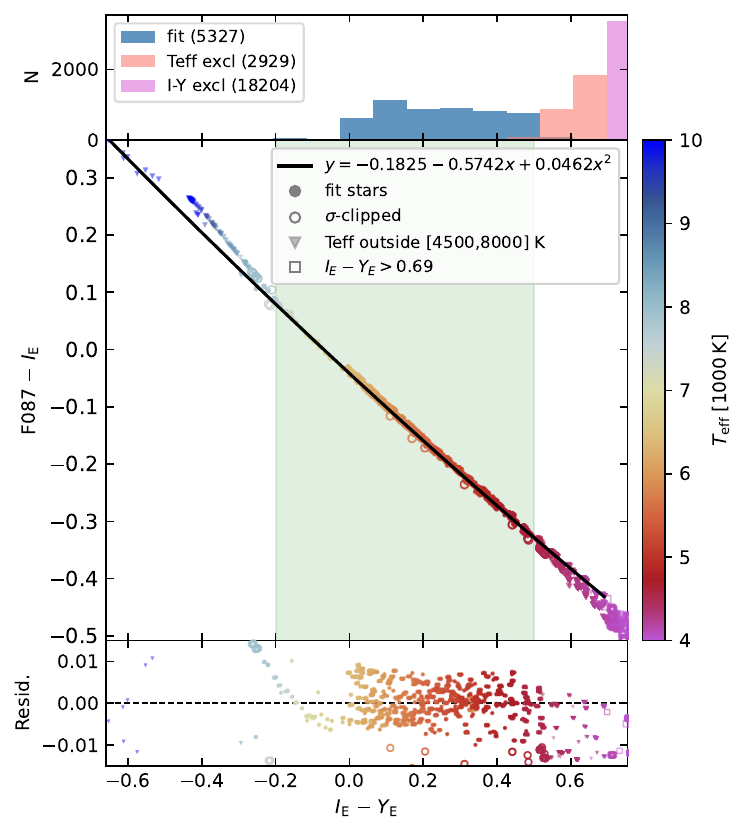}%
\includegraphics[width=0.245\textwidth]{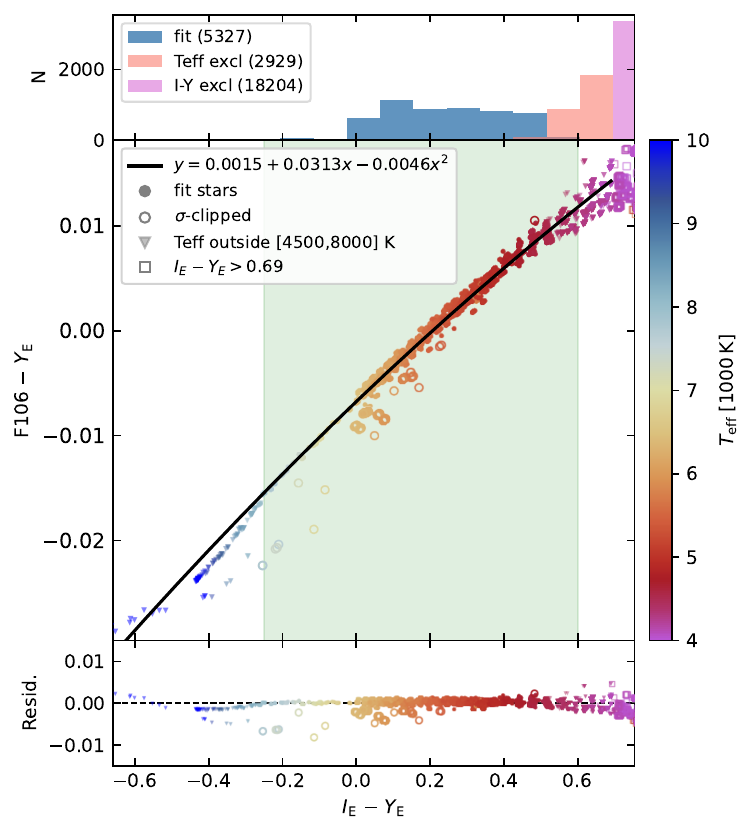}%
\includegraphics[width=0.245\textwidth]{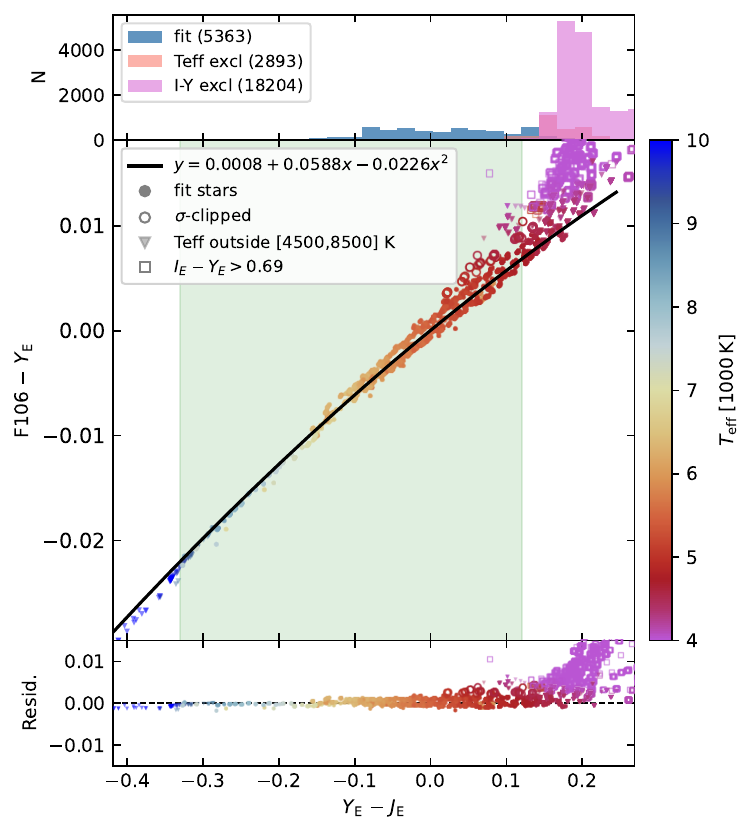}%
\\[2pt]
\includegraphics[width=0.245\textwidth]{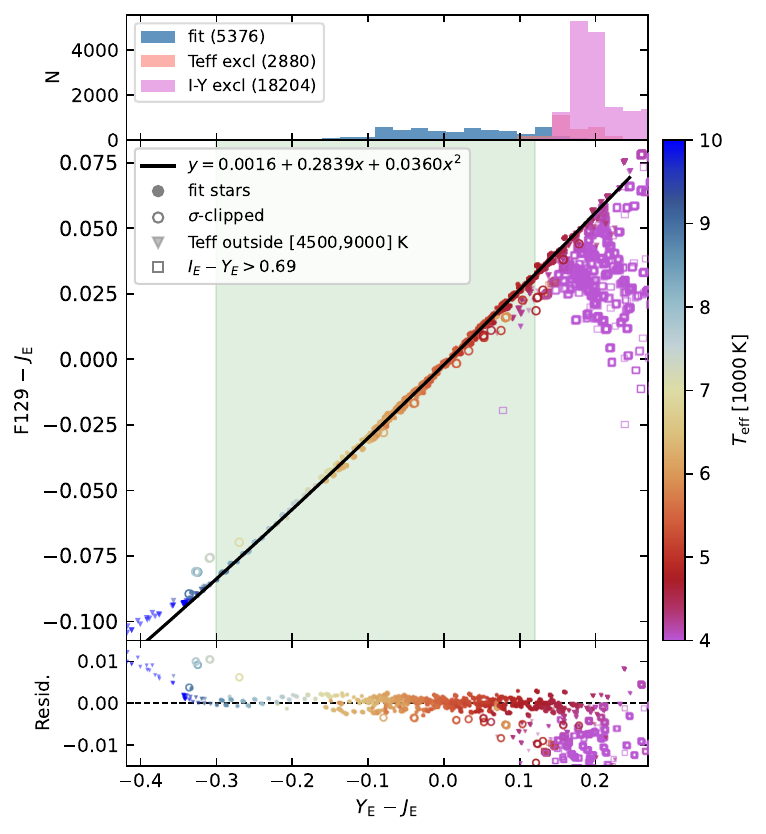}%
\includegraphics[width=0.245\textwidth]{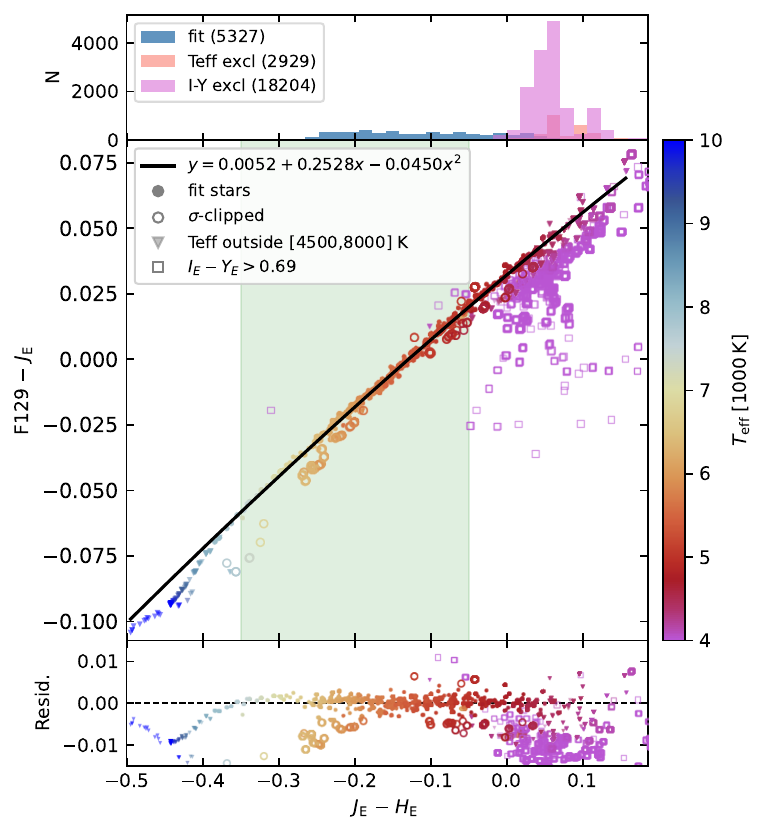}%
\includegraphics[width=0.245\textwidth]{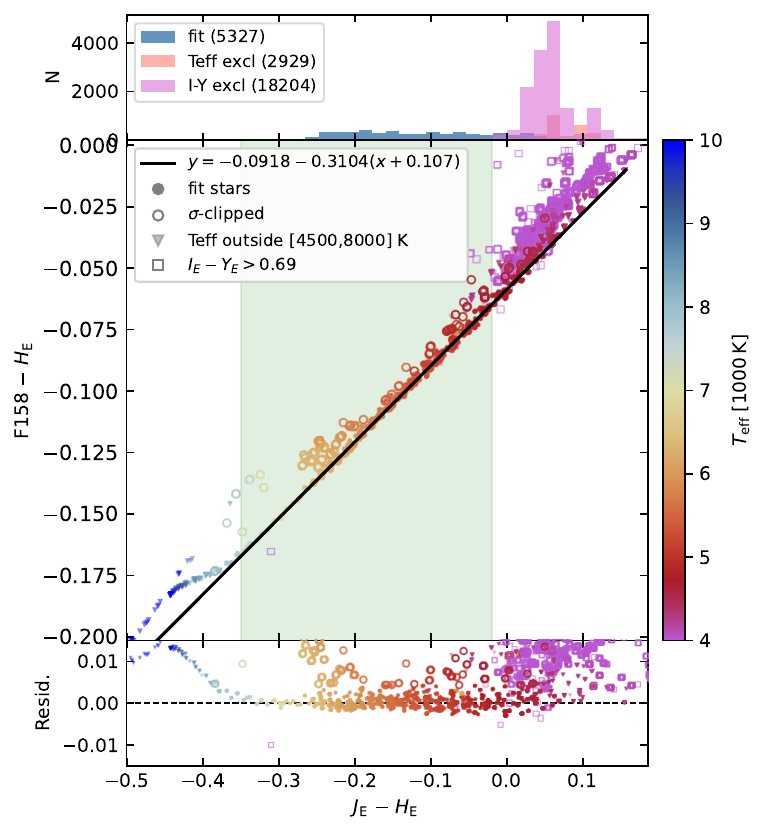}%
\includegraphics[width=0.245\textwidth]{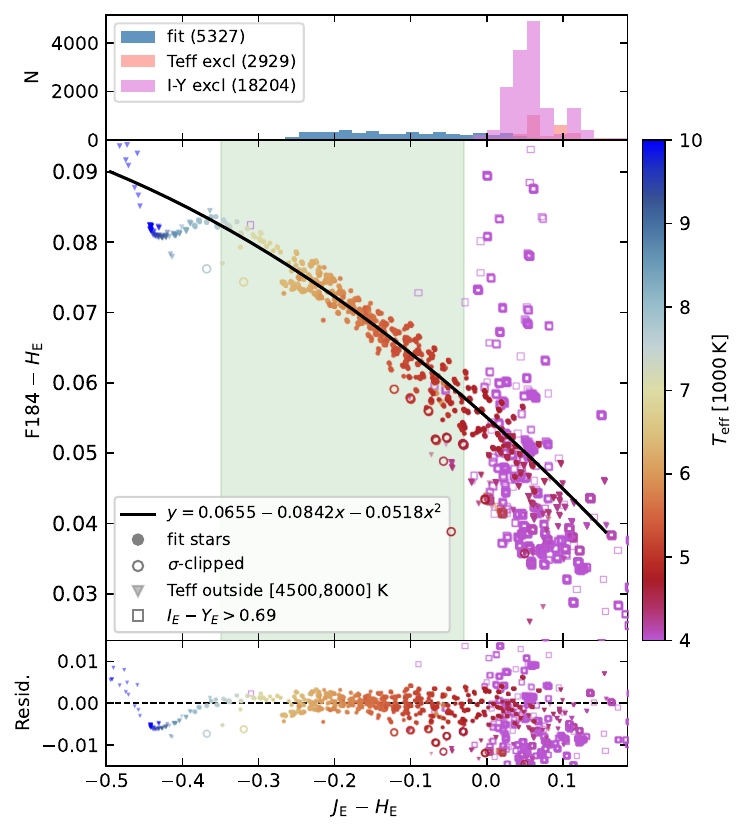}%
\caption{Photometric transformations from \Euclid to \textit{Roman}. Explanation as in \cref{fig:transformation_euclid_roman}.}
\label{fig:e2r}
\end{figure*}

\begin{figure*}
\centering
\includegraphics[width=0.325\textwidth]{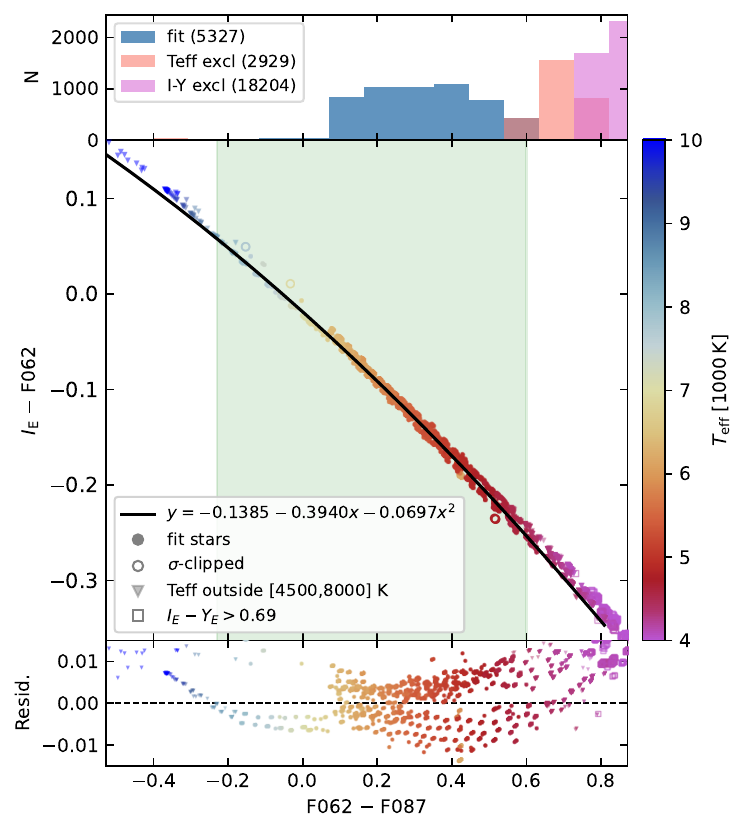}%
\includegraphics[width=0.325\textwidth]{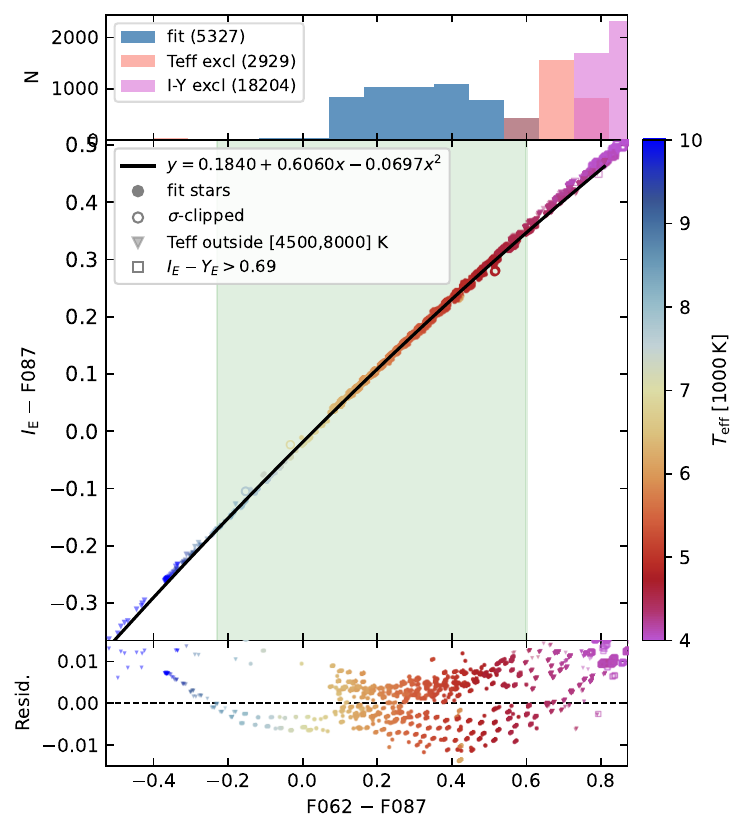}%
\includegraphics[width=0.325\textwidth]{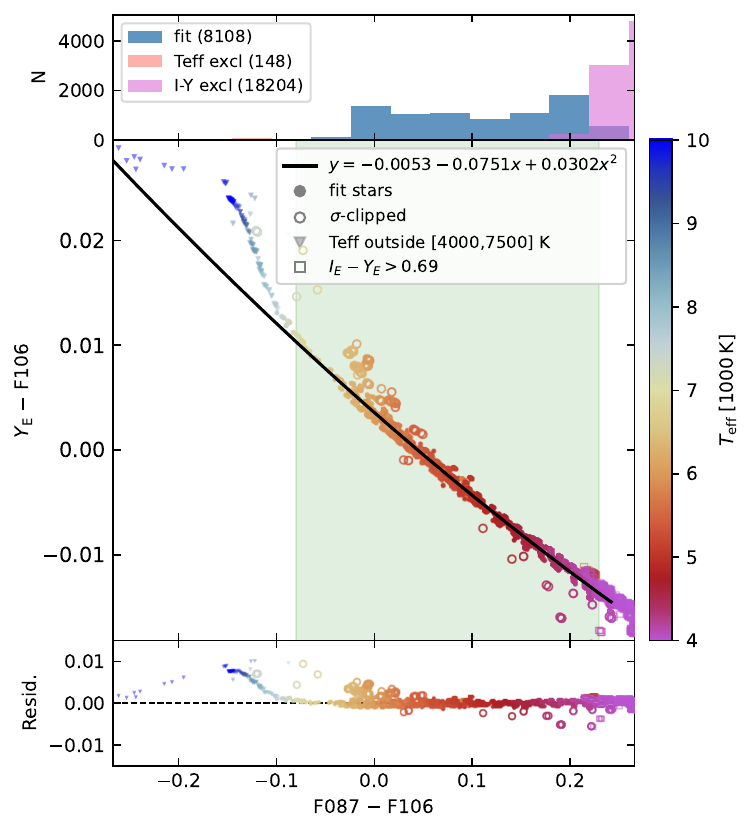}%
\\[2pt]
\includegraphics[width=0.325\textwidth]{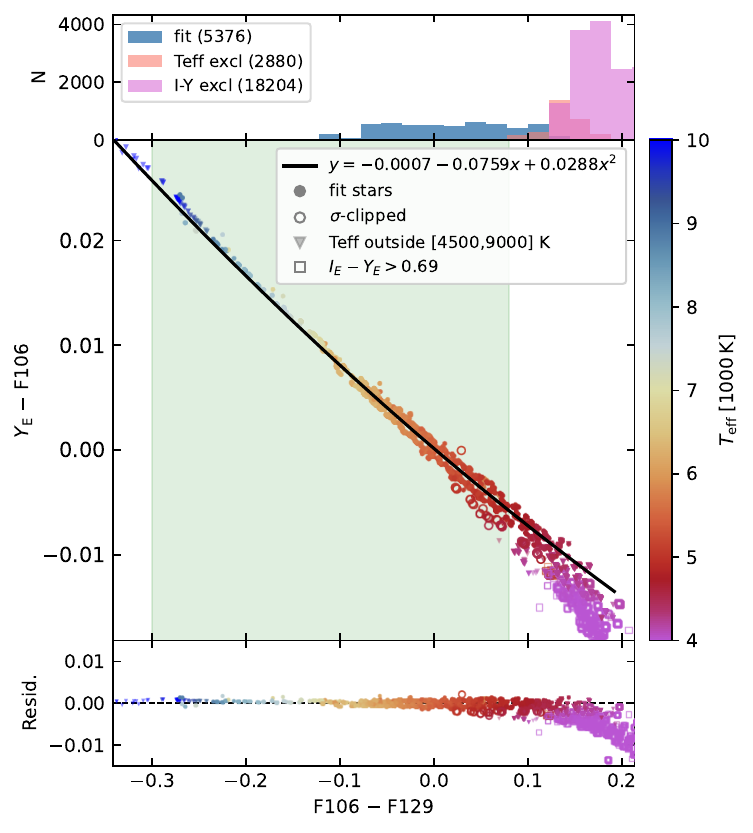}%
\includegraphics[width=0.325\textwidth]{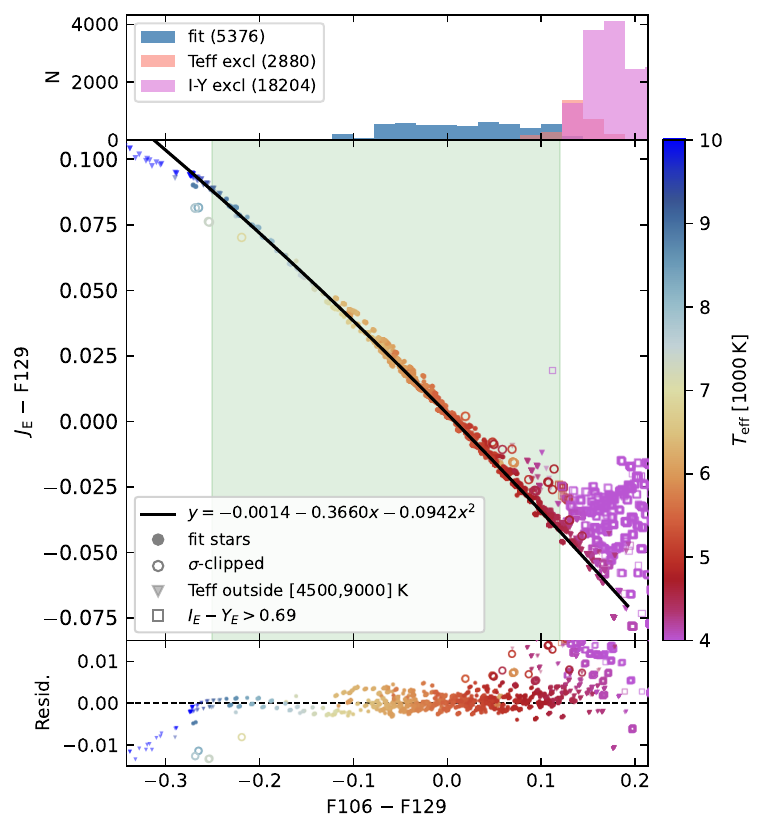}%
\includegraphics[width=0.325\textwidth]{transform_figures/transform_JE_F129_vs_F129_F158}%
\\[2pt]
\includegraphics[width=0.325\textwidth]{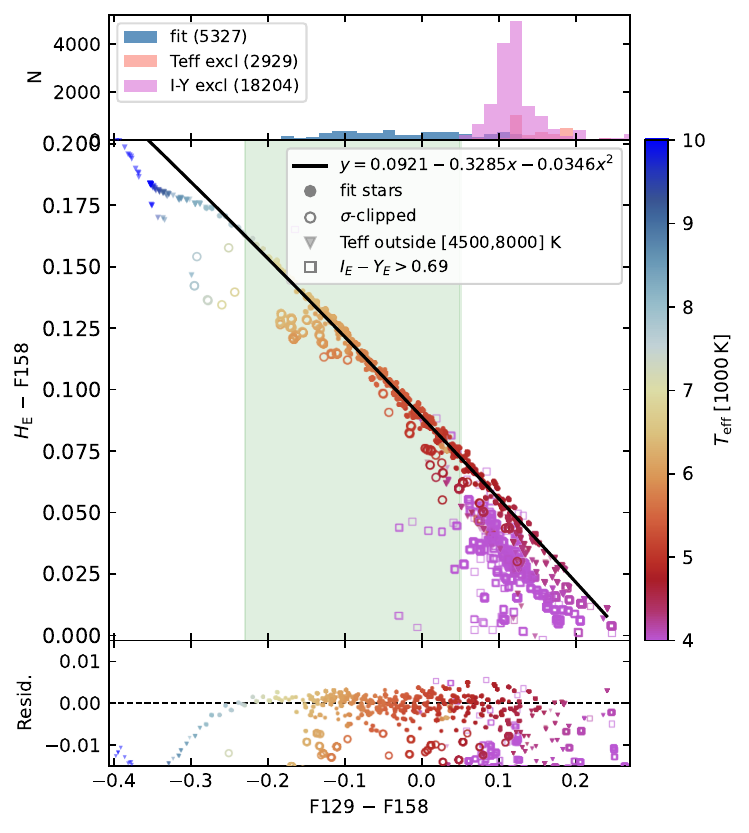}%
\includegraphics[width=0.325\textwidth]{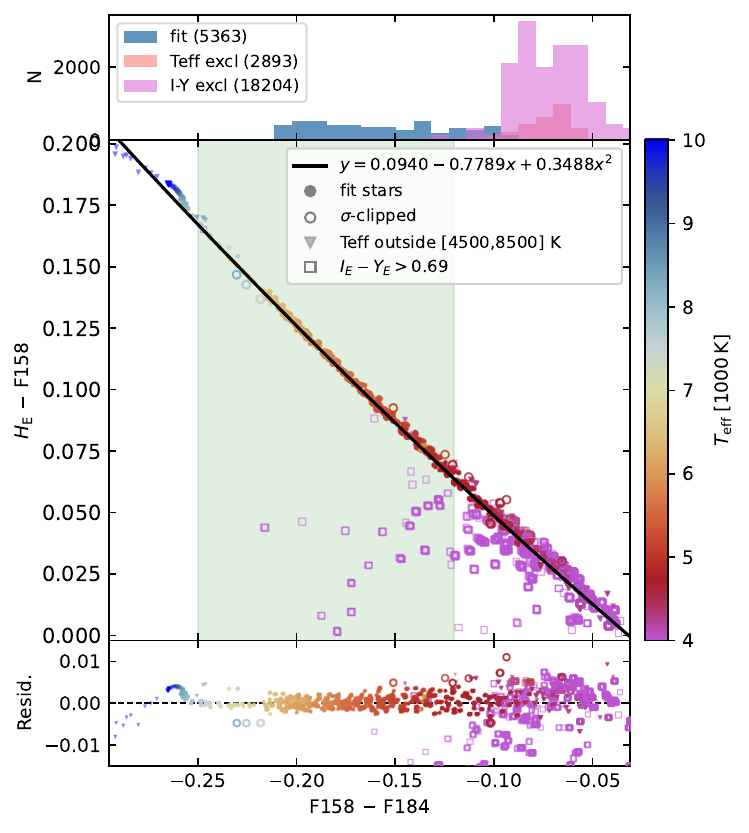}%
\includegraphics[width=0.325\textwidth]{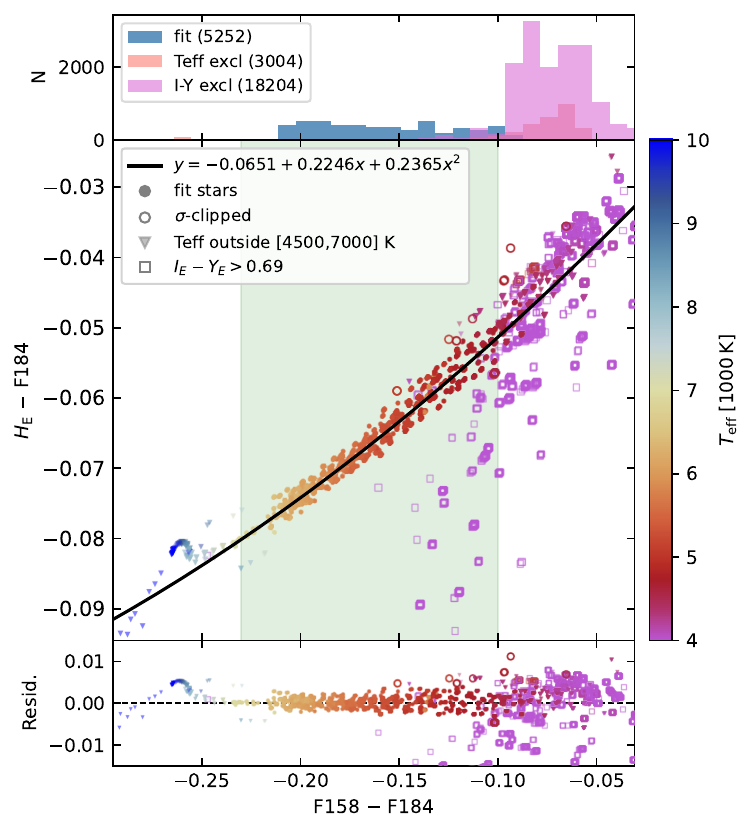}%
\caption{Photometric transformations from \textit{Roman} to \Euclid.}
\label{fig:r2e}
\end{figure*}

\begin{figure*}
\centering
\includegraphics[width=0.245\textwidth]{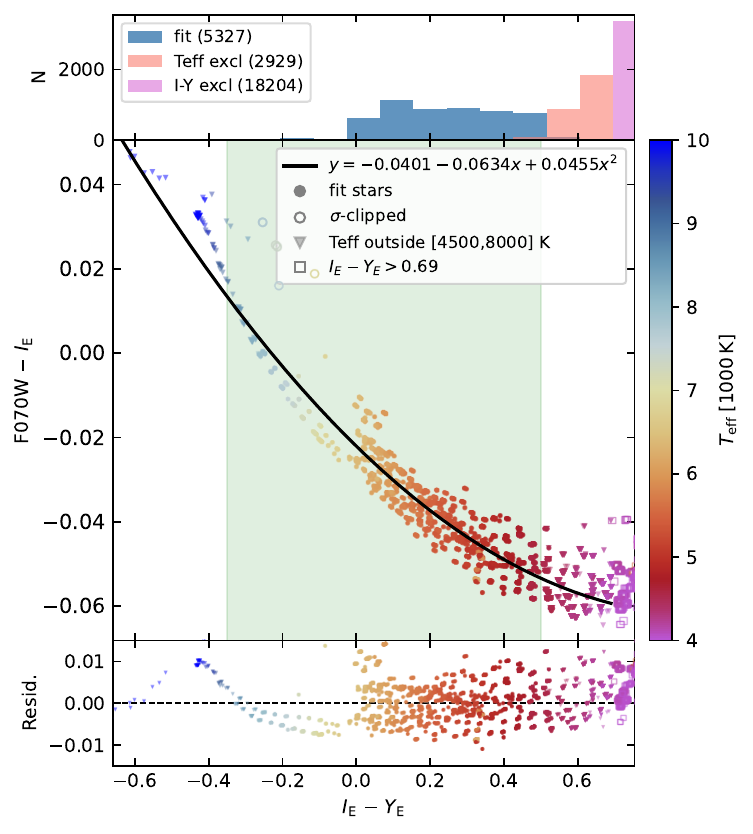}%
\includegraphics[width=0.245\textwidth]{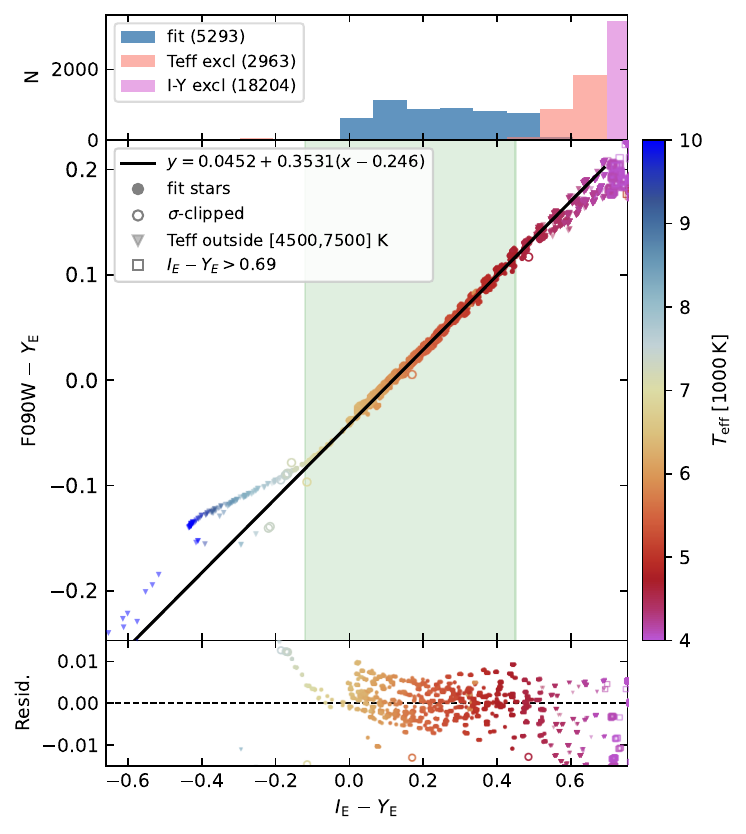}%
\includegraphics[width=0.245\textwidth]{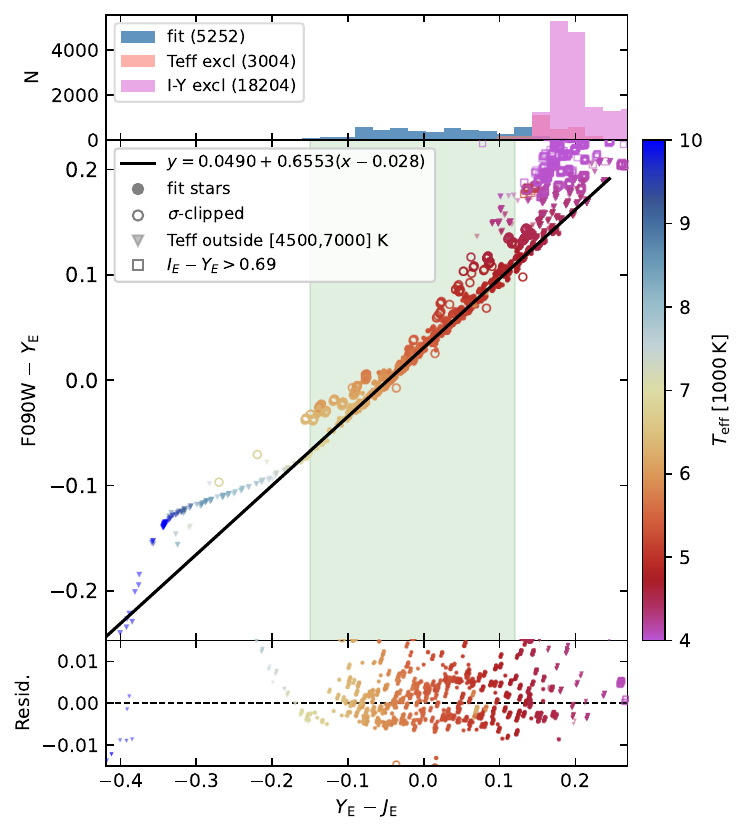}%
\includegraphics[width=0.245\textwidth]{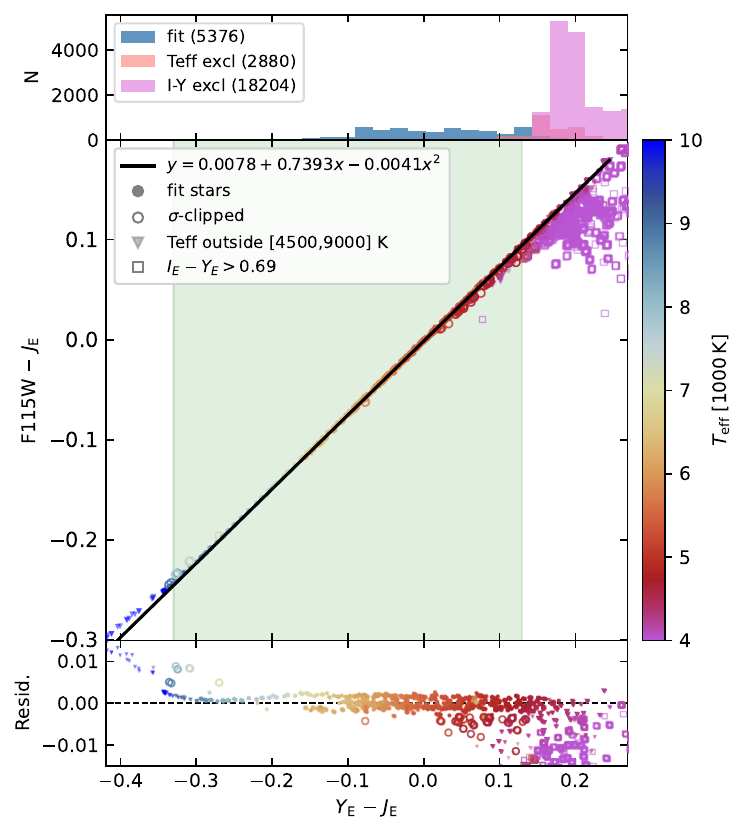}%
\\[2pt]
\includegraphics[width=0.245\textwidth]{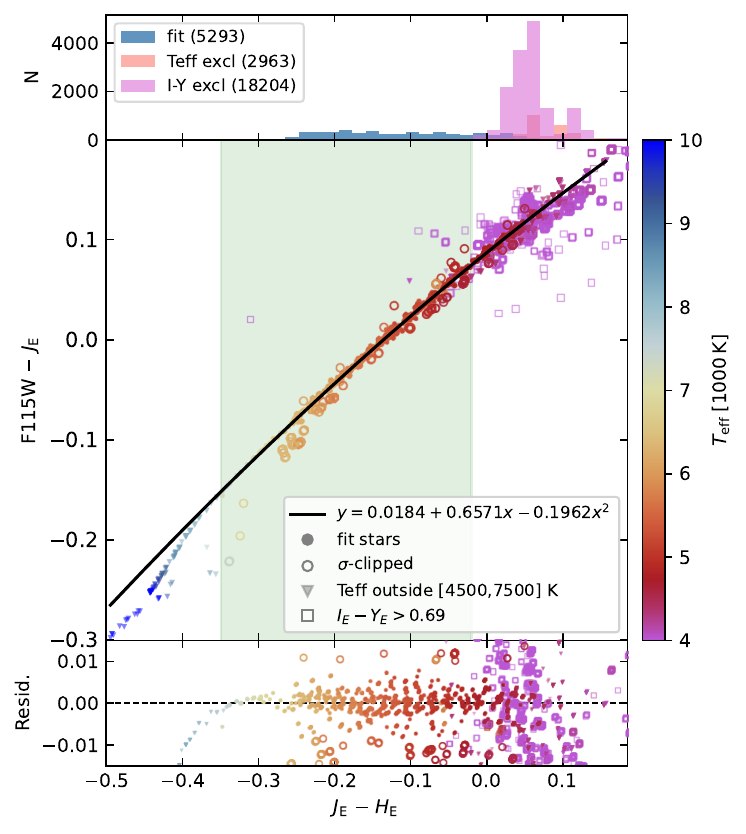}%
\includegraphics[width=0.245\textwidth]{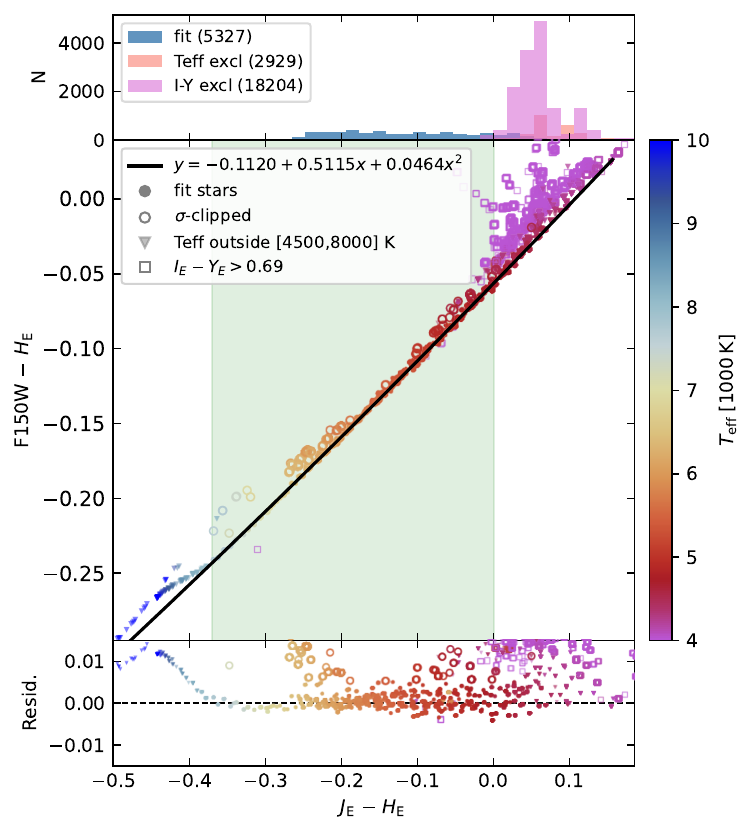}%
\includegraphics[width=0.245\textwidth]{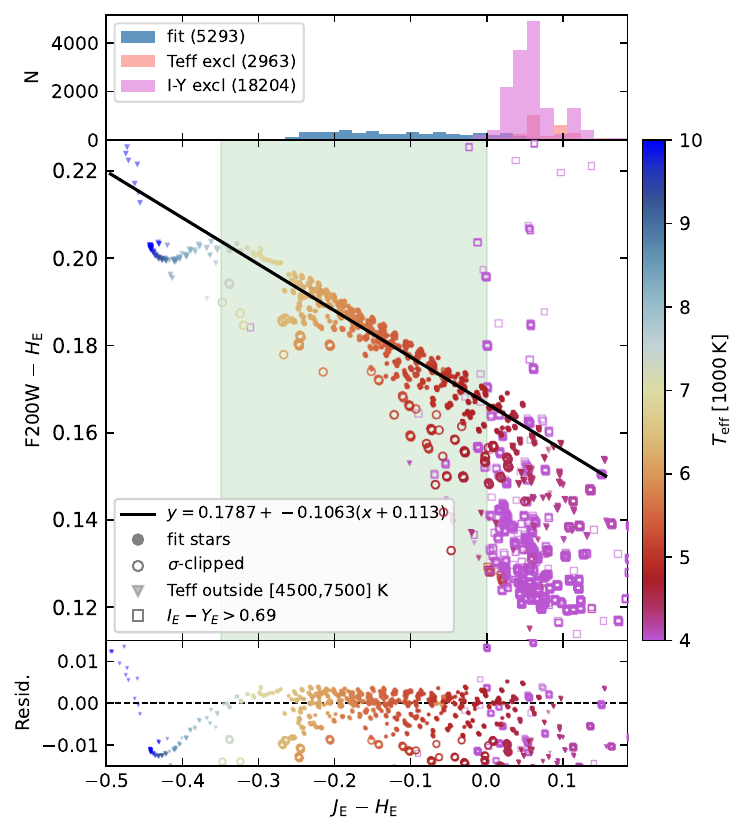}%
\caption{Photometric transformations from \Euclid to \ac{JWST}.}
\label{fig:e2j}
\end{figure*}

\begin{figure*}
\centering
\includegraphics[width=0.245\textwidth]{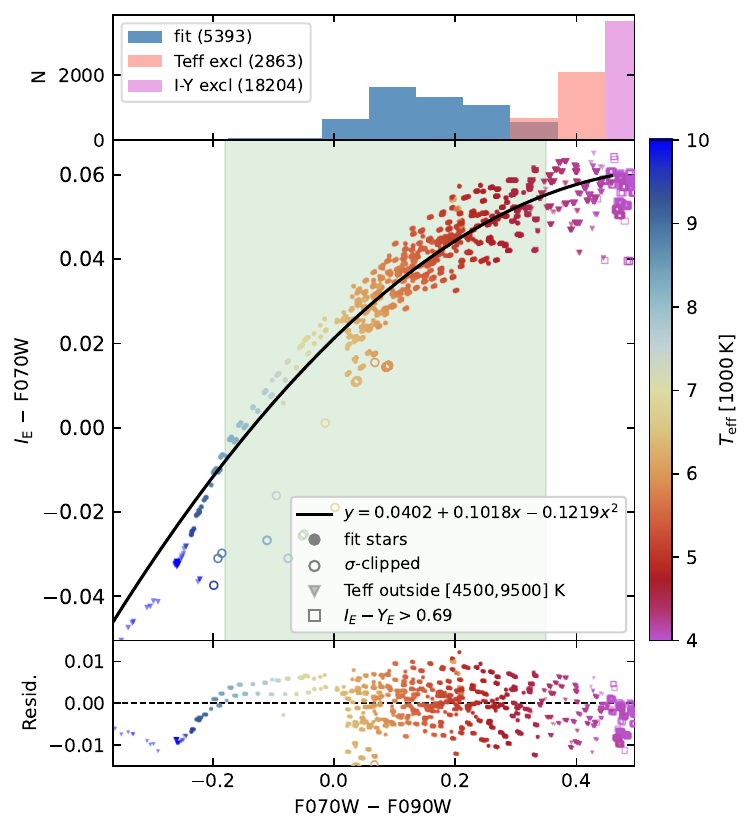}%
\includegraphics[width=0.245\textwidth]{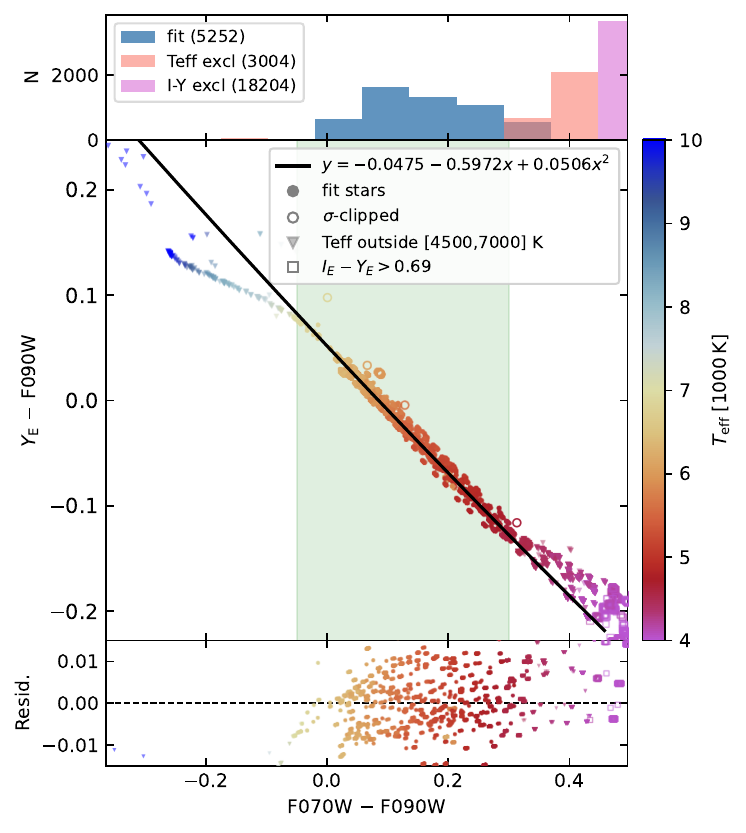}%
\includegraphics[width=0.245\textwidth]{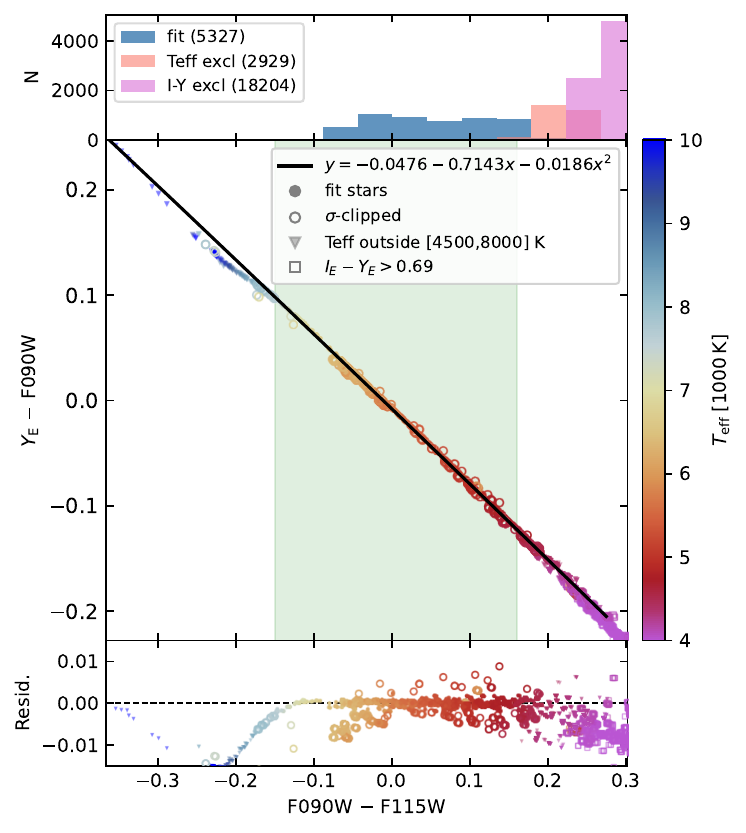}%
\includegraphics[width=0.245\textwidth]{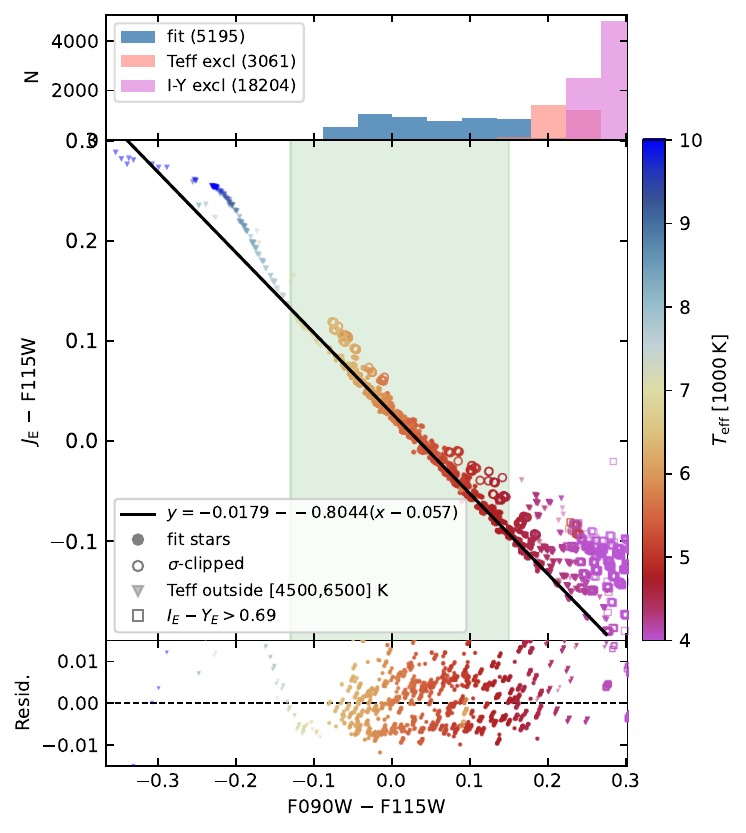}%
\\[2pt]
\includegraphics[width=0.245\textwidth]{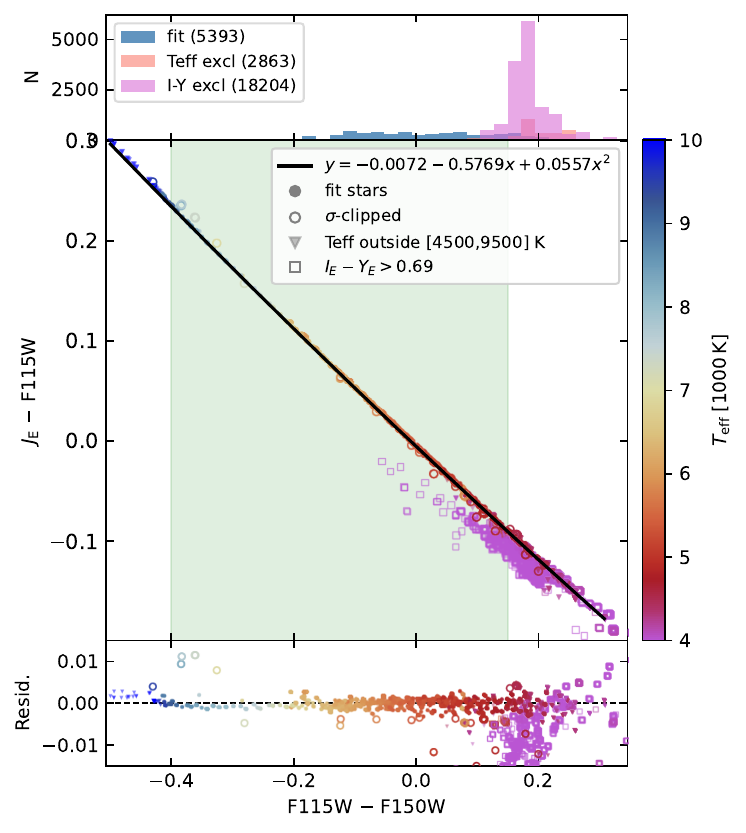}%
\includegraphics[width=0.245\textwidth]{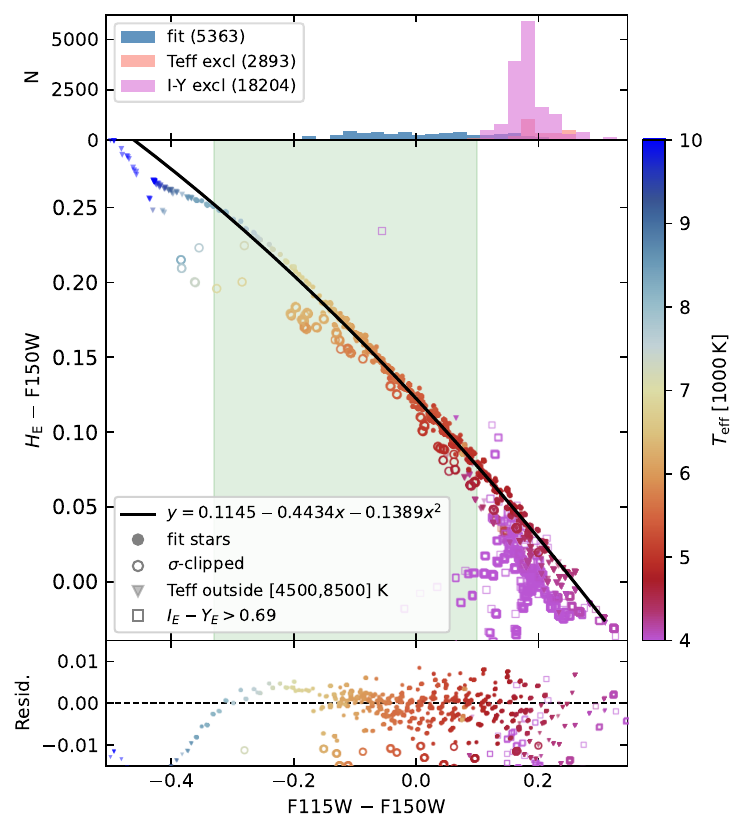}%
\includegraphics[width=0.245\textwidth]{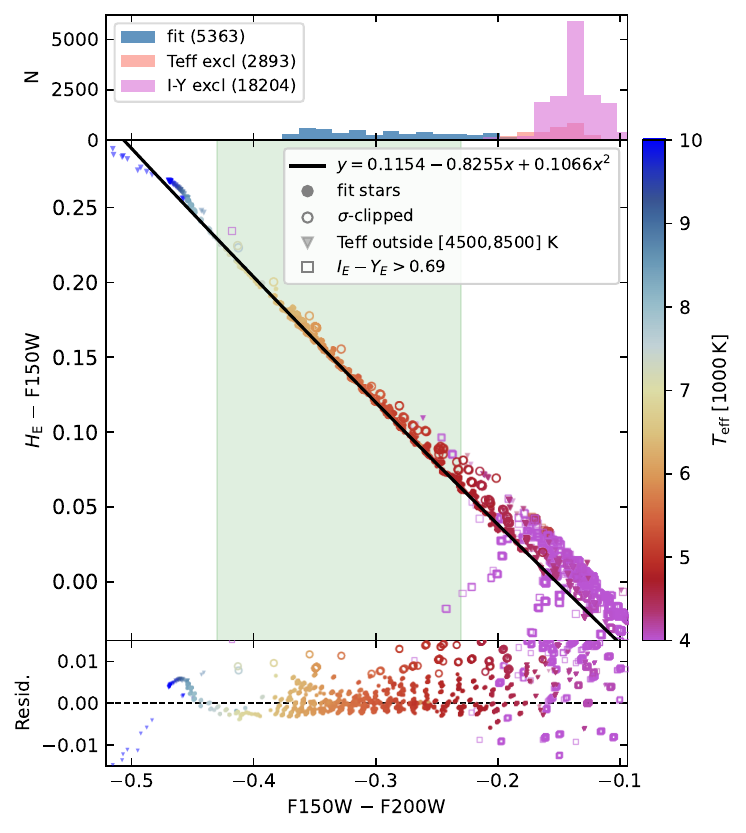}%
\includegraphics[width=0.245\textwidth]{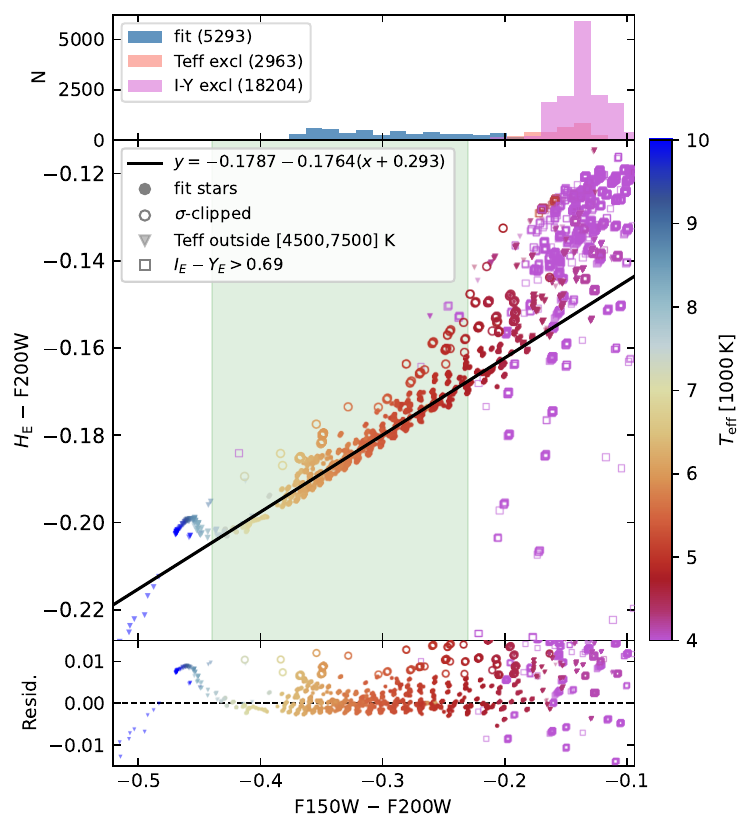}%
\caption{Photometric transformations from \ac{JWST} to \Euclid.}
\label{fig:j2e}
\end{figure*}

\begin{figure*}
\centering
\includegraphics[width=0.325\textwidth]{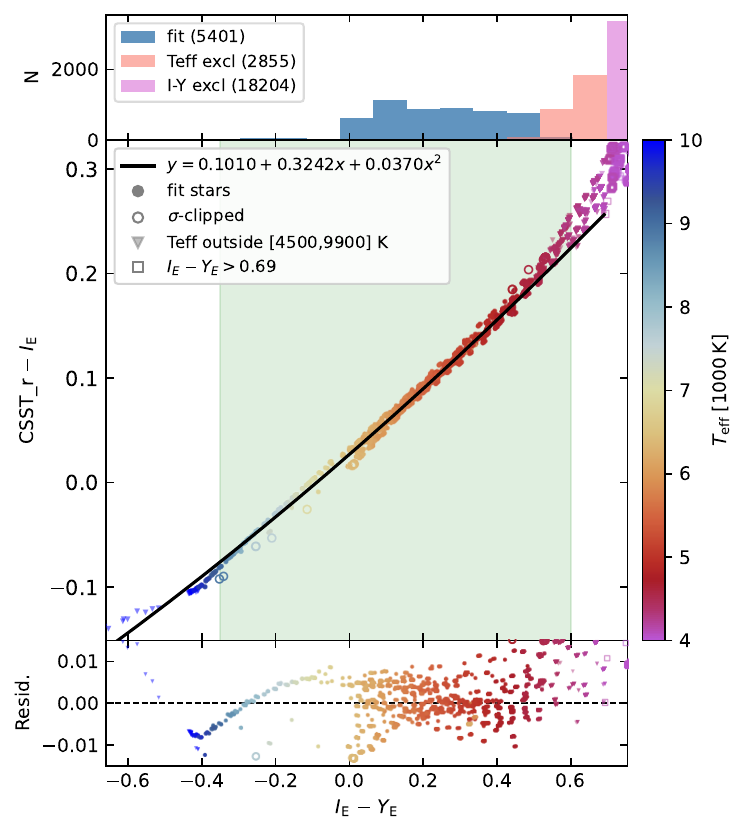}%
\includegraphics[width=0.325\textwidth]{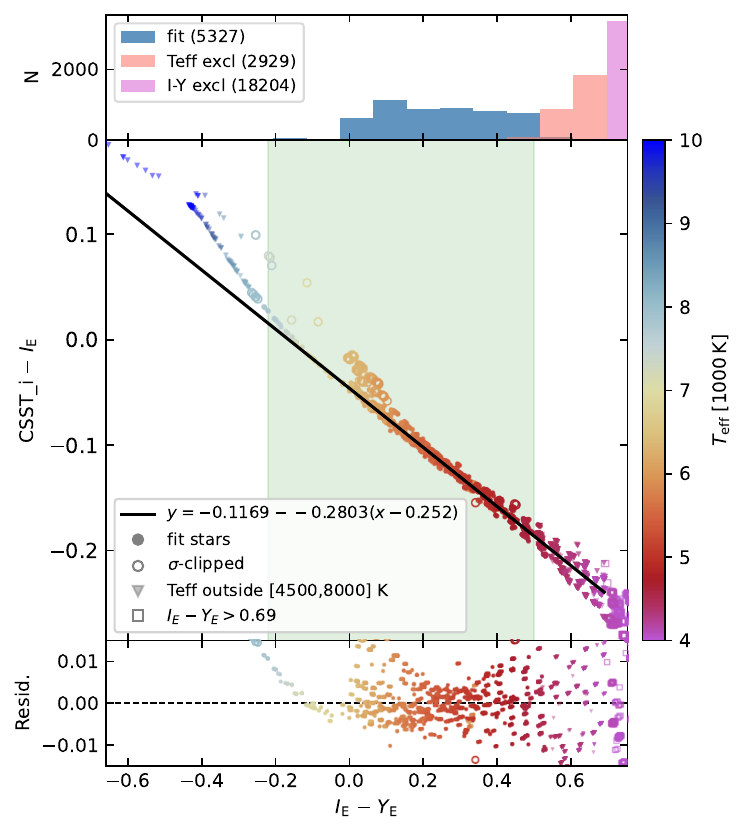}%
\includegraphics[width=0.325\textwidth]{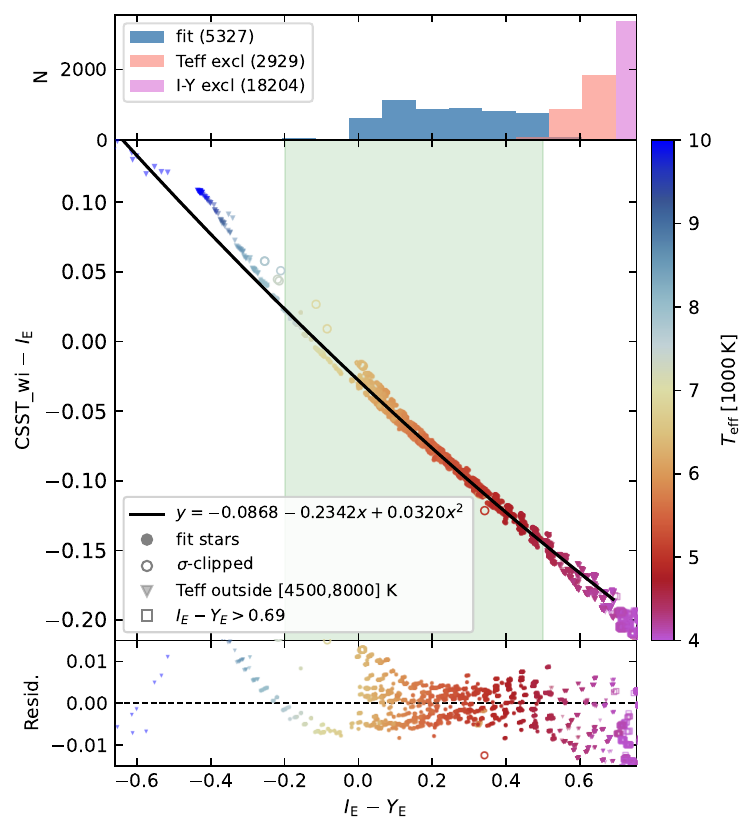}%
\\[2pt]
\includegraphics[width=0.325\textwidth]{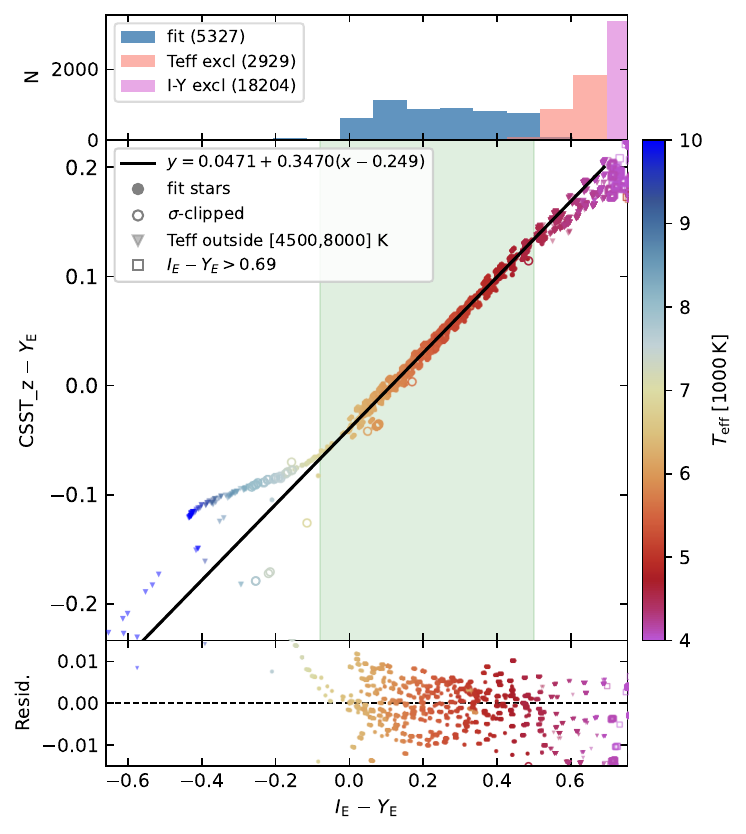}%
\includegraphics[width=0.325\textwidth]{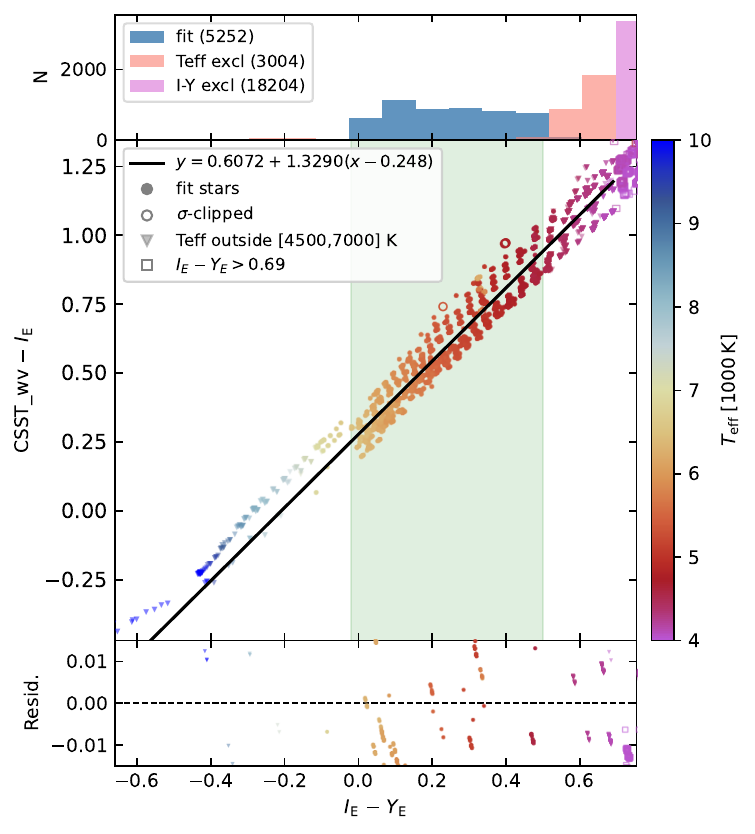}%
\parbox[b]{0.325\textwidth}{}
\caption{Photometric transformations from \Euclid to \ac{CSST}}
\label{fig:e2c}
\end{figure*}

\begin{figure*}
\centering
\includegraphics[width=0.325\textwidth]{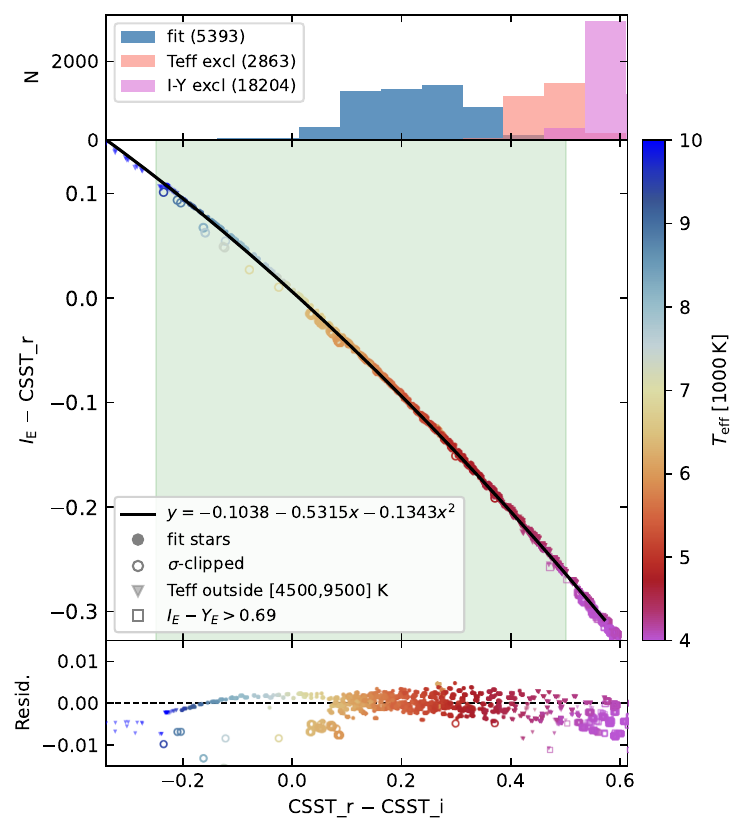}%
\includegraphics[width=0.325\textwidth]{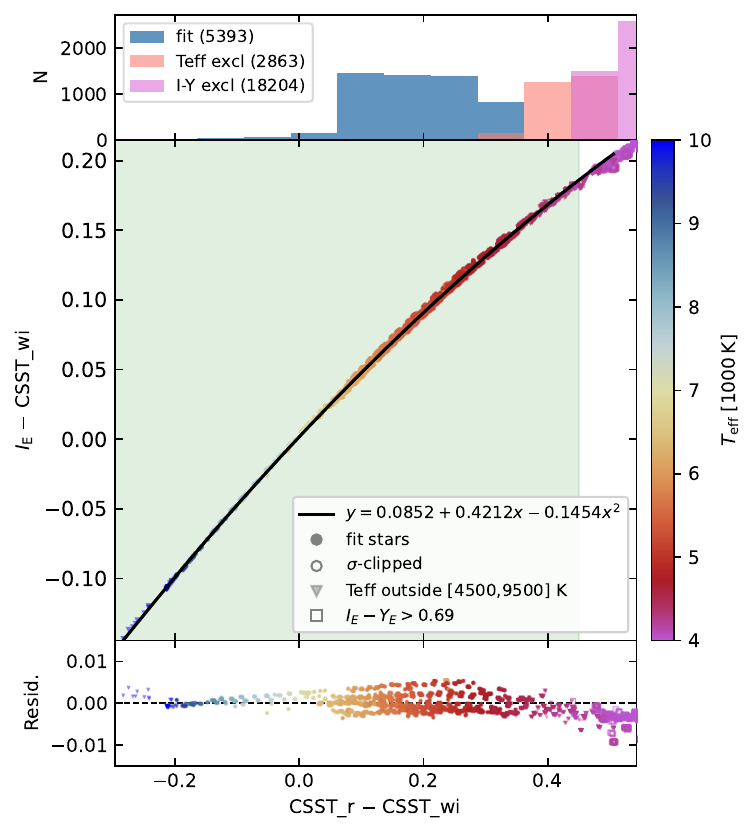}%
\includegraphics[width=0.325\textwidth]{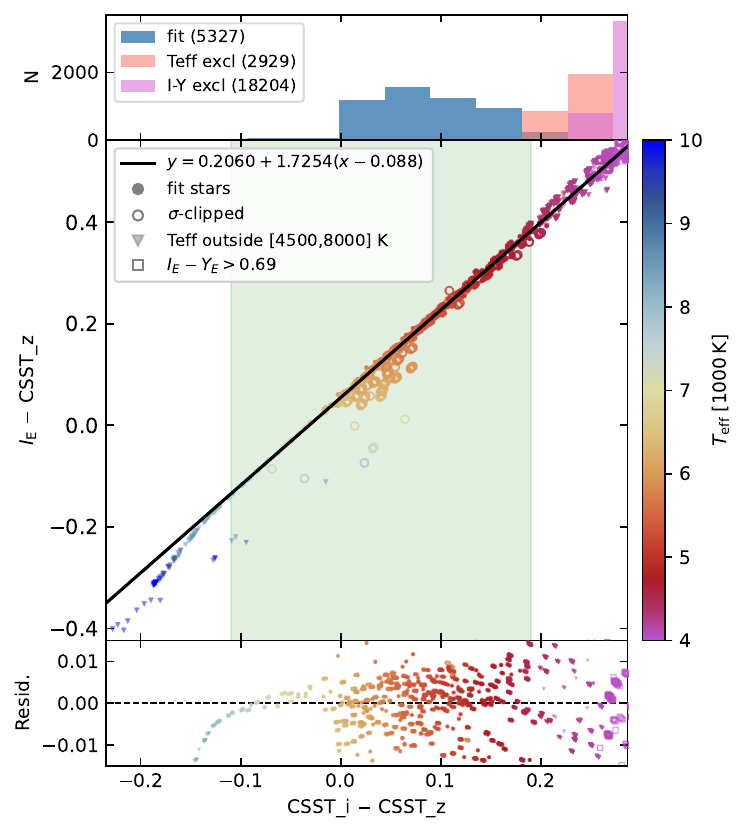}%
\\[2pt]
\includegraphics[width=0.325\textwidth]{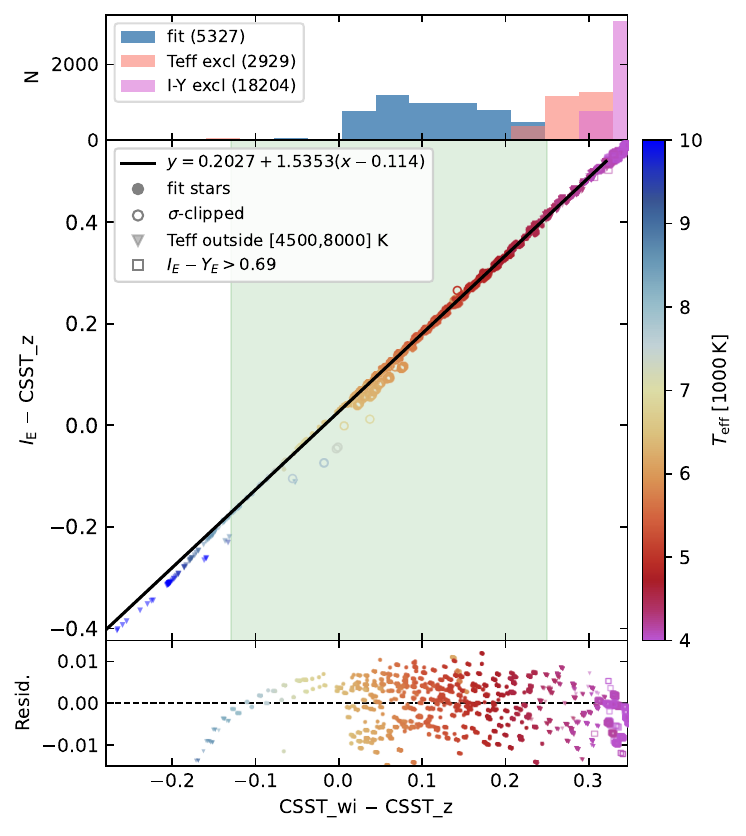}%
\includegraphics[width=0.325\textwidth]{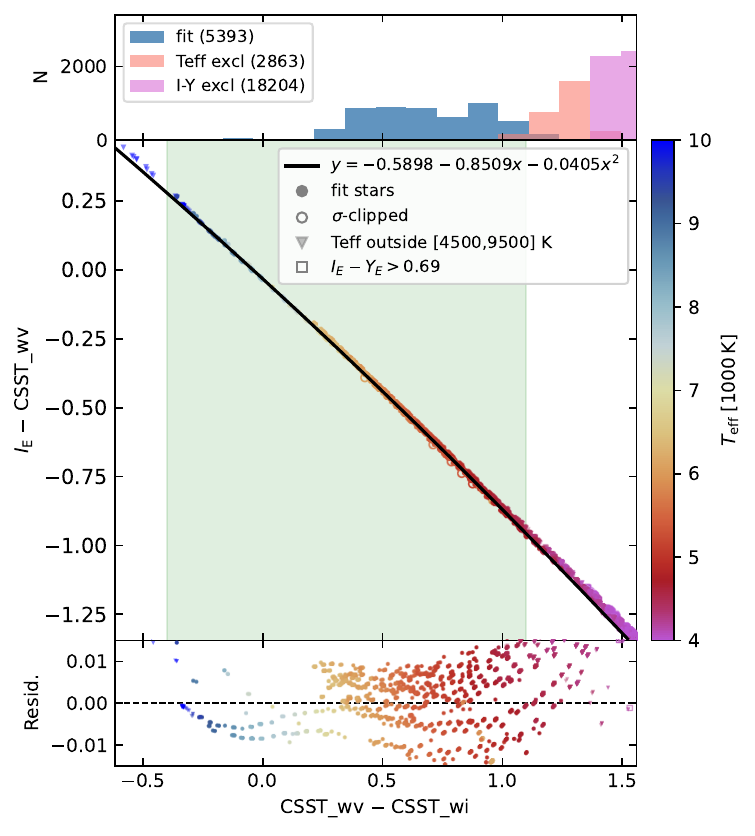}%
\parbox[b]{0.325\textwidth}{}
\caption{Photometric transformations from \ac{CSST} to \Euclid.}
\label{fig:c2e}
\end{figure*}

\section{Survey and system characterisations\label{apdx:characterisations}}

\begingroup
\small
\setlength\extrarowheight{2pt}
\begin{longtable}{|>{\columncolor{gray!15}}p{0.12\textwidth}|p{0.82\textwidth}|}
\caption{Survey and system characterisations in the calibration framework ({\tt CalF}, see \cref{sec:calf}).}\label{tab:characterisations}\\
\hline
\textbf{ID} & \textbf{Description} \\
\hline
\endfirsthead
\multicolumn{2}{c}{\tablename\ \thetable{} -- continued from previous page} \\
\hline
\textbf{ID} & \textbf{Description} \\
\hline
\endhead
\hline
\multicolumn{2}{r}{\textit{Continued on next page}} \\
\endfoot
\hline
\endlastfoot
{\bf SURV-001}\label{CHAR-SURV-001} & \textbf{Depth}\newline Determines whether \Euclid achieves the required survey depths and number counts, serving as a go/no-go criterion at the end of PV. One complete pass of the EDF-N and -- if visible -- the EDF-S provides sufficient data. \\
\hline
{\bf SURV-002}\label{CHAR-SURV-002} & \textbf{Relative photometric accuracy}\newline Assesses whether \Euclid can achieve the required photometric precision for multiply detected sources across different images, serving as a go/no-go criterion at the end of PV. Estimated by analysing the consistency of flux measurements for sources detected multiple times in different exposures of the EDF-N during PV. \\
\hline
{\bf SURV-003}\label{CHAR-SURV-003} & \textbf{Sensitivity}\newline Uses the single pass of the EDF-N area in PV to verify the pre-launch estimates of the VIS and NISP detection limits for the chosen exposure times. This is another critical go/no-go criterion to commence the survey. \\
\hline
{\bf SURV-004}\label{CHAR-SURV-004} & \textbf{Zodiacal model}\newline Validates the zodiacal background model that is based on previous space missions, a go/no-go criterion to commence the survey. Pre-launch uncertainties in absolute flux calibration, intrinsic zodiacal variability, straylight modelling, and passband transformations to \Euclid passbands limit the zodiacal uncertainties to the 5--10\% range. The zodiacal background would need to be 30\% higher than expected before necessitating changes to the exposure times in the \ac{ROS}. \\
\hline
{\bf TEL-001}\label{CHAR-TEL-001} & \textbf{Telescope thermal timescale}\newline Determines the typical time-scale required for the telescope to reach sufficiently stable thermo-optical equilibrium for weak lensing measurements after considerable changes in attitude relative to the Sun. As a metric, we use the focus term derived from measurements of the diffraction spikes, together with thermistor readings from the \ac{PLM}. \\
\hline
{\bf TEL-002}\label{CHAR-TEL-002} & \textbf{PLM contamination}\newline Aims to detect water ice on optical surfaces, using temporal changes in stellar photometry in the self-calibration field as well as comparisons of \ac{EWS} images with \textit{Gaia} photometry. This is a by-product of nominal calibration processing. \\
\hline
{\bf TEL-003}\label{CHAR-TEL-003} & \textbf{External straylight levels}\newline Must be obtained early in PV to map straylight contamination levels at various spacecraft attitudes, following the commissioning discovery that Sunlight enters the instrument cavity -- among others -- through a scattering chain originating from an illuminated thruster nozzle. Straylight levels are largely mitigated by rotating the spacecraft to move the nozzle into shadow. This characterisation determines the `dark' \ac{AA} and \ac{SAA} subspace usable for survey operations. \\
\hline
{\bf TARGET-001}\label{CHAR-TARGET-001} & \textbf{Colour cradient calibration fields}\newline Observes five CANDELS fields (AEGIS, GOODS-NORTH, COSMOS, UDS, GOODS-CDFS) to calibrate biases in VIS shear measurements caused by spatially variable \acp{SED} within galaxies in conjunction with the chromatic VIS \ac{PSF}. The AEGIS field must be completed in the first year due to its extensive \ac{HST} observations; at least one field must be observed each following year. \\
\hline
{\bf TARGET-002}\label{CHAR-TARGET-002} & \textbf{Photo-z calibration fields}\newline Observes four deep fields (COSMOS-wide, VVDS-Deep, CDFS, SXDS) with \ac{SNR} 5 times better than the \ac{EWS}. This yields NISP photometry in the \YE, \JE, and \HE bands for approximately 20\,000 galaxies with existing ground-based spectroscopy, enabling calibration of the photo-$z$ methods. The time-critical observation strategy requires the COSMOS and CDFS fields to be observed early-on, since these data also fulfill colour-gradient calibration requirements. \\
\hline
{\bf TARGET-003}\label{CHAR-TARGET-003} & \textbf{Noise bias calibration fields}\newline Observes approximately 200\,000 galaxies with \ac{SNR} 6 times that of the wide survey across at least 40\,deg$^2$. These data enable the characterisation and calibration of noise-induced biases in shape measurements, which must be corrected for in the shallower \ac{EWS} data. The data are automatically provided by the \ac{EDS}. \\
\hline
{\bf TARGET-004}\label{CHAR-TARGET-004} & \textbf{Completeness and purity calibration}\newline Generates a spectroscopic subsample of at least 120\,000 galaxies with 99\% purity selected using identical criteria as in the \ac{EWS}. Requires a survey area of at least 40\,deg$^2$ based on expected galaxy number. The observations require multiple orientations (ten visits with minimum \ang{5;;} separation angles plus four angles from the \ac{ROS}) to enable optimal decontamination of overlapping slitless spectra. The full area is observed using all four red grisms and with the blue grism. A 3$\times$3 field subset must be observed with the blue grism already during PV. The data enable the correction of selection biases in the \ac{EWS} and ensure that the spectroscopic sample meets the requirements for galaxy clustering and redshift distribution measurements. \\
\end{longtable}
\endgroup
  
\section{In-flight calibration observations (\CALBLOCK s)\label{sec:apdx_calblocks}}
\begingroup
\small
\setlength\extrarowheight{2pt}
\begin{longtable}{|>{\columncolor{teal!15}}p{0.06\textwidth}|p{0.88\textwidth}|}
\caption{In-flight calibration observations in the {\tt CalF}  (\cref{sec:calf}).}\label{tab:calblocks-f}\\
\hline
\textbf{ID} & \textbf{Description} \\
\hline
\endfirsthead
\multicolumn{2}{c}{\tablename\ \thetable{} -- continued from previous page} \\
\hline
\textbf{ID} & \textbf{Description} \\
\hline
\endhead
\hline
\multicolumn{2}{r}{\textit{Continued on next page}} \\
\endfoot
\hline
\endlastfoot
\hypertarget{target:F-001}{\bf F-001} & \textbf{Self-calibration}\newline {\bf Objective:} Provides baseline monitoring of Euclid's spectrophotometric response to space weathering and outgassing through multiple dithered exposures of the self-calibration field near the North Ecliptic Pole. {\bf Observations:} Executes the ROS\_D1 at 60 pointings with specific grisms; four visits are required to observe all grisms at all dithers, including blue-grism observations with \ac{AA} offsets of at least $\pm\ang{1.4;;}$ for spectral decontamination. {\bf Cadence:} Every 25--40 days. {\bf Duration:} Approximately 23.5 hours per visit (94 hours total for four visits). The self-calibration observations support a large number of calibration products: NISP \ac{BFE} (\cref{CALPRODUCT-NI-003}), NISP-P Absolute Photometric Standards (\cref{CALPRODUCT-NP-001}), NISP-P Focal Plane Astrometric Solution (\cref{CALPRODUCT-NP-002}), NISP-P Large-Scale Flat (\cref{CALPRODUCT-NP-003}), NISP-P PSF Model (\cref{CALPRODUCT-NP-007}), NISP-P Dichroic Ghost Model (\cref{CALPRODUCT-NP-004}), NISP-S Dichroic Ghost Model (\cref{CALPRODUCT-NS-011}), NISP-S 0th Order Centroid SED Dependence Calibration (\cref{CALPRODUCT-NS-001}), NISP-S Absolute Photometric Standards (\cref{CALPRODUCT-NS-002}), NISP-S Focal Plane Distortion (\cref{CALPRODUCT-NS-003}), NISP-S Large-Scale Flat (\cref{CALPRODUCT-NS-004}), NISP-S Spectral Trace Calibration (\cref{CALPRODUCT-NS-006}), NISP-S Wavelength Dispersion Check (\cref{CALPRODUCT-NS-007}), VIS Absolute Photometric Standards (\cref{CALPRODUCT-VS-001}), VIS BFE (\cref{CALPRODUCT-VS-004}), VIS Cross-talk (\cref{CALPRODUCT-VS-008}), VIS Focal Plane Astrometric Solution (\cref{CALPRODUCT-VS-010}), VIS Large-Scale Flat (\cref{CALPRODUCT-VS-013}), VIS Dichroic Ghost Model (\cref{CALPRODUCT-VS-012}), and VIS Shuttermaps (\cref{CALPRODUCT-VS-028}). \\
\hline
\hypertarget{target:F-002}{\bf F-002} & \textbf{Noise-bias fields (Euclid Deep Fields)}\newline {\bf Objective:} Obtains a sample of 200\,000 galaxies with \ac{SNR} six times that of the \ac{EWS} from a 40\,deg$^2$ area (EDF-North, EDF-South, and EDF-Fornax) to quantify noise bias in weak lensing measurements. {\bf Observations:} Uses the ROS with a blue:red grism exposure-time ratio of 5:3 to account for different zodiacal background and photon flux; regular grid dithering with spiral offsets between visits. {\bf Cadence:} Continuous replanning throughout the mission to account for space-weather losses etc. {\bf Duration:} Variable, depends on total area coverage required, and is partially fulfilled by CALBLOCK-F-003. \\
\hline
\hypertarget{target:F-003}{\bf F-003} & \textbf{Completeness and purity calibration (CPC)}\newline {\bf Objective:} Generates a spectroscopic subsample of at least 120\,000 galaxies with 99\% purity selected with the same criteria as the \ac{EWS}, requiring observations with sufficiently many orientations for optimal slitless spectra decontamination. {\bf Observations:} Obtained from EDF-N and EDF-S (43\,deg$^2$ total); requires ten visits per field with minimum \ang{5;;} angle separation and the K-sequence angles \citep[see][]{Scaramella-EP1}; The blue:red grism ratio is the same as for CALBLOCK-F-002, that is 5:3. {\bf Cadence:} Continuous replanning throughout the mission. {\bf Duration:} Variable based on completion requirements; partially fulfilled by CALBLOCK-F-002. \\
\hline
\hypertarget{target:F-004}{\bf F-004} & \textbf{Colour-gradient calibration fields}\newline {\bf Objective:} VIS and NISP observe five fields (\cref{table:EAF}) with extensive \ac{HST} ancillary data to four times the mean \ac{EWS} \ac{SNR} to calibrate for colour gradient effects in galaxies that bias shear measurements. {\bf Observations:} Uses the ROS with a blue:red grism exposure-time ratio approximating 5:3; spiral dithering pattern; three of the fields are already covered by the photo-$z$ calibration observations (CALBLOCK-F-005). {\bf Cadence:} The AEGIS field must be completed in the first year, followed by at least one additional field per following year. {\bf Duration:} 63.8 hours (excluding 125.4 hours already covered by the photo-$z$ fields). \\
\hline
\hypertarget{target:F-005}{\bf F-005} & \textbf{\texorpdfstring{Photo-$z$}{Photo-z} calibration fields}\newline {\bf Objective:} Obtains NISP photometry in the \YE, \JE, and \HE bands for 20\,000 galaxies with existing ground-based spectroscopy at five times the \ac{EWS} \ac{SNR} to calibrate \Euclid photometric-redshift methods. {\bf Observations:} Four fields (COSMOS-wide, VVDS-Deep, CDFS, SXDS) are observed with the ROS and the standard 5:3 blue:red grism exposure-time ratio; a spiral dithering is employed; the COSMOS field spanning a $2\times2$ mosaic has priority with a staged depth-completion strategy. {\bf Cadence:} COSMOS started in first year with specific completion milestones; other fields distributed over following years. {\bf Duration:} 745.3 hours total. \\
\hline
\hypertarget{target:F-006}{\bf F-006} & \textbf{Wide Survey}\newline {\bf Objective:} The \ac{EWS} provides inline calibration products as part of the ROS including VIS flats, biases, darks, \ac{CTI} measurements, trap pumping, and short science exposures that validate and update numerous calibration products. {\bf Observations:} Executes the \ac{EWS} with cycling VIS \ac{ROS}-inline calibrations, the cycle repeats after 20 \ac{ROS}; NISP acquires slew darks (\cref{sec:ROS}) during field slews. {\bf Cadence:} Continuous survey operations. {\bf Duration:} Not applicable, as these are the intrinsic survey operations. For reference, one ROS lasts about 4300 seconds. \\
\hline
\hypertarget{target:F-007}{\bf F-007} & \textbf{VIS nonlinearity, PTC, BFE and NISP LED flats}\newline {\bf Objective:} VIS: Monitors nonlinearity, \ac{PTC}, and \ac{BFE} using on-sky stellar fields with special dither patterns and exposure times. NISP: Acquires \ac{LED} flats at 30\,kADU fluence for monthly small-scale flat calibration and persistence monitoring. {\bf Observations:} VIS executes 16 four-exposure nonlinearity sequences plus flat/short-science pairs at the self-calibration field; NISP takes 30 photo-MACC exposures per \ac{LED} for all five \acp{LED}, followed by 4.5 hours of darks for persistence decay. {\bf Cadence:} Every 25--40 days, alternating between LED1 and LED4 variants (reference to CalF PDF excerpt). {\bf Duration:} 10.2 hours. \\
\hline
\hypertarget{target:F-008}{\bf F-008} & \textbf{VIS PSF calibration and NISP photo-nonlinearity}\newline {\bf Objective:} VIS: Calibrates and validates the \ac{PSF} model using stars in magnitude range $\IE\in[17.5;21.0]$ at five times the \ac{EWS} \ac{SNR}. NISP: Takes parallel MACC(4,16,4) nonlinearity sequences in ``engineering RAW'' mode \citep[see][]{EuclidSkyNISP} at various \ac{LED} flux levels cycling through twelve configurations. {\bf Observations:} VIS observes one of 24 \ac{PSF} calibration fields 100 times with 289\,s exposures using spiral dithering; NISP cycles through twelve \ac{LED} flux variants (8, 20, 80, 200, 450, 550, 12, 40, 150, 350, 750, 950\,e$^-$\,s$^{-1}$) taking 10--30 ramps per detector pair followed by darks for persistence decay. {\bf Cadence:} Approximately every 2--6 weeks, tied to survey thermal states to guarantee a stable \ac{PSF} for VIS. {\bf Duration:} 15 hours. \\
\hline
\hypertarget{target:F-009}{\bf F-009} & \textbf{NISP spectro-nonlinearity}\newline {\bf Objective:} Determines the nonlinearity for spectro-MACC(15,16,11) mode using \ac{LED} exposures in engineering RAW mode; validates the model through observations with different exposure times. {\bf Observations:} Validation: observes the self-calibration field in all filters with both photo-MACC(4,16,4) and photo-MACC(15,16,11). Measurement: two variants acquiring 15--30 spectro-MACC ramps per detector pair at five flux levels each (variant 1, having 8, 16, 30, 70, 150\,e$^-$\,s$^{-1}$; variant 2, having 12, 20, 50, 110, 175\,e$^-$\,s$^{-1}$) using \ac{LED} C, followed by persistence decay darks. {\bf Cadence:} Once per year for each variant, including validation. {\bf Duration:} 340.1 hours (14.2 days) for both variants. \\
\hline
\hypertarget{target:F-010}{\bf F-010} & \textbf{NISP wavelength dispersion and VIS BFE}\newline {\bf Objective:} NISP: Observes a compact \ac{PN} \citep{Paterson-EP32} with strong emission lines to obtain the field-dependent dispersion solution; transfers the solution to the self-calibration field for drift monitoring using stellar absorption features. VIS: Obtains flat-field pairs for \ac{BFE} calibration. {\bf Observations:} Establishes an astrometric reference with four ROS, then observes the \ac{PN} at 80 positions (a quincunx pattern per detector) taking two spectro-MACC exposures per position, immediately followed by a slew to the self-calibration field; the grism wheel is not moved until all exposures of the \ac{PN} and the self-calibration field are completed for a grism. VIS takes flat pairs at 10\,kADU using LED1 and LED4 during the NISP spectro exposures. {\bf Cadence:} Once after 11--16\,months of science operations. {\bf Duration:} 160.8 hours (6.7 days). \\
\hline
\hypertarget{target:F-011}{\bf F-011} & \textbf{NISP reciprocity failure}\newline {\bf Objective:} Validates reciprocity failure (or flux-dependent nonlinearity from charge trapping) for each detector by observing the self-calibration field with and without additional \ac{LED} background illumination to test trap-passivation effects. {\bf Observations:} Repeats the survey's standard S-dither pattern \citep{Scaramella-EP1} 18 times at the self-calibration field with small random offsets; odd repetitions without \ac{LED}, even repetitions with \ac{LED} D at increasing fluxes (2, 4, 8, \ldots, 512\,e$^-$\,s$^{-1}$\,pixel$^{-1}$) providing 220--56\,000\,e$^-$\,pixel$^{-1}$ additional background in \HE-band. {\bf Cadence:} Yearly, beginning one year into the survey. {\bf Duration:} 5.3 hours. \\
\hline
\hypertarget{target:F-012}{\bf F-012} & \textbf{VIS flatband voltage shift}\newline {\bf Objective:} Monitors the flatband voltage shift in \ac{CCD} \ac{MOS} capacitors caused by ionising radiation. It measures the charge-injection turn-on/off point as a function of \ac{IDL} voltage, which increases with accumulated radiation dose. {\bf Observations:} Uses a dedicated VIS commanding sequence, stepping the \ac{IDL} voltage from 10.5\,V to 12.5\,V in 0.05--0.20\,V increments (34 steps total), injecting into 100 lines and reading a windowed region after each voltage change; all data are taken with the  shutter closed. {\bf Cadence:} Every 2--6 weeks, jointly with CALBLOCK-F-008 and CALBLOCK-F-015. {\bf Duration:} 0.2 hours. \\
\hline
\hypertarget{target:F-013}{\bf F-013} & \textbf{Phase-diversity campaign (PDC)}\newline {\bf Objective:} Determines the VIS wavefront error for \ac{PSF} modelling as a function of various thermal states (\ac{AA} and \ac{SAA} combinations), observing stellar fields in defocused and focused states after thermal stabilisation. {\bf Observations:} Requires a 5--7 day thermal stabilisation at each of ten different \ac{SAA}/\ac{AA} positions; three defocused \ac{AA} values for low-polarisation fields plus one high-polarisation field; remainder varies \ac{SAA} and \ac{AA} in-focus. Executes eight ROS\_D1 for photometric classification; molecular contamination is monitored at the self-calibration field via executions of CALBLOCK-F-014; the observations include defocused shutter maps (see also \cref{sec:dr2_visshutter}). {\bf Cadence:} Once, immediately after PV-phase and before the survey starts; if necessary, scheduled again during future unallocated time. {\bf Duration:} 60--105 days total depending on 5--10 day thermal-stabilisation periods. \\
\hline
\hypertarget{target:F-014}{\bf F-014} & \textbf{Contamination scan}\newline {\bf Objective:} Provides quick photometric monitoring of the self-calibration field at high cadence to track contamination evolution during the phase diversity campaign (CALBLOCK-F-013) when thermal stability is critical. {\bf Observations:} Executes a single \ac{ROS} at the centre of the self-calibration field. {\bf Cadence:} Approximately weekly during \ac{PV} (first two weeks) and PDC; must not disturb thermal stabilisation periods. {\bf Duration:} 1.2 hours. \\
\hline
\hypertarget{target:F-015}{\bf F-015} & \textbf{VIS Serial trap pumping}\newline {\bf Objective:} Executes \ac{STP}, which requires a readout-electronics soft reset that causes a short-term bias instability; provides all twelve \ac{STP} variants for trap characterisation but must be followed immediately by \ac{PSF} calibration (CALBLOCK-F-008) as subsequent VIS images will show the bias offset for several hours; the \ac{PSF} calibration is insensitive to the transient bias instability. {\bf Observations:} Acquires a full set of twelve different \ac{STP} configurations (dwell times, shuffle parameters, etc), followed by three trash biases to clear the bias perturbation. {\bf Cadence:} Scheduled every 2--6 weeks immediately prior to CALBLOCK-F-008, then followed by CALBLOCK-F-012. {\bf Duration:} 1.9 hours plus 3 dithers. \\
\hline
\hypertarget{target:F-016}{\bf F-016} & \textbf{VIS charge-injection timing}\newline {\bf Objective:} Gains statistics on charge-injection temporal consistency and adds information about trap capture/release times by varying the charge-injection speed. {\bf Observations:} Executes a single block with six different charge-injection configurations. {\bf Cadence:} Every 5--7 months. {\bf Duration:} 2.5 hours. \\
\hline
\hypertarget{target:F-017}{\bf F-017} & \textbf{Grism sanity}\newline {\bf Objective:} This is a technical observation consisting of a single red-grism \ac{ROS}. It has to be scheduled on rare occasions to resolve a look-ahead issue concerning red/blue grism that can occur in the survey. {\bf Observations:} Executes a single red-grism \ac{ROS}. {\bf Cadence:} When needed. {\bf Duration:} 1.2 hours. \\
\hline
\hypertarget{target:F-018}{\bf F-018} & \textbf{NISP persistence}\newline {\bf Objective:} Monthly observations to update persistence parameters due to considerable time variability in the persistence behaviour. {\bf Observations:} Takes three trash darks for persistence decay from preceding observations, then repeats twice: a single \ac{LED} E exposure with a fluence of 30\,kADU (690\,e$^-$\,s$^{-1}$) followed by five MACC(4,16,4) darks and four MACC(15,16,11) darks. {\bf Cadence:} Monthly, scheduled between CALBLOCK-F-001 and CALBLOCK-F-007 to avoid affecting either and allow persistence decay before the survey resumes. {\bf Duration:} 1.8 hours. \\
\end{longtable}
\endgroup

\section{NISP common calibration products\label{sec:nisp_common_calproducts}}
\begingroup
\small
\setlength\extrarowheight{2pt}
\begin{longtable}{|>{\columncolor{red!15}}p{0.12\textwidth}|p{0.82\textwidth}|}
\caption{CalProduct: NISP Common Calibration Products in the {\tt CalF} (\cref{sec:calf}).}\label{tab:nisp-common-calproducts}\\
\hline
\textbf{ID / Cadence} & \textbf{Description} \\
\hline
\endfirsthead
\multicolumn{2}{c}{\tablename\ \thetable{} -- continued from previous page} \\
\hline
\textbf{ID / Cadence} & \textbf{Description} \\
\hline
\endhead
\hline
\multicolumn{2}{r}{\textit{Continued on next page}} \\
\endfoot
\hline
\endlastfoot
\textbf{NI-001\label{CALPRODUCT-NI-001}}\par\medskip Continuous & \textbf{NISP bad pixel map}\newline Continuously updated, aggregating information from baseline maps, dark current, flats, slew darks, and permanently disconnected pixels. Both intermittent and permanently bad pixels enter the map. \ac{RTN} pixels are currently not covered by processing, and pixels affected by persistence are masked while exceeding thresholds.\\
\hline
\textbf{NI-002\label{CALPRODUCT-NI-002}}\par\medskip Once at commissioning; once during PV & \textbf{NISP baseline map}\newline Measures the bias value for each pixel after reset and is affected by kTC noise from thermodynamically created charges in (disconnected) capacitors. Exhibits a temperature dependence of approximately $-100$\,ADU\,K$^{-1}$, with reference pixels set to 5.5\,kADU and science pixels to 10\,kADU. Updated during commissioning and PV to account for temperature changes. The baseline map ensures that the vast majority of pixels operate within the linear range of the analogue-digital chain; it is stable for each up-the-ramp acquisition and thus removed through ramp fitting. Potential baseline drifts can be detected in the slew darks in conjunction with five raw lines whose \ac{MACC} groups are always fully downlinked. \\
\hline
\textbf{NI-003\label{CALPRODUCT-NI-003}}\par\medskip Monthly & \textbf{NISP brighter-fatter effect}\newline \ac{BFE} can be characterised from flat fields, and validated by comparing observed \ac{FWHM} of stars at different magnitudes; BFE is partially masked by interpixel capacitance and charge diffusion. We presume the \ac{BFE} is constant. \\
\hline
\textbf{NI-004\label{CALPRODUCT-NI-004}} & \textbf{NISP colour-gradient calibration fields}\newline Replaced by CHAR-TARGET-001. \\
\hline
\textbf{NI-005\label{CALPRODUCT-NI-005}} & \textbf{NISP crosstalk}\newline Channel crosstalk has an amplitude of about 0.001\%, detectable only for super-saturated stars. This calibration identifies the threshold at which bright stars cause a detectable signal in the form of dark rings in neighbouring channels. Crosstalk is unproblematic in the \ac{EWS}, but masking will be needed for DR2 in the \ac{EDF}. \\
\hline
\textbf{NI-006\label{CALPRODUCT-NI-006}}\par\medskip Once during PV & \textbf{NISP dark current}\newline Determines the true inflight amplitude of the dark current, which could not be measured accurately on ground. The dark current is very low (approximately 0.02\,e$^-$/s), and requires hundreds of exposures in the two \ac{MACC} readout modes. It was measured once during PV, after permitting for sufficient persistence decay prior to the measurements. \\
\hline
\textbf{NI-007\label{CALPRODUCT-NI-007}} & \textbf{NISP intrapixel sensitivity variation}\newline IPSV was characterised on ground using spot scan measurements with flight-like detectors. In flight, a mean \ac{IPSV} profile can be determined by measuring flux variations of stable sources in the self-calibration field as a function sub-pixel position known from \textit{Gaia} astrometry. This has not yet been performed. \\
\hline
\textbf{NI-008\label{CALPRODUCT-NI-008}} & \textbf{NISP inter-pixel capacitance verification}\newline IPC is local electronic crosstalk caused by capacitive coupling between adjacent pixels that spreads charge prior to readout. This broadens the \ac{PSF} and smooths Poisson noise, requiring correction before deriving, for example, the gain. The \ac{IPC} distribution is narrow and expected to change minimally with detector position and over time \citep[see][]{LeGraet22}. It is wavelength-independent and must be distinguished from charge diffusion; the latter has a similar appearance but depends on the wavelength-dependent absorption-length of photons. \\
\hline
\textbf{NI-009\label{CALPRODUCT-NI-009}} & \textbf{NISP inoculation}\newline {\bf Deactivated.} Deep saturation of detectors has been shown to inactivate charge traps on time scales of weeks and months \citep{mcleod2016}. This `inoculation' effect has been demonstrated on ground for the NISP H2RGs. However, it is even less understood than charge persistence itself. There is a concern that massive flooding of all pixels could alter the detector's diode junction through strong electrostatic fields, causing effects not seen in normal survey data and possibly voiding other calibration data. Deliberate inoculation was therefore never attempted with NISP. \\
\hline
\textbf{NI-010\label{CALPRODUCT-NI-010}}\par\medskip Monthly & \textbf{NISP LED flats}\newline 30 flats per \ac{LED}, with a flux of 90\,e\,s$^{-1}$ yielding 4\,kADU with the standard MACC(4,16,4) readout. Higher fluences tested during PV phase resulted in poorly understood flats in some configurations, and have not been pursued further. \\
\hline
\textbf{NI-011\label{CALPRODUCT-NI-011}}\par\medskip Monthly: MACC(4,16,4); Yearly: MACC(15,16,11) & \textbf{NISP nonlinearity verification}\newline Nonlinearity data for the MACC(4,16,4) mode are obtained at twelve flux levels ranging from 8\,e\,s$^{-1}$ to 950\,e\,s$^{-1}$. The fluxes are interleaved and distributed across two blocks that are observed monthly in an alternating sequence. To sample the nonlinearity as charge accumulates in the detector, all groups in the ramp must be downloaded. Due to bandwidth limitations, only two detectors can be calibrated simultaneously, meaning we need 8 times as many exposures as normally. This makes nonlinearity calibration one of the more time-consuming in-orbit calibration activities. The nonlinearity of the MACC(15,16,11) mode is characterised in an analogous manner. We use ten flux levels between 8\,e\,s$^{-1}$ and 175\,e\,s$^{-1}$, distributed over two blocks that must be executed once per year, but not necessarily back-to-back since the nonlinearity appears to be very stable. Validation is performed by taking undispersed images in the self-calibration field with both MACC(4,16,4) and MACC(15,16,11) modes. After nonlinearity calibration, non-saturated stars within the overlapping dynamic range of the two exposure modes should have consistent fluxes within the measurement uncertainties. \\
\hline
\textbf{NI-012\label{CALPRODUCT-NI-012}}\par\medskip Once during PV (initial); monthly thereafter & \textbf{NISP persistence}\newline This calibration provides the time constants and amplitudes of charge traps. They were obtained for five different fluence levels in PV using the reddest \ac{LED} E. In the early phase of the nominal survey, it became evident that the persistence parameters change slowly depending on their illumination history, and eventually return to their nominal state. We update the persistence parameters now on a monthly basis, using a single fluence of 30\,kADU. \\
\hline
\textbf{NI-013\label{CALPRODUCT-NI-013}}\par\medskip Yearly & \textbf{NISP reciprocity failure}\newline Reciprocity failure describes flux-nonlinearity from charge-trapping effects. Basically, a brighter source fills traps more quickly and therefore enters a linear regime sooner. We calibrate this on a yearly basis by observing the self-calibration field without and with additional, different light levels from \ac{LED} D. \\
\hline
\hypertarget{target:NI-014}{\textbf{NI-014\label{CALPRODUCT-NI-014}}}\par\medskip Per field slew & \textbf{NISP slew darks}\newline A single slew dark is taken during field slews. It provides continuous monitoring of detector performance throughout the mission without dedicated observing time. Slew darks are contaminated by persistence from both the zodiacal background, cosmic ray hits, and astrophysical sources in the previous exposures. While median combination removes the persistence from compact sources, the spatially uniform zodiacal contribution remains. \\
\hline
\textbf{NI-015\label{CALPRODUCT-NI-015}} & \textbf{NISP snowballs}\newline Snowballs are highly energetic events that appear as radially symmetric or elliptic, extended features that may saturate the detectors. They likely originate from energetic cosmic rays, and possibly radioactive decay in or near the detectors. Snowballs may be surrounded by a number a numbers of hot pixels. Affected pixels need to be masked in consecutive exposures. This is not a calibration product as such, and was introduced to track this effect. \\
\end{longtable}
\endgroup

\section{NISP-P calibration products\label{sec:nisp-p-calproducts}}

\begingroup
\small
\setlength\extrarowheight{2pt}
\begin{longtable}{|>{\columncolor{yellow!15}}p{0.18\textwidth}|p{0.76\textwidth}|}
\caption{NISP-P calibration products in the {\tt CalF} (\cref{sec:calf}).}\label{tab:nisp-p-calproducts}\\
\hline
\textbf{ID / Cadence} & \textbf{Description} \\
\hline
\endfirsthead
\multicolumn{2}{c}{\tablename\ \thetable{} -- continued from previous page} \\
\hline
\textbf{ID / Cadence} & \textbf{Description} \\
\hline
\endhead
\hline
\multicolumn{2}{r}{\textit{Continued on next page}} \\
\endfoot
\hline
\endlastfoot
\textbf{NP-001\label{CALPRODUCT-NP-001}}\par\medskip Once during PV; optional repeats in routine phase & \textbf{Absolute photometric standards}\newline This calibration obtains absolute flux standard measurements across the NISP focal plane by observing spectro-photometric standard stars on a $4\times4$ grid centred on each detector. The observations characterise the entire optical path and provide independent verification of global system response, including persistence and trap filling effects. Executed once during PV-phase with high spatial sampling using \YE, \JE, and \HE filters, with optional low-sampling repeats during routine phase if photometric discrepancies arise. \\
\hline
\textbf{NP-002\label{CALPRODUCT-NP-002}}\par\medskip Monthly & \textbf{Focal plane astrometric solution}\newline This product determines relative detector positions, orientations, and optical distortion by comparing source positions against \textit{Gaia} astrometric references in multiply dithered self-calibration observations. While optical distortion is expected to remain stable, relative detector positions may drift due to outgassing, mechanical relaxation, and thermal effects, requiring monthly monitoring through self-calibration observations with approximately 140 unsaturated \textit{Gaia} stars per detector providing sufficient accuracy. \\
\hline
\textbf{NP-003\label{CALPRODUCT-NP-003}}\par\medskip Monthly & \textbf{Large-Scale Flat}\newline Monthly reconstruction of matching broad-band large-scale flat fields for \YE, \JE, and \HE filters using stellar photometry from self-calibration observations. These flats remove residual large-scale non-uniformity in the \ac{LED} flats and any large-scale illumination effects from the payload module, ensuring photometric consistency across the focal plane. \\
\hline
\textbf{NP-004\label{CALPRODUCT-NP-004}} & \textbf{Dichroic ghost model}\newline In-orbit validation of the ground-based dichroic ghost model (based on {\tt Zemax} simulations) to confirm location and amplitude of ghosts requiring masking for sources brighter than 14.6\,AB\,mag with amplitude ratio of $8\times10^{-8}$. While size, shape, and wavelength dependence are expected to match simulations, any inaccuracies will be evident in masked pixel data allowing model updates, with optical ghosts from NISP internal optics considered negligible. \\
\hline
\textbf{NP-005\label{CALPRODUCT-NP-005}} & \textbf{Photo-z calibration fields}\newline Replaced by CHAR-TARGET-002, this involved observing five fields (COSMOS, VVDS-Deep, CDFS, SXDS) to obtain NISP-P photometry in \YE, \JE, and \HE bands for 20\,000 galaxies with existing spectroscopy at 5 times the \ac{EWS} \ac{SNR}. VIS data at minimum two times the \ac{EWS} depth aids target detection and deblending, with all observations conducted using the ROS for practical and legacy purposes, achieving 5 times \ac{EWS} depth for both instruments. \\
\hline
\textbf{NP-006\label{CALPRODUCT-NP-006}} & \textbf{PN reference photometric catalogue}\newline This product provides a reference photometric catalogue supporting NISP-S spectral observations of planetary nebulae, particularly for determining wavelength zeropoints in the spectral dispersion solution. Observations use the ROS at each dither position across five positions per detector for each grism, with NISP-P data used to determine accurate stellar positions since the bright target will be saturated in NISP images. \\
\hline
\textbf{NP-007\label{CALPRODUCT-NP-007}}\par\medskip Monthly & \textbf{PSF model}\newline Monthly calibration and validation of the NISP \ac{PSF} model using stars in the 16--21\,AB\,mag range, with approximately 12\,000 stars per detector per filter after 60 dithers in the self-calibration field providing sufficient sampling. The \ac{PSF} model must be robust against undersampling and intra-pixel sensitivity variations to achieve position accuracy better than 0.1 pixel for brighter NISP sources, though no formalised requirements currently exist. \\
\hline
\textbf{NP-008\label{CALPRODUCT-NP-008}}\par\medskip Once during PV & \textbf{Straylight check}\newline Once during PV-phase, this calibration tests for glints (parasitic reflections from mechanical features like detector edges) and validates the normalised detector irradiance (NDI) model for unbaffled scattered light. The glint check places a medium-bright star (around 6\,AB\,mag) adjacent to the focal plane using the ROS\_D1, while NDI validation places very bright stars (0--2 AB mag) at logarithmically decreasing separations from the focal-plane edge to verify the bright-star avoidance radius.\\
\hline
\textbf{NP-009\label{CALPRODUCT-NP-009}}\par\medskip Monthly & \textbf{Small-scale flat}\newline Monthly reconstruction of matching broad-band small-scale flat fields for \YE, \JE, and \HE filters from NISP \ac{LED} flats using ground-based quantum efficiency maps. These flats correct \ac{PRNU} and small-scale features in photometric data, complementing the large-scale flats derived from self-calibration observations. \\
\hline
\textbf{NP-010\label{CALPRODUCT-NP-010}} & \textbf{Zodiacal model}\newline Replaced by CHAR-SURV-004, this involved updating the zodiacal background model based on routine wide survey fields sampling the regularly observed ecliptic longitude and latitude range. While large deviations are not expected given well-characterised zodiacal backgrounds from other missions, validation ensures backgrounds aren't substantially higher than predicted, which could impact survey depth and require operational adjustments. \\
\hline
\textbf{NP-011\label{CALPRODUCT-NP-011}} & \textbf{Colour-gradient calibration fields}\newline Replaced by CHAR-TARGET-001. \\
\end{longtable}
\endgroup

\section{NISP-S calibration products\label{sec:nisp-s-calproducts}}
\begingroup
\small
\setlength\extrarowheight{2pt}
\begin{longtable}{|>{\columncolor{orange!15}}p{0.14\textwidth}|p{0.80\textwidth}|}
\caption{NISP-S calibration products in the {\tt CalF} (\cref{sec:calf}).}\label{tab:nisp-s-calproducts}\\
\hline
\textbf{ID / Cadence} & \textbf{Description} \\
\hline
\endfirsthead
\multicolumn{2}{c}{\tablename\ \thetable{} -- continued from previous page} \\
\hline
\textbf{ID / Cadence} & \textbf{Description} \\
\hline
\endhead
\hline
\multicolumn{2}{r}{\textit{Continued on next page}} \\
\endfoot
\hline
\endlastfoot
\textbf{NS-001\label{CALPRODUCT-NS-001}} & \textbf{0th-order centroid SED dependence}\newline This calibration characterises how source \acp{SED} affect the 0th-order centroid position, which appears as a double-peaked peanut-like shape due to prism dispersion and grating blaze effects. Since the main \ac{SED} impact changes the ratio of the two peaks (and thus the inferred mean position), this dependence must be quantified using wide survey data as sources with different \acp{SED} are mapped at different \ac{FPA} positions, with no immediate PV calibration required. \\
\hline
\textbf{NS-002\label{CALPRODUCT-NS-002}}\par\medskip Once during PV & \textbf{Absolute photometric standards}\newline Absolute flux standard measurements obtained once during PV by observing \ac{HST} CALSPEC spectro-photometric standards as a function of detector position for each grism, with subsequent self-calibration field monitoring providing relative throughput. The star is stepped across five positions per detector using ROS\_D1 observations. The grisms are scheduled to minimise spectra order contamination from other sources. \\
\hline
\textbf{NS-003\label{CALPRODUCT-NS-003}}\par\medskip $\geq$ every 6 months (self-calibration); once during PV and once after 1\,yr (PN) & \textbf{Focal-plane distortion}\newline This calibration determines the distortion pattern across the NISP spectroscopic focal plane through three mappings: sky (\textit{Gaia} references) to 0th-order using bright stars; 0th-order to reference wavelength in the 1st order; and reconstruction of the wavelength dispersion solution from \ac{PN}observations. Monitoring occurs at least every six months using secondary wavelength calibrators (absorption line systems) in the self-calibration field, with full \ac{PN}-based wavelength dispersion determined once during PV and once after one year. \\
\hline
\textbf{NS-004\label{CALPRODUCT-NS-004}}\par\medskip Monthly & \textbf{Large-scale flat}\newline Monthly reconstruction of matching spectroscopic large-scale flat fields for all four grisms from self-calibration observations. These flats correct for large-scale illumination variations and vignetting effects across the focal plane, ensuring consistent flux calibration across the spectroscopic field of view. \\
\hline
\textbf{NS-005\label{CALPRODUCT-NS-005}}\par\medskip Monthly & \textbf{Small-scale flat}\newline Monthly reconstruction of matching spectroscopic small-scale flat fields for all four grisms from NISP \ac{LED} flats using ground-based characterisation data. These flats correct pixel-to-pixel response variations and small-scale sensitivity features in the spectroscopic data, complementing the large-scale corrections. \\
\hline
\textbf{NS-006\label{CALPRODUCT-NS-006}}\par\medskip Every 6 months & \textbf{Spectral trace calibration}\newline This product determines the relative positions of 0th, 1st, and 2nd spectral orders as a function of detector position, monitored every 6\,months using self-calibration observations. The trace calibration ensures accurate extraction of spectroscopic data by characterising how spectral orders are positioned and oriented across the \ac{FPA}. \\
\hline
\textbf{NS-007\label{CALPRODUCT-NS-007}}\par\medskip Monthly & \textbf{Wavelength dispersion check}\newline Monthly validation of the wavelength dispersion model using secondary wavelength calibrators (stellar absorption-line systems) in the self-calibration field, identified during PV-phase when primary dispersion was determined from \ac{PN} observations. While full dispersion solutions may not be derivable from secondary standards due to insufficient spatial density, they can reveal considerable drifts indicating need for \ac{PN}-based recalibration. \\
\hline
\textbf{NS-008\label{CALPRODUCT-NS-008}}\par\medskip Once during PV; once after 1\,yr & \textbf{Wavelength dispersion with Planetary Nebulae}\newline This calibration observes compact PNe with strong emission lines at 80 positions across each detector to obtain wavelength dispersion solutions for all grisms, executed once during PV and once after one year. The dispersion solution is immediately transferred to the self-calibration field without intermediate grism-wheel movements to maintain thermal stability, establishing secondary standards for routine drift monitoring.\\
\hline
\textbf{NS-009\label{CALPRODUCT-NS-009}} & \textbf{Wavelength Transfer to self-calibration field}\newline The wavelength dispersion established using the \ac{PN} observations is transferred to the self-calibration field by identifying suitable secondary standards (point-like emission-line galaxies and stellar absorption-line systems) for regular monitoring. Self-calibration observations must immediately follow or precede planetary nebula observations to avoid thermal state changes that could alter the wavelength dispersion before transfer. \\
\hline
\textbf{NS-010\label{CALPRODUCT-NS-010}} & \textbf{Completeness and Purity Calibration}\newline Replaced by CHAR-TARGET-004. \\
\hline
\textbf{NS-011\label{CALPRODUCT-NS-011}}\par\medskip Once during PV & \textbf{Dichroic Ghost Model}\newline Once during PV phase, this validates the on-ground dichroic ghost model (based on {\tt Zemax} simulations) by confirming ghost location and amplitude for sources requiring masking, with amplitude ratio requirement to be determined ($8\times10^{-8}$ for NISP-P provides reference). While size, shape, and wavelength dependence are expected to match simulations, any inaccuracies will be evident in masked pixel data allowing model updates, with optical ghosts from NISP internal optics considered negligible. \\
\hline
\textbf{NS-012\label{CALPRODUCT-NS-012}}\par\medskip Once during PV & \textbf{Grism Centers of Rotation}\newline This calibration determines the grism-wheel rotation centres by observing a stellar field five times per grism setup with independent random realisations of the jitter angle (typically \ang{0.1;;} \ac{RMS}, occasionally up to \ang{0.5;;}), executed once during PV. While the dispersion pattern does not purely rotate due to optics chromaticity, to first order in small jitter angles the pattern can be derived from reference calibration as simple rotation, requiring full grism-wheel rotation between exposures to ensure independent angle realisations. \\
\end{longtable}
\endgroup

\section{VIS calibration products\label{sec:vis-calproducts}}
\begingroup
\small
\setlength\extrarowheight{2pt} 
\begin{longtable}{|>{\columncolor{blue!20}}p{0.18\textwidth}|p{0.76\textwidth}|}
\caption{VIS calibration products in the {\tt CalF}  (\cref{sec:calf}).}\label{tab:vis-calproducts}\\
\hline
\textbf{ID / Cadence} & \textbf{Description} \\
\hline
\endfirsthead
\multicolumn{2}{c}{\tablename\ \thetable{} -- continued from previous page} \\
\hline
\textbf{ID / Cadence} & \textbf{Description} \\
\hline
\endhead
\hline
\multicolumn{2}{r}{\textit{Continued on next page}} \\
\endfoot
\hline
\endlastfoot
\textbf{VS-001\label{CALPRODUCT-VS-001}}\par\medskip Once during PV; optional low-sampling repeats in routine phase & \textbf{Absolute photometric standards}\newline Obtains absolute flux calibrations across the VIS focal plane by observing spectro-photometric standards on a $4\times4$ grid centred on each NISP detector (yielding approximately seven positions per VIS \ac{CCD} due to different \ac{FPA} layouts). It characterises the entire optical path and checks system response including persistence and trap filling. Executed once during PV with high spatial sampling using short-science exposures to avoid saturation, with optional low-sampling repeats during routine phase should photometric discrepancies arise.\\
\hline
\textbf{VS-002\label{CALPRODUCT-VS-002}} & \textbf{Passband variations}\newline Deactivated. Determines dichroic passband variations from stellar observations with known \acp{SED}, though VIS passband variations are quantified on ground using optical raytracing and dichroic reflectance measurements.\\
\hline
\textbf{VS-003\label{CALPRODUCT-VS-003}}\par\medskip Once early in PV; monthly in routine ops & \textbf{Bias}\newline The VIS detector bias level may vary over time and exhibits spatial variations at readout beginning (bias stabilisation), with the spatial pattern expected to be stable within each quadrant and accurately computed from series of individual bias exposures. One master bias is created early in PV-phase, with routine operations acquiring bias frames during the \ac{ROS} for monthly master bias computation.\\
\hline
\textbf{VS-004\label{CALPRODUCT-VS-004}}\par\medskip Once during PV & \textbf{Brighter-fatter effect}\newline This calibration characterises the \ac{BFE} where increasing charges in pixel potential wells change physical pixel boundaries, causing photons arriving late in exposures to create photo-electrons in neighbouring pixels. This is detectable as increasing stellar \ac{FWHM} with flux and decreasing pixel-to-pixel variance in high-fluence flats. Executed once during PV using self-calibration stellar field observations and flat-field pairs, this achromatic effect directly impacts \acp{PSF} models and shear measurements and may be subject to radiation damage.\\
\hline
\textbf{VS-005\label{CALPRODUCT-VS-005}}\par\medskip 6$\times$ early in PV; every 1--2 weeks during PV; daily in routine ops & \textbf{CTI charge-injection lines}\newline This calibration provides the \ac{CTI} model for correcting charge transfer inefficiency due to radiation damage, based on \ac{CIL} data and flat fields. The full sequence is executed six times early in PV with 42 \ac{CIL} patterns plus once every 1--2 weeks during PV, while daily routine operations acquire eight \ac{CIL} images through the \ac{ROS} with 32 different patterns cycling completely over approximately four days.\\
\hline
\textbf{VS-006\label{CALPRODUCT-VS-006}} & \textbf{Colour-gradient calibration fields}\newline Replaced by CHAR-TARGET-001.\\
\hline
\textbf{VS-007\label{CALPRODUCT-VS-007}}\par\medskip Monthly & \textbf{Cosmetic map}\newline Monthly update of the cosmetic map/bad pixel map based on various data sources acquired during routine operations and dedicated calibrations. This continuously maintained product tracks detector defects, hot pixels, dead pixels, and other cosmetic issues affecting data quality.\\
\hline
\textbf{VS-008\label{CALPRODUCT-VS-008}}\par\medskip Monthly & \textbf{Crosstalk}\newline Monthly determination and updating of the VIS crosstalk matrix from self-calibration and wide-survey observations, allowing correction of crosstalk from unsaturated stars and masking of cross-talk from saturated sources in the science pipeline. The calibration monitors electronic crosstalk between detector readout channels and components.\\
\hline
\textbf{VS-009\label{CALPRODUCT-VS-009}}\par\medskip Once during PV; daily in routine ops & \textbf{Darks}\newline The VIS dark current is expected to be extremely low ($\leq0.01$\,e$^{-}$\,s$^{-1}$\,pix$^{-1}$) and could not be fully characterised on ground, though tests showed $<0.1$ ADU over 350\,s integration. Measured once during PV with routine operations acquiring four darks daily ($2\times325$\,s, $1\times200$\,s, $1\times95$\,s) through the \ac{ROS}, these exposures primarily measure stray sunlight (which never completely disappears even at operational attitudes). They also probe cosmic-ray flux, removal routines, and are used to update the cosmetic map.\\
\hline
\textbf{VS-010\label{CALPRODUCT-VS-010}}\par\medskip Once (stitch alignment); monthly (self-calibration positions); during PV (distortion) & \textbf{Focal-plane astrometric solution}\newline This calibration has three goals: determining relative stitch block alignment ($256\times512$ pixel blocks, 128 per detector, measured once using $\geq1000$--2000 \textit{Gaia} stars per detector at 25--50\,arcmin$^{-2}$ density); determining relative detector positions/orientations subject to drift (monitored monthly via self-calibration); determining the optical distortion with high angular resolution during PV. The astrometric solution may vary due to detector drift, mechanical relaxation, and thermal effects.\\
\hline
\textbf{VS-011\label{CALPRODUCT-VS-011}}\par\medskip Weekly in routine phase; twice during PV (4 weeks apart) & \textbf{Photon transfer curve}\newline Weekly calibration during routine phase obtaining the \ac{PTC} for every detector quadrant to determine the gain, executed twice during PV phase four weeks apart. The \ac{PTC} curve is affected by electronic nonlinearity (independent of differential nonlinearity) and can yield nonlinearity estimates for comparison with primary methods from on-sky and flat-field measurements.\\
\hline
\textbf{VS-012\label{CALPRODUCT-VS-012}}\par\medskip Once during PV & \textbf{Dichroic ghost model}\newline Once during PV phase, this validates the ground-based dichroic ghost model (based on {\tt Zemax} simulations and \ac{PLM} end-to-end tests) by confirming ghost location and amplitude for sources brighter than 15 AB mag, with the main ghost source being dichroic backside reflection having amplitude ratio requirement $<2\times10^{-6}$. While size, shape, and wavelength dependence are expected to match simulations, any inaccuracies will be evident in masked pixel data, with focal plane reflection ghosts already stopped by design and NISP reflection ghosts about four orders of magnitude lower.\\
\hline
\textbf{VS-013\label{CALPRODUCT-VS-013}}\par\medskip Monthly & \textbf{Large-scale flat}\newline Monthly reconstruction of matching broad-band large-scale flats for VIS from self-calibration stellar photometry, removing simultaneously any residual large-scale non-uniformity in \ac{LED} flats and large-scale illumination effects from baffle vignetting, mirror edges, coating variations on folding mirrors, molecular contamination, VIS shutter map, and apodisation. The large-scale flat characterises illumination variations that cannot be captured by the \ac{LED} flat-field system alone.\\
\hline
\textbf{VS-014\label{CALPRODUCT-VS-014}}\par\medskip Once during PV; regularly in routine ops & \textbf{LED pixel response non-uniformity}\newline This calibration obtains zero-signal \ac{PRNU} for each of six \ac{LED} wavelengths by extrapolation to zero signal or brighter-fatter effect correction, executed once during PV phase and regularly during \ac{ROS} routine operations. The calibration flags cold/dead pixels and hot pixels while characterizing pixel-to-pixel response variations across the focal plane using the VIS calibration unit's monochromatic \acp{LED}.\\
\hline
\textbf{VS-015\label{CALPRODUCT-VS-015}} & \textbf{Noise bias calibration fields}\newline Replaced by CHAR-TARGET-003.\\
\hline
\textbf{VS-016\label{CALPRODUCT-VS-016}} & \textbf{Nonlinearity}\newline This calibration characterises the nonlinearity of the VIS \ac{CCD} response over the full dynamic range. It validates and updates the ground-derived nonlinearity correction using in-flight flat-field data acquired at multiple fluence levels. Accurate nonlinearity correction is essential to preserve photometric accuracy and PSF stability.\\
\hline
\textbf{VS-017\label{CALPRODUCT-VS-017}} & \textbf{Phase diversity}\newline This product derives wavefront-error information from focussed and defocussed images. The resulting phase diversity analysis constrains low-order optical aberrations and supports the validation of the VIS \ac{PSF} model. It also provides diagnostic input for monitoring optical stability over time.\\
\hline
\textbf{VS-018\label{CALPRODUCT-VS-018}} & \textbf{Extended pixel edge response (CTI-EPER)}\newline This calibration measures charge trailing caused by \ac{CTI} using extended pixel-edge response (EPER) techniques. The data quantify deferred charge and trap-related effects that impact photometry, astrometry, and shape measurements. The results are used to validate and refine the VIS \ac{CTI} correction models.\\
\hline
\textbf{VS-019\label{CALPRODUCT-VS-019}} & \textbf{PSF calibration}\newline This product provides an empirical characterisation of the VIS \ac{PSF} using stellar observations. It captures spatial and temporal \ac{PSF} variations across the \ac{FPA} under nominal observing conditions.\\
\hline
\textbf{VS-020\label{CALPRODUCT-VS-020}} & \textbf{Blooming threshold}\newline This calibration determines the signal level at which blooming occurs, using high-fluence flat fields and, optionally, stellar observations.\\
\hline
\textbf{VS-021\label{CALPRODUCT-VS-021}} & \textbf{Straylight check}\newline This product assesses the level of straylight contamination in VIS images. It checks for glints from stars falling onto physical features in the \ac{FPA}, such as detector or baffle edges. It also assesses the \ac{NDI} reaching the detectors from in-field and out-field sources.\\
\hline
\textbf{VS-022\label{CALPRODUCT-VS-022}} & \textbf{Small-scale flat field}\newline This calibration provides the \ac{PRNU} and other small-scale variations that are not captured by the \ac{LSF}. The \ac{PRNU} map is computed from different \ac{LED} flats to be representative for the broad VIS passband.\\
\hline
\textbf{VS-023\label{CALPRODUCT-VS-023}} & \textbf{Zodiacal model}\newline Replaced by CHAR-SURV-004.\\
\hline
\textbf{VS-024\label{CALPRODUCT-VS-024}} & \textbf{Flatband voltage shift}\newline This calibration assesses radiation damage in the detectors' metal-oxide semiconductors, leading to a shift in the flat-band voltage.\\
\hline
\textbf{VS-025\label{CALPRODUCT-VS-025}} & \textbf{Trap pumping}\newline This product identifies and characterises charge traps using trap-pumping techniques. It measures trap densities, release timescales, and spatial distributions. The results feed directly into \ac{CTI} correction models and radiation damage monitoring.\\
\hline
\textbf{VS-026\label{CALPRODUCT-VS-026}} & \textbf{CTI-induced covariance}\newline This calibration quantifies pixel-to-pixel noise correlations induced by \ac{CTI} effects. Such covariances impact shape measurements and noise modelling if left uncorrected.\\
\hline
\textbf{VS-027\label{CALPRODUCT-VS-027}} & \textbf{Persistence}\newline This determines any residual signal persistence following illumination by bright sources. It measures persistence amplitudes and decay timescales under in-flight conditions.\\
\hline
\textbf{VS-028\label{CALPRODUCT-VS-028}} & \textbf{Shuttermap}\newline Provides a map of the effective exposure-time modulation by the VIS blade shutter. This map is required to separate mechanical from optical effects in the \ac{LSF}, and to achieve consistent photometric zeropoints across exposures with different integration time.\\
\end{longtable}
\endgroup

\end{appendix}

\end{document}